\documentclass[12pt]{article}
\usepackage{amssymb,amsmath}
\makeatletter

\usepackage{arydshln}

\newcount\comment@nesting

\begingroup
\xdef\comment@begincomment{\string\\begin\string\{comment\string\}}
\xdef\comment@endcomment{\string\\end\string\{comment\string\}}
\endgroup

\begingroup
\catcode`\^^M=\active
\def\@temp{\endgroup\def\comment@processline##1^^M}%
\@temp{%
    \def\comment@curline{#1}%
    \let\@next=\comment@processline
    \ifx\comment@curline\comment@endcomment
        \ifnum\comment@nesting=0
            \def\@next{\end{comment}}%
        \else
            \global\advance\comment@nesting by -1
        \fi
    \else
        \ifx\comment@curline\comment@begincomment
            \global\advance\comment@nesting by 1
        \fi
    \fi
    \@next
}

\makeatother

\usepackage[dvipdfmx]{graphicx}
\usepackage{bm}
\usepackage{color}
\usepackage{here}
\usepackage{enumerate}
\usepackage{here}
\usepackage{subfigure}
\usepackage{tikz}
\usepackage[T1]{fontenc}
\usepackage{hyperref}

\newcommand{\Zb}{\mathbb{Z}}

\newcommand{\Nb}{\mathbb{N}}
\newcommand{\Acal}{\mathcal{A}}
\newcommand{\Ccal}{\mathcal{C}}
\newcommand{\Dcal}{\mathcal{D}}
\newcommand{\Mcal}{\mathcal{M}}
\newcommand{\Ncal}{\mathcal{N}}
\newcommand{\Wcal}{\mathcal{W}}

\DeclareMathOperator*{\Tr}{{\rm Tr}}

\newcommand{\II}{\mathbb{II}}

\newcommand{\Pf}{\operatorname{Pf}}

\numberwithin{equation}{section}

\allowdisplaybreaks[1]

\definecolor{mygreen}{rgb}{0,0.714,0.286}

\begin{document}

\thispagestyle{empty}
\begin{flushright}

\end{flushright}
\vskip1.5cm
\begin{center}
{\Large \bf 
Boundary confining dualities
\\
\vskip0.75cm
and Askey-Wilson type $q$-beta integrals}

\vskip1.5cm
Tadashi Okazaki\footnote{tokazaki@seu.edu.cn}

\bigskip
{\it Shing-Tung Yau Center of Southeast University,\\
Yifu Architecture Building, No.2 Sipailou, Xuanwu district, \\
Nanjing, Jiangsu, 210096, China
}

\bigskip
and
\\
\bigskip
Douglas J. Smith\footnote{douglas.smith@durham.ac.uk}

\bigskip
{\it Department of Mathematical Sciences, Durham University,\\
Upper Mountjoy, Stockton Road, Durham DH1 3LE, UK}

\end{center}

\vskip1cm
\begin{abstract}
We propose confining dualities of $\mathcal{N}=(0,2)$ half-BPS boundary conditions in 3d $\mathcal{N}=2$ supersymmetric $SU(N)$, $USp(2n)$ and $SO(N)$ gauge theories. Some of these dualities have the novel feature that one (anti)fundamental chiral has Dirichlet boundary condition while the rest have Neumann boundary conditions. While some of the dualities can be extended to 3d bulk dualities, others should be understood intrinsically as 2d dualities as they seem to hold only at the boundary. 
The gauge theory Neumann half-indices are well-defined even for theories which contain monopole operators with non-positive scaling dimensions and they are given by Askey-Wilson type $q$-beta integrals. As a consequence of the confining dualities, new conjectural identities of such integrals are found. 
\end{abstract}

\newpage
\setcounter{tocdepth}{3}
\tableofcontents

\section{Introduction and conclusion}
\label{sec_intro}

Confining duality is a phenomenon where 
the infrared physics of a gauge theory can be described in terms of gauge invariant composites and their interactions. 
Typically in a confining duality the matter fields of the dual theory transforming in rank-$2$ tensor representations of the flavor symmetry group can be viewed as mesons of a gauge theory, while higher rank representations correspond to baryons. 
Confining dualities are found 
in several supersymmetric gauge theories including 
4d $\mathcal{N}=1$ supersymmetric gauge theories 
\cite{Intriligator:1995ne,Berkooz:1995km,Pouliot:1995me,Luty:1996cg,Csaki:1996sm,Csaki:1996zb,Terning:1997jj,Garcia-Etxebarria:2012ypj,Garcia-Etxebarria:2013tba,Etxebarria:2021lmq,Bajeot:2022lah,Bajeot:2022kwt,Bottini:2022vpy}, 
3d $\mathcal{N}=2$ supersymmetric gauge theories 
\cite{Amariti:2015kha,Nii:2016jzi,Pasquetti:2019uop,Pasquetti:2019tix,Benvenuti:2020gvy,Benvenuti:2021nwt,Bajeot:2022lah} 
and 2d $\mathcal{N}=(0,2)$ supersymmetric gauge theories \cite{Sacchi:2020pet}. 
Confining dualities are useful to construct a sequence of dual theories and to find an essential or basic part of dualities. 

In this paper we propose new confining dualities for 3d $\mathcal{N}=2$ gauge theories with $\mathcal{N}= (0,2)$ boundary conditions. 
While several confining dualities can be obtained from Seiberg-like dualities 
\cite{
Aharony:1997gp,
Giveon:2008zn,Niarchos:2008jb,
Kapustin:2011vz,Kapustin:2011gh,Benini:2011mf,Hwang:2011ht,Aharony:2011ci,
Aharony:2013dha,Park:2013wta,Aharony:2013kma,Kim:2013cma,Csaki:2014cwa,
Aharony:2014uya,Nii:2014jsa,Amariti:2014lla,
Hwang:2015wna,
Amariti:2017gsm,Benini:2017dud, 
Hwang:2018uyj,Amariti:2018wht, Amariti:2018gdc,
Nii:2019qdx,Nii:2019wjz,
Benvenuti:2020wpc,Nii:2020ikd,Nii:2020xgd,Amariti:2020xqm,Nii:2020eui,
Kubo:2021ecs,
Hwang:2022jjs,Amariti:2022iaz,Amariti:2022lbw,
Amariti:2022iaz} 
and tested by computing the IR protected data, e.g.\ superconformal index (or full-index) \cite{Bhattacharya:2008zy, Kim:2009wb, Imamura:2011su, Kapustin:2011jm, Dimofte:2011py}, 
our strategy is to consider the theories in the presence of a boundary with boundary conditions preserving the gauge group and to study the half-indices 
which enumerate the gauge invariant BPS local operators obeying the half-BPS $\mathcal{N}=(0,2)$ boundary conditions \cite{Gadde:2013wq,Okazaki:2013kaa,Gadde:2013sca,Yoshida:2014ssa,Dimofte:2017tpi,Brunner:2019qyf,Costello:2020ndc,Sugiyama:2020uqh,Alekseev:2022gnr,Dedushenko:2022fmc}. 
These half-indices \cite{Gadde:2013wq,Gadde:2013sca,Yoshida:2014ssa,Dimofte:2017tpi} are powerful tools to test the dualities of $\mathcal{N}=(0,2)$ supersymmetric boundary conditions. While the full-indices are not well-behaved for the theories involving monopole operators with non-positive dimensions, the half-indices encoding the Neumann boundary conditions for gauge fields are well-defined even for such theories. We propose confining dualities of $\mathcal{N}=(0,2)$ boundary conditions for 3d $\mathcal{N}=2$ $SU(N)$, $USp(2n)$ and $SO(N)$ gauge theories from several identities of half-indices as well as the precise matching of the boundary 't Hooft anomalies. We refer to them as \textit{boundary confining dualities}. 
It turns out that the Neumann half-indices of 3d $\mathcal{N}=2$ gauge theories which have confining descriptions take the form of Askey-Wilson type $q$-beta integrals \cite{MR783216}. From these integrals we conjecture new boundary confining dualities by checking the agreement of the corresponding half- and full-indices. Some of these cases are standard examples of 3d Seiberg-like dualities in the bulk while other cases correspond to 4d Seiberg-like dualities compactified to 3d, specifically those discussed in subsections \ref{sec_GAW_SUN_integrals}, \ref{sec_GNR_USp6_2AS_integral}, \ref{sec_GR_integral_AS}, \ref{sec_G_integral_ASAS} and \ref{sec_New_SUN_integral_AS}. 
They include examples of Seiberg-like boundary dualities with different boundary conditions for chiral multiplets transforming in the same representation of the gauge group -- as far as we are aware the only similar cases of mixed boundary conditions have been considered for SQED \cite{Dimofte:2017tpi} -- 
as well as dualities of theories with chiral multiplets in the rank-$2$ antisymmetric representation of the gauge group. 
However, in the cases considered in sections \ref{sec_GNR_USp4_3AS_integral}, \ref{sec_GNR_USp6_2AS_integral}, \ref{sec_GAW_SO_integrals} and \ref{sec_SON_Nm2_discrete_integrals} the bulk theories have dimension-zero monopoles and chiral multiplets of non-positive dimensions, so we have examples where there is only a boundary duality. Similarly, in the case of section~\ref{sec_SON_N_integrals}, although the bulk monopoles would have positive dimension it is not possible to choose the R-charges so that all chirals have positive dimension.

Our results demonstrate the interplay between physics and mathematics. On the one hand we rephrase mathematically proven identities as half-index identities, giving an interpretation as known or conjectured new dualities of 3d $\mathcal{N}=2$ theories with boundaries. On the other hand, similar dualities are proposed and the corresponding matching of half-indices leads to new mathematical identities which can be proven using similar techniques to the previously known generalized Askey-Wilson identities. Therefore, as 
a result of the expected confining dualities, we also find new conjectural identities of Askey-Wilson type $q$-beta integrals.

\subsection{Structure}
\label{sec_structure}
The paper is organized as follows. 
In section \ref{sec_Askey_Wilson} 
we discuss the boundary confining duality of Neumann boundary conditions for the $SU(2)$ gauge theory with four flavors. 
The half-index realizes the Askey-Wilson $q$-beta integral \cite{MR783216} and the duality stems from the Seiberg-like dualities \cite{Aharony:2013dha,Park:2013wta}. In section \ref{sec_NR_integral} we propose the boundary confining duality for the $SU(2)$ gauge theory with six fundamental chirals where one of the fundamental chirals has Dirichlet boundary condition while the others have Neumann boundary condition. The half-index is identified with the Nassrallah-Rahman integral \cite{MR772878,MR845667} and as we discuss, the duality also holds in the bulk. In section \ref{sec_G_integrals} we discuss the higher-rank generalization of sections \ref{sec_Askey_Wilson} and \ref{sec_NR_integral}. The half-indices are identified with integrals studied by Gustafson \cite{MR1139492}. Consequently, we propose new confining dualities for $SU(N)$, $USp(2n)$ and $SO(N)$ gauge theories, including cases with rank-$2$ antisymmetric chirals. 
In addition, we propose boundary confining dualities for specific $USp(4)$ and $USp(6)$ gauge theories with one or more rank-$2$ antisymmetric chiral multiplets. Except for the boundary confining dualities in subsections \ref{sec_GNR_USp4_3AS_integral}, \ref{sec_GNR_USp6_2AS_integral}, \ref{sec_GAW_SO_integrals} and z\ref{sec_SON_Nm2_discrete_integrals}, they can be generalized to bulk confining dualities. In section \ref{sec_GR_integral} we find the confining dualities of more general $SU(N)$ and $USp(2n)$ gauge theories with fundamental and rank-$2$ antisymmetric chirals. The Neumann half-indices can be identified with the Gustafson-Rakha integrals \cite{MR1266569}. In section \ref{sec_New_integral_identities} we find new Askey-Wilson type $q$-beta integrals which arise from boundary confining dualities. 

\subsection{Future works}
\label{sec_future}
\begin{itemize}

    \item It would be interesting to generalize the boundary confining dualities to exceptional gauge groups and explore their Askey-Wilson type $q$-beta integrals associated with the exceptional Lie algebra. We also expect further boundary confining dualities based on bulk dualities for theories with classical Lie algebras. We hope to report results in our upcoming work \cite{except:2023}.
    
        \item The Seiberg-like duality of 3d $\mathcal{N}=2$ $SU(N)$ gauge theory can be derived from $U(N)$ Seiberg-like duality by gauging manipulation \cite{Aharony:2013dha,Park:2013wta}. While the boundary dualities work in the case with $N_f=N_a=N$ (where there is no dual gauge group) as we discussed in this paper, more general cases seem to need more delicate treatment of gauging the Dirichlet boundary conditions by introducing the 2d vector multiplet. It would be nice to figure out the gauging and ungauging manipulations for the case with boundary and we hope to report on this in future work. 

    \item While the full-indices of 3d $\mathcal{N}=4$ ugly or bad theories \cite{Gaiotto:2008ak} are not well-defined due to the non-positive dimensions of monopoles, the Neumann half-indices can be computed even for such theories. It would be tempting to explore boundary confining dualities for $\mathcal{N}\ge 4$ theories. \footnote{See \cite{Okazaki:2019bok} for the boundary confining duality for $\mathcal{N}=(0,4)$ boundary conditions.}
    
    \item While some of the dualities of $\mathcal{N}=(0,2)$ boundary conditions discussed in this paper can be extended to the bulk, others may not originate from the bulk 3d $\mathcal{N}=2$ theories so that they should be interpreted as ``$\mathcal{N}=(0,2)$ boundary dualities". Such dualities potentially enlarge the web of known dualities. It would be intriguing to explore the ``boundary dualities" in other setups. 
    
    \item The Neumann half-indices can be generalized by introducing the Chern-Simons (CS) coupling and Wilson line operators. It would be interesting to generalize the Askey-Wilson type $q$-beta integrals by introducing additional functions of gauge fugacities in the integrand to figure out more dualities with CS terms and line defects. The duality appetizer  \cite{Jafferis:2011ns,Kapustin:2011vz,Benvenuti:2021nwt} with boundary conditions would translate into the Askey-Wilson type $q$-beta integrals. Also we should be able to examine such decorated half-indices by studying the difference equations which encode the algebra of line  operators. 
        
    \item Brane construction of the $\mathcal{N}=(0,2)$ boundary conditions can be useful 
    to find more boundary confining dualities as well as the holographic dual descriptions. It should be possible to address this from the T-dual configuration in \cite{Okazaki:2020pbb}. 
\end{itemize}

\section{Askey-Wilson integral}
\label{sec_Askey_Wilson}
Askey and Wilson evaluated an integral which can be understood as a $q$-analog of the classical beta integral \cite{MR783216}. We argue that it can be physically understood as a half-index of the 3d $\mathcal{N}=2$ $SU(2)$ gauge theory with four flavors and that it demonstrates the confining duality. 

\subsection{$SU(2)$ with $N_f = 4$ flavors}
\label{sec_su2nf2na2}
Consider theory A as a 3d $\mathcal{N}=2$ $SU(2)$ gauge theory with 
$N_f=4$ fundamental chirals $Q_I$, $I=1,2,3,4$ with R-charge $r_a$. When we later consider $SU(N)$ gauge group, 
this corresponds to $N_f = N = 2$ fundamental and $N_a = N = 2$ antifundamental chirals.

Unlike the case of $U(2)$ gauge theory, 
the $SU(2)$ gauge theory with $4$ fundamental flavors is free from gauge anomaly, without a 2d Fermi multiplet in the determinant representation of gauge group,
when we impose the $\mathcal{N}=(0,2)$ Neumann boundary conditions for the $SU(N)$ vector multiplet and the chiral multiplets. \footnote{See \cite{Dimofte:2017tpi} for full details of Seiberg-like boundary dualities of $U(N)$ gauge theories with fundamental and antifundamental chirals}  
The half-index is evaluated as\footnote{
See \cite{Okazaki:2021pnc,Okazaki:2021gkk} for the notations and conventions of half-indices. 
}
\begin{align}
\label{su2_4a}
\mathbb{II}^{A}_{(\mathcal{N},N,N)}
&=\frac{(q)_{\infty}}{2}
\oint \frac{ds}{2\pi is}
(s^{\pm2};q)_{\infty}
\prod_{\alpha=1}^{4}
\frac{1}{(q^{\frac{r_a}{2}} s^{\pm} a x_{\alpha};q)_{\infty}}
\end{align}
where $\prod_{\alpha = 1}^4 x_{\alpha} = 1$. 
Turning off the fugacities $x_{\alpha}$ and setting $r_a=\frac12$, 
the half-index (\ref{su2_4a}) has the $q$-series expansion 
\begin{align}
&\mathbb{II}^{A}_{(\mathcal{N},N,N)}
\nonumber\\
&=1+6a^2q^{1/2}
+20a^4q+(6a^2+50a^6)q^{3/2}+35(a^4+3 a^8)q^2
\nonumber\\
&+2a^2(3+57a^4+98a^8)q^{5/2}
+\cdots
\end{align}

The integral (\ref{su2_4a}) is identified with the Askey-Wilson integral\footnote{In the mathematical literature the expressions are usually given in terms of unconstrained fugacities $X_{\alpha} = a x_{\alpha}$ and with $r_a = 0$. In our notation the parameter $r_a$ is introduced by the replacement $a \rightarrow q^{r_a/2} a$.} \cite{MR783216} 
which is a $q$-extension of the classical beta integral. 
It is equal to 
\begin{align}
\label{su2_4b}
\prod_{\alpha<\beta}^{4}
\frac{(q^{2r_a} a^4;q)_{\infty}}
{(q^{r_a} a^2 x_{\alpha} x_{\beta};q)_{\infty}}. 
\end{align}
We observe that 
the expression (\ref{su2_4b}) can be interpreted as the half-index of 
theory B which consists of free chiral multiplets, specifically
a chiral multiplet $M_{\alpha \beta}$ with Neumann boundary conditions transforming as the rank-$2$ antisymmetric representation of
the $SU(4)$ flavor symmetry and a chiral $V$ with $U(1)_a$ charge $4$ with Dirichlet boundary conditions.
The content of the two theories is summarized as
\begin{align}
\label{SU2_4_charges}
\begin{array}{c|c|c|c|c|c}
& \textrm{bc} & SU(2) & 
SU(4)& U(1)_a & U(1)_R \\ \hline
\textrm{VM} & \mathcal{N} & {\bf Adj} & {\bf 1} & 0 & 0 \\
Q_{\alpha} & \textrm{N} & {\bf 2} & {\bf 4} & 1 & r_a \\
 \hline
M_{\alpha \beta} & \textrm{N} & {\bf 1} & {\bf 6} & 2 & 2r_a \\
V & \textrm{D} & {\bf 1} & {\bf 1} & -4 & 2 - 4r_a
\end{array}
\end{align}
Note here that the parameter $r_a$ corresponds to shifting the R-charge by $r_a$ times the $U(1)_a$ charge. In our conventions the scaling dimension is half the R-charge so we require all chirals in theory B to have positive R-charge for Neumann boundary conditions and R-charge less than $2$ for Dirichlet boundary conditions. In the case here this means $r_a > 0$. This ensures the unitarity bound is not violated
\footnote{Note that, assuming the duality, requiring this for theory B is equivalent to ensuring that no gauge-invariant operators in theory A violate the unitarity bound.}
and the half-indices do not contain negative powers of $q$. For a 3d bulk duality the corresponding requirement would be that all R-charges lie in the interval $(0,2)$.

The 't Hooft anomalies match for these theories and boundary conditions since
\begin{align}
\label{bdy_AN_su2_2}
\Acal = & \underbrace{2\Tr(s^2) + \frac{3}{2}r^2}_{\textrm{VM}, \; \Ncal}
 - \underbrace{\left( 2\Tr(s^2) + \Tr(x^2) + 4(a-r)^2 \right)}_{Q_{\alpha}, \; N}
  \nonumber \\
  = & - \underbrace{\left( \Tr(x^2) + 3(2a-r)^2 \right)}_{M_{\alpha \beta}, \; N}
   + \underbrace{\frac{1}{2}(-4a+r)^2}_{V, \; D}
  \nonumber \\
  = & - \Tr(x^2) - 4a^2 + 8ar - \frac{5}{2}r^2
\end{align}
where $s$ is the field strength for the $SU(2)$ gauge group, 
$x$ for the $SU(4)$ flavor symmetry group, 
$a$ for the $U(1)_a$ axial symmetry group, 
$r$ for the $U(1)_R$ R-symmetry group. \footnote{
See \cite{Dimofte:2017tpi} for the calculation of the boundary anomaly polynomial. 
}
Note that we have taken $r_a$ for simplicity here since matching of anomalies for any value of $r_a$ guarantees matching for all values of $r_a$.

To summarize, we propose the following confining duality of $\mathcal{N}=(0,2)$ boundary conditions: 
\begin{align}
\label{bcdual_su2nf2na2}
&\textrm{$SU(2)$ $+$ $4$ fund. chirals $Q_{\alpha}$ with b.c. $(\mathcal{N},N)$}
\nonumber\\
&\Leftrightarrow 
\textrm{
$SU(4)$ antisym. chiral $M_{\alpha \beta}$ $+$ a single chiral $V$ with b.c. $(N,D)$}. 
\end{align}

The boundary confining duality (\ref{bcdual_su2nf2na2}) has a counterpart in the 3d bulk. 
In fact, the full-index of theory A is calculated as
\begin{align}
\label{su2_4afull}
I^A&=
\frac12 \sum_{m\in \mathbb{Z}}
\oint \frac{ds}{2\pi is}
(1-q^{|m|}s^{\pm 2})
\prod_{\alpha=1}^{4}
\frac{
(q^{1-\frac{r_a}{2}+\frac{|m|}{2}}s^{\mp}a^{-1}x_{\alpha}^{-1};q)_{\infty}
}
{
(q^{\frac{r_a+|m|}{2}}s^{\pm}ax_{\alpha};q)_{\infty}
}
q^{(1-2r_a)|m|} a^{-4|m|}
\end{align}
and that of theory B is 
\begin{align}
\label{su2_4bfull}
I^B&=\prod_{\alpha<\beta}
\frac{(q^{2r_a}a^4;q)_{\infty}}
{(q^{1-2r_a}a^{-4};q)_{\infty}}
\frac{(q^{1-r_a}a^{-2}x_{\alpha}^{-1}x_{\beta}^{-1};q)_{\infty}}
{(q^{r_a}a^{2}x_{\alpha}x_{\beta};q)_{\infty}}. 
\end{align}
The full-indices (\ref{su2_4afull}) and (\ref{su2_4bfull}) precisely agree with each other as a special case of the $SU(N)$ Seiberg-like duality \cite{Aharony:2013kma,Park:2013wta}. 
The operators can be identified as $M_{\alpha \beta} \sim Q_{\alpha} Q_{\beta}$ (mesons in theory A) and $V$ corresponds to the minimal monopole in theory A.
The superpotentials are
\begin{align}
    \Wcal_A = & 0 \\
    \Wcal_B = & V \Pf M
\end{align}
where here and throughout this article the theory B superpotential is easily interpreted in terms of a Lagrange multiplier (or in other examples several) imposing a constraint which through the operator mapping is seen to be an identity in theory A. E.g.\ in this case the antisymmetric product of the $4$ $Q_{\alpha}$ chirals must be antisymmetric in the $4$ $SU(2)$ gauge indices which is obviously not possible, hence it must vanish.

\section{Nassrallah-Rahman integral}
\label{sec_NR_integral}
We propose a new type of the Seiberg-like duality of $SU(2)$ gauge theory with fundamental and antifuindamental chiral multiplets where one of the chirals has a different boundary condition. We argue that the half-index of the Neumann boundary condition for the vector multiplet is identified with the Nassrallah-Rahman integral \cite{MR772878,MR845667} which generalizes the Askey-Wilson $q$-integral with an additional parameter. In addition, we discuss that the duality can be generalized to bulk theories. 

\subsection{$SU(2)$ with $N_f=3, N_a=2(+1)$ flavors}
\label{sec_su2nf3na3}
Next consider theory A as a 3d $\mathcal{N}=2$ $SU(2)$ gauge theory 
with $3$ fundamental chirals $Q_I$, $I=1,2,3$ of R-charge $0$, 
$2$ antifundamental chirals $\overline{Q}_{\alpha}$, $\alpha=1,2$ of R-charge $0$ 
and an antifundamental chiral $\tilde{Q}$ of R-charge $2$. 
This can be interpreted as a special case of $SU(N)$ gauge theory with $N_f=N+1$ and $N_a=N$ plus another antifundamental chiral discussed later. Since we have gauge group $SU(2)$, we could also describe this as $SU(2)$ with $5 + 1$ fundamental chirals, but here we describe independently fundamental and antifundamental chirals for easier comparison with later generalizations.

We consider the Neumann boundary condition for the $SU(2)$ vector multiplet 
while $\tilde{Q}$ obeys the Dirichlet boundary condition, 
and the other chirals satisfy Neumann boundary conditions. 

The half-index of Neumann boundary conditions for Theory A is calculated as
\begin{align}
\label{Nassrallah-Rahman}
&
\II_{(\mathcal{N},N;N,D)}^A
\nonumber\\
&=
\frac{(q)_{\infty}}{2!} \oint \frac{ds}{2\pi i s}
(s^{\pm 2}; q)_{\infty}
\frac{(q^{(3r_a + 2r_b)/2} a^3 b^2 s^{\pm}; q)_{\infty}}{\left( \prod_{I = 1}^3 (q^{r_a/2} a s^{\pm} x_{I}; q)_{\infty} \right) \left( \prod_{\alpha = 1}^2 (q^{r_b/2} b s^{\mp} \tilde{x}_{\alpha}; q)_{\infty} \right)}
\end{align}
with $\prod_{I = 1}^3 x_I = \prod_{\alpha = 1}^2 \tilde{x}_{\alpha} = 1$.
When we choose $r_a=\frac12$ and $r_b=\frac12$ 
and turn off the fugacities $x_{I}$ and $\tilde{x}_{\alpha}$, 
the half-index (\ref{Nassrallah-Rahman}) can be evaluated as
\begin{align}
&
\II_{(\mathcal{N},N;N,D)}^A
=1+(3a^2+6ab+b^2)q^{1/2}
+(6a^4+16a^3b+21a^2b^2+6ab^3+b^4)q
\nonumber\\
&+\Bigl[
10a^6+30 a^5 b+b^2+48a^4 b^2+54 a^3 b^3+b^6
+3ab(2+2b^4)
\nonumber\\
&+a^2\Bigl(1+13b^4+2 (1 + 4 b^4)\Bigr)
\Bigr]q^{3/2}+\cdots. 
\end{align}
The half-index (\ref{Nassrallah-Rahman}) is known as 
the Nassrallah-Rahman integral \cite{MR772878,MR845667}. 
It is shown to be equal to
\begin{align}
\label{su2_6hindexb}
\frac{\left( \prod_{I = 1}^3 \left( q^{(2r_a + 2r_b)/2} a^2 b^2 x_I^{-1}; q \right)_{\infty} \right) \prod_{\alpha = 1}^2 (q^{(3r_a + r_b)/2} a^3 b \tilde{x}_{\alpha}^{-1}; q)_{\infty}}{\left( \prod_{I = 1}^3 \left( q^{r_a} a^2 x_I^{-1}; q \right)_{\infty} \right) \left( q^{r_b} b^2; q \right)_{\infty} \prod_{I = 1}^3 \prod_{\alpha = 1}^2 (q^{(r_a + r_b)/2} ab x_I \tilde{x}_{\alpha}; q)_{\infty}}
\end{align}
We see that 
the expression (\ref{su2_6hindexb}) can be interpreted as 
the half-index of the theory consisting of five kinds of chiral multiplets 
$M_{I\alpha}$, $B^I$, $\overline{B}$, $\widetilde{B}_{\alpha}$ and $\widetilde{M}_I$ 
obeying Neumann, Neumann, Neumann, Dirichlet and Dirichlet boundary conditions. 

The field content and boundary conditions are summarized as
\begin{align}
\label{SUN_6_charges}
\begin{array}{c|c|c|c|c|c|c|c}
& \textrm{bc} & SU(2) & SU(3) & SU(2) & U(1)_a & U(1)_b & U(1)_R \\ \hline
\textrm{VM} & \mathcal{N} & {\bf Adj} & {\bf 1} & {\bf 1} & 0 & 0 & 0 \\
Q_I & \textrm{N} & {\bf 2} & {\bf 3} & {\bf 1} & 1 & 0 & r_a \\
\overline{Q}_{\alpha} & \textrm{N} & {\bf 2} & {\bf 1} & {\bf 2} & 0 & 1 & r_b \\
\widetilde{Q} & \textrm{D} & {\bf 2} & {\bf 1} & {\bf 1} & -3 & -2 & 2 -3r_a-2r_b \\
 \hline
M_{I\alpha} & \textrm{N} & {\bf 1} & {\bf 3} & {\bf 2} & 1 & 1 & r_a+r_b \\
B^I & \textrm{N} & {\bf 1} & {\bf \overline{3}} & {\bf 1} & 2 & 0 & 2r_a \\
\overline{B} & \textrm{N} & {\bf 1} & {\bf 1} & {\bf 1} & 0 & 2 & 2r_b \\
\widetilde{B}_{\alpha} & \textrm{D} & {\bf 1} & {\bf 1} & {\bf 2} & -3 & -1 & 2 - 3r_a-r_b \\
\widetilde{M}_I & \textrm{D} & {\bf 1} & {\bf 3} & {\bf 1} & -2 & -2 & 2 - 2(r_a+r_b) \\
\end{array}
\end{align}
Here we have explicitly written the shifted $R$-charges by shifting the $R$-charge by $r_a$ times the $U(1)_a$ charge and by $r_b$ times the $U(1)_b$ charge. Suitable choices of $r_a$ and $r_b$ (in this case simply $r_A > 0$ and $r_b > 0$) are required to ensure the indices are convergent. However, to save space we will not explicitly present the shifted $R$-charges in later tables.

The operators can be identified as: mesons in theory A, $M_{I \alpha} \sim Q_I \overline{Q}_{\alpha}$ and $\widetilde{M}_I \sim Q_I \widetilde{Q}$; baryons in theory A, $B^I \sim \epsilon Q_J Q_K \epsilon^{IJK}$, $\overline{B} \sim \epsilon \overline{Q}_1 \overline{Q}_2$ and $\widetilde{B}_{\alpha} \sim \epsilon \overline{Q}_{\alpha} \widetilde{Q}$ where $\epsilon$ without indices indicates antisymmetric contraction of the gauge indices.

The anomalies match for these theories and boundary conditions since
\begin{align}
\label{bdy_SU2_3_3_anom}
\Acal & = \underbrace{2\Tr(s^2) + \frac{3}{2}r^2}_{\textrm{VM}, \; \Ncal}
 - \underbrace{\left( \frac{3}{2} \Tr(s^2) + \frac{2}{2} \Tr(x^2) + 3(a-r)^2 \right)}_{Q_I, \; N}
  \nonumber \\
 & - \underbrace{\left( \Tr(s^2) + \Tr(\tilde{x}^2) + 2(b-r)^2 \right)}_{\overline{Q}_{\alpha}, \; N}
 + \underbrace{\frac{1}{2} \Tr(s^2) + \big( -3a-2b+r \big)^2}_{\widetilde{Q}, \; D}
  \nonumber \\
  = & - \underbrace{\left( \Tr(x^2) + \frac{3}{2} \Tr(\tilde{x}^2) + 3(a+b-r)^2 \right)}_{M_{I \alpha}, \; N}
   - \underbrace{\left( \frac{1}{2} \Tr(x^2) + \frac{3}{2}(2a-r)^2 \right)}_{B^I, \; N}
   - \underbrace{\frac{1}{2}(2b-r)^2}_{\overline{B}, \; N}
   \nonumber \\
 & + \underbrace{\left( \frac{1}{2} \Tr(\tilde{x}^2) + \big( -3a-b+r \big)^2 \right)}_{\widetilde{B}^{\alpha}, \; D}
 + \underbrace{\left( \frac{1}{2} \Tr(x^2) + \frac{3}{2}(-2a-2b+r)^2 \right)}_{\widetilde{M}_I, \; D}
  \nonumber \\
  = & - \Tr(x^2) - \Tr(\tilde{x}^2) + 6a^2 + 2b^2 + 12ab - \frac{5}{2}r^2. 
\end{align}

Therefore we conjecture that the $\mathcal{N}=(0,2)$ Neumann boundary condition  
in $SU(2)$ gauge theory with 6 flavors has the following confinement: 
\begin{align}
\label{bcdual_su2nf3na3}
&\textrm{$SU(2)$ $+$ $3$ fund. $+2$ antifund. $+1$ antifund. with b.c. $(\mathcal{N},N;N,D)$}
\nonumber\\
&\Leftrightarrow 
\textrm{
$SU(3)\times SU(2)$ bifund. chiral $M_{I\alpha}$ $+$ an $SU(3)$ antifund. chiral $B^I$}
\nonumber\\
&
\textrm{$+$ a singlet $\overline{B}$ 
$+$ an $SU(2)$ antifund. chiral $\widetilde{B}_{\alpha}$ 
$+$ an $SU(3)$ fund. chiral $\widetilde{M}_I$ 
}
\nonumber\\
&\textrm{
with b.c. $(N,N,N,D,D)$}. 
\end{align}

Moreover, there is a corresponding confining duality in the bulk by generalizing the boundary confining duality (\ref{bcdual_su2nf3na3})
\begin{align}
\label{dual_su2nf3na3}
&\textrm{$SU(2)$ $+$ $3$ fund. $+2$ antifund. $+1$ antifund. }
\nonumber\\
&\Leftrightarrow 
\textrm{
$SU(3)\times SU(2)$ bifund. chiral $M_{I\alpha}$ $+$ an $SU(3)$ antifund. chiral $B^I$}
\nonumber\\
&
\textrm{$+$ a singlet $\overline{B}$ 
$+$ an $SU(2)$ antifund. chiral $\widetilde{B}_{\alpha}$ 
$+$ an $SU(3)$ fund. chiral $\widetilde{M}_I$ 
}.
\end{align}

In particular the theory A index is well-defined as all monopoles have positive dimension. In fact, given the charges of $\tilde{Q}$, the monopole dimensions (R-charges) are given by
\begin{align}
\label{Mono_Ch_su2_2p1}
    2 |m| \ge 2
\end{align}
with $m \in \Zb$, $m \ne 0$ and there is no dependence on the values of $r_a$ and $r_b$ indicating that the monopoles have no other charges.
Both theories have non-zero superpotentials
\begin{align}
    \Wcal_A = & V \\
    \Wcal_B = & B^I M_{I \alpha} \widetilde{B}^{\alpha} + \overline{B} \widetilde{M}_I B^I
\end{align}
where $V$ is the minimal monopole operator in theory A having R-charge $2$ and no other charges. The theory A superpotential explains the global charges in theory A which do not include a potential $U(1)_c$ symmetry under which $\widetilde{Q}$ would have charge $1$. Including such a symmetry would result in $V$ having non-zero $U(1)_c$ charge, hence the monopole superpotential excludes such a $U(1)_c$.

As explained in \cite{Aharony:2013dha}, the linear monopole superpotential arises from compactification of a 4d theory to 3d. Hence, with this superpotential for theory A, this is the 3d realization of the $N_f = N_A = N$ case of the 4d $SU(N) \leftrightarrow SU(N_f - N)$ Seiberg duality, which in this case is a confining duality as the theory B gauge group is the trivial $SU(1)$.

The full-indices of theory A and B are evaluated as
\begin{align}
\label{NR_indexA}
I^A&=
\frac12 \sum_{m\in \mathbb{Z}}
\oint \frac{ds}{2\pi is}
(1-q^{|m|}s^{\pm2})
\prod_{I=1}^3 
\frac{
(q^{1-\frac{r_a}{2}+\frac{|m|}{2}}a^{-1}s^{\mp}x_{I}^{-1};q)_{\infty}
}
{
(q^{\frac{r_a+|m|}{2}}as^{\pm}x_{I};q)_{\infty}
}
\nonumber\\
&\times 
\prod_{\alpha=1}^2
\frac{
(q^{1-\frac{r_b}{2}+\frac{|m|}{2}}b^{-1}s^{\pm}\tilde{x}_{\alpha}^{-1};q)_{\infty}
}
{
(q^{\frac{r_b+|m|}{2}}bs^{\mp}\tilde{x}_{\alpha};q)_{\infty}
}
\frac{
(q^{\frac{3r_a+2r_b+|m|}{2}}a^3 b^2 s^{\pm};q)_{\infty}
}
{
(q^{1-\frac{3r_a+2r_b}{2}+\frac{|m|}{2}}a^{-3}b^{-2}s^{\mp};q)_{\infty}
}q^{|m|}
\end{align}
and 
\begin{align}
\label{NR_indexB}
I^B&=
\prod_{I=1}^3 
\frac{
(q^{\frac{2r_a+2r_b}{2}}a^2b^2x_I^{-1};q)_{\infty}
}
{
(q^{1-\frac{2r_a+2r_b}{2}}a^{-2}b^{-2}x_I;q)_{\infty}
}
\prod_{\alpha=1}^2
\frac{
(q^{\frac{3r_a+r_b}{2}}a^3b\tilde{x}_\alpha^{-1};q)_{\infty}
}
{
(q^{1-\frac{3r_a+r_b}{2}}a^{-3}b^{-1}\tilde{x}_\alpha;q)_{\infty}
}
\nonumber\\
&\times 
\prod_{I=1}^3 
\frac{
(q^{1-r_a}a^{-2}x_I;q)_{\infty}
}
{
(q^{r_a}a^2x_I^{-1};q)_{\infty}
}
\frac{
(q^{1-r_b}b^{-2};q)_{\infty}
}
{
(q^{r_b}b^2;q)_{\infty}
}
\prod_{I=1}^3
\prod_{\alpha=1}^2
\frac{
(q^{1-\frac{r_a+r_b}{2}}a^{-1}b^{-1}x_I^{-1}\tilde{x}_{\alpha}^{-1};q)_{\infty}
}
{
(q^{\frac{r_a+r_b}{2}}abx_I\tilde{x}_{\alpha};q)_{\infty}
}. 
\end{align}
In fact, we have found the precise matching of full-indices (\ref{NR_indexA}) and (\ref{NR_indexB}) as strong evidence of the duality (\ref{dual_su2nf3na3}). 
For example, when we set $r_a$ $=$ $r_b$ $=1/4$ and switch off the flavored fugacities, we find 
\begin{align}
I^A&=I^B
\nonumber\\
&=1+(3a^2+6ab+b^2)q^{1/4}
+(6a^4+3a^{-2}b^{-2}+2a^{-3}b^{-1}
\nonumber\\
&+16a^3b+21a^2b^2+6ab^3+b^4)q^{1/2}
+(12a^{-2}+10a^6+8b^{-2}+18a^{-1}b^{-1}
\nonumber\\
&+2a^{-3}b+30a^5b+48a^4b^2+54a^3b^3+21a^2b^4+6ab^5+b^6)q^{3/4}
+\cdots.
\end{align}

\section{Gustafson integrals}
\label{sec_G_integrals}
Gustafson derived several higher-rank generalization of the Askey-Wilson and Nassrallah-Rahman integral identities including for various $SU(N)$ \cite{MR1139492, MR1266569}, $USp(2n)$ \cite{MR1139492, MR1147876, MR1266569} and $SO(N)$ \cite{MR1139492} groups. Here we give an interpretation of these results as half-index identities arising from boundary dual theories. Several of these boundary dualities correspond to bulk dualities and we provide some checks of matching indices. However, others do not seem to arise from bulk dualities since the natural bulk theories have non-positive dimension chirals or monopoles.

\subsection{$SU(N)$ with $N_f = N_a = N$ flavors}
\label{sec_GAW_SUN_integrals}
We consider a 3d $\mathcal{N}=2$ gauge theory with gauge group $SU(N)$, 
$N_f=N$ fundamental chirals $Q_{I}$ and $N_a=N$ antifundamental chirals $\overline{Q}_{\alpha}$. 
We choose the $\mathcal{N}=(0,2)$ half-BPS Neumann boundary condition for the $SU(N)$ vector multiplet 
and Neumann boundary conditions for the chiral multiplets. 
This can be viewed as a higher-rank $SU(N)$ generalization of the theory discussed in section \ref{sec_Askey_Wilson}. 

We propose that 
theory A is dual to theory B which has $N_f \times N_a$ bifundamental chirals $M_{I \alpha}$ of the non-Abelian flavor symmetry group $SU(N_f) \times SU(N_a)$ and
two singlet chirals $B$, $\overline{B}$ with Neumann boundary conditions. 
We also have a singlet chiral $V$ with Dirichlet boundary conditions. 
In fact, this is an example of dual boundary conditions for a known bulk duality \cite{Aharony:1997bx}. 
This is summarized
\footnote{From now on we save space by not explicitly parameterising the R-charge by the possible shifts proportional to other global $U(1)$ charges, although suitable shifts are required to ensure the unitarity bound is satisfied.}
as
\begin{align}
\label{SUN_N_charges}
\begin{array}{c|c|c|c|c|c|c|c}
& \textrm{bc} & SU(N) & SU(N_f = N) & SU(N_a = N) & U(1)_a & U(1)_b & U(1)_R \\ \hline
\textrm{VM} & \mathcal{N} & {\bf Adj} & {\bf 1} & {\bf 1} & 0 & 0 & 0 \\
Q_I & \textrm{N} & {\bf N} & {\bf N_f} & {\bf 1} & 1 & 0 & 0 \\
\overline{Q}_{\alpha} & \textrm{N} & {\bf \overline{N}} & {\bf 1} & {\bf N_a} & 0 & 1 & 0 \\
 \hline
M_{I \alpha} & \textrm{N} & {\bf 1} & {\bf N_f} & {\bf N_a} & 1 & 1 & 0 \\
B & \textrm{N} & {\bf 1} & {\bf 1} & {\bf 1} & N & 0 & 0 \\
\overline{B} & \textrm{N} & {\bf 1} & {\bf 1} & {\bf 1} & 0 & N & 0 \\
V & \textrm{D} & {\bf 1} & {\bf 1} & {\bf 1} & -N & -N & 2
\end{array}
\end{align}

The anomalies match for these theories and boundary conditions since
\begin{align}
\label{bdy_AN_su1}
\Acal = & \underbrace{N\Tr(s^2) + \frac{N^2 - 1}{2}r^2}_{\textrm{VM}, \; \Ncal}
 - \underbrace{\left( \frac{N}{2} \Tr(s^2) + \frac{N}{2} \Tr(x^2) + \frac{N^2}{2}(a-r)^2 \right)}_{Q_I, \; N}
 \nonumber \\
 & - \underbrace{\left( \frac{N}{2} \Tr(s^2) + \frac{N}{2} \Tr(\tilde{x}^2) + \frac{N^2}{2}(b-r)^2 \right)}_{\overline{Q}_{\alpha}, \; N}
  \nonumber \\
  = & - \underbrace{\left( \frac{N}{2} \Tr(x^2) + \frac{N}{2} \Tr(\tilde{x}^2) + \frac{N^2}{2}(a+b-r)^2 \right)}_{M_{I \alpha}, \; N}
   - \underbrace{\frac{1}{2}(Na-r)^2}_{B, \; N}
   - \underbrace{\frac{1}{2}(Nb-r)^2}_{\overline{B}, \; N}
   \nonumber \\
 &  + \underbrace{\frac{1}{2}(-Na-Nb+r)^2}_{V, \; D}
  \nonumber \\
  = & - \frac{N}{2} \Tr(x^2) - \frac{N}{2} \Tr(\tilde{x}^2) - \frac{N^2}{2}a^2 - \frac{N^2}{2}b^2 + N^2 ar + N^2 br - \frac{N^2 + 1}{2}r^2. 
\end{align}

The theory A half-index is 
\begin{align}
\label{SUN_N_halfindex}
&
\mathbb{II}_{(\mathcal{N},N,N)}^A
=
\frac{(q)_{\infty}^{N-1}}{N!} \prod_{i=1}^N \oint \frac{ds_i}{2\pi i s_i}
\prod_{i \ne j}^N (s_i s_j^{-1}; q)_{\infty}
\nonumber\\
&\times 
\frac{1}{\prod_{i = 1}^N \left( \prod_{i = 1}^N (q^{r_a/2} a s_i x_I; q)_{\infty} \right) \left( \prod_{\alpha = 1}^{N} (q^{r_b/2} b s_i^{-1} \tilde{x}_{\alpha}; q)_{\infty} \right)}. 
\end{align}
For example, setting $r_a=r_b=\frac12$, $x_I = 1$ and $\tilde{x}_{\alpha} = 1$ we have
\begin{align}
&N=3: 
\nonumber\\
&\mathbb{II}_{(\mathcal{N},N,N)}^A=
1+9abq^{1/2}+(a^3+b^3)q^{3/4}
+45a^2b^2q+8ab(a^3+b^3)q^{5/4}
\nonumber\\
&+(a^6+9ab+165a^3b^3+b^6)q^{3/2}+(a^3+45a^5b^2+b^3+45a^2b^5)q^{7/4}+\cdots
,\\
&N=4: 
\nonumber\\
&\mathbb{II}_{(\mathcal{N},N,N)}^A=
1+16abq^{1/2}+(a^4+136a^2b^2+b^4)q+16ab(1+a^4+51a^2 b^2+ b^4)q^{3/2}
\nonumber\\
&+(a^8+136a^6b^2+b^4+b^8+8a^2b^2(32+17b^4)+a^4(1+3876b^4))q^2+\cdots.
\end{align}

The half-index (\ref{SUN_N_halfindex}) can be viewed as a higher-rank Askey-Wilson integral that generalizes (\ref{su2_4a}). 
Gustafson \cite{MR1139492} showed that it is given by
\begin{align}
\label{SUN_N_halfindexD}
\frac{\left( q^\frac{Nr_a+Nr_b}{2} a^N b^N; q \right)_{\infty}}{\left( q^{Nr_a/2} a^N; q \right)_{\infty} \left( q^{Nr_b/2} b^N; q \right)_{\infty} \prod_{I = 1}^N \prod_{\alpha = 1}^{N} (q^{(r_a + r_b)/2} ab x_I \tilde{x}_{\alpha}; q)_{\infty}}. 
\end{align}
We can exactly identify equation (\ref{SUN_N_halfindexD}) with the half-index of theory B with the specific boundary conditions described above.

Note that the bulk theory A also has a Seiberg-like dual description \cite{Aharony:2013dha,Park:2013wta}. 
The mapping of operators between theory A and theory B is given by:
$M_{I \alpha} \sim Q_I \overline{Q}_{\alpha}$, $B \sim \epsilon Q_{1} \cdots Q_{N} = \det Q$ and
$\overline{B} \sim \epsilon \overline{Q}_1 \cdots \overline{Q}_N = \det \overline{Q}$ where the antisymmetric tensors are used to contact the gauge indices. 
Theory A has zero superpotential while theory B has a superpotential
\begin{align}
\label{SUN_N_B_superpot}
\Wcal_B = -V \left( \det M - B \overline{B} \right)
\end{align}
which imposes the chiral ring relation $\det M = B \overline{B}$ in theory B which corresponds to
$\det (Q \overline{Q}) = \det Q \det \overline{Q}$ in theory A. It can easily be seen that the
contribution of $V$ with Dirichlet boundary condition does exactly this in the theory B half-index. 

Through two dualities this gives an alternative derivation of the above duality. 
The first is the Seiberg dual of a $SU(N)$ theory with $N_f = N_F$ fundamental and $N_a = N_F$ antifundamental chirals which is a $U(N_F - N) \times U(1)_y$ theory with $N_f = N_F$ fundamental and $N_a = N_F$ antifundamental chirals, two chirals $v_{\pm}$ with charges $\pm 1$ under $U(1)_y$ and chirals $M_{I \alpha}$ in the bifundamental representation of the flavor symmetry group $SU(N_f) \times SU(N_a)$. In the case here with $N_F = N$ the dual gauge group is just $U(1)_y$ and there are hence no fundamental or antifundamental chirals (of $U(0)$). As the chirals $M_{I \alpha}$ are not charged under $U(1)_y$ we can consider them as spectators and dualize the $U(1)_y$ gauge theory with the two chirals $v_{\pm}$ which is dual to the $XYZ$ model, where in our notation we identify $V$, $B$ and $\overline{B}$ with $X$, $Y$ and $Z$.

We can extend the above discussion to include boundary conditions as shown in (\ref{SUN_N_charges_ACB}). Here $U(1)_A$ is the axial symmetry in theory A and $U(1)_G$ is the global $U(1)$ arising from the gauging/ungauging process to derive the $SU(N)$ Seiberg duality from the $U(N)$ duality \cite{Aharony:2013dha, Park:2013wta}. 
\begin{align}
\label{SUN_N_charges_ACB}
\begin{array}{c|c|c|c|c|c|c|c|c}
& \textrm{bc} & SU(N) & SU(N_f = N) & SU(N_a = N) & U(1)_A & U(1)_G & U(1)_y & U(1)_R \\ \hline
\textrm{VM} & \mathcal{N} & {\bf Adj} & {\bf 1} & {\bf 1} & 0 & 0 & 0 & 0 \\
Q_I & \textrm{N} & {\bf N} & {\bf N_f} & {\bf 1} & 1 & 1 & 0 & 0 \\
\overline{Q}_{\alpha} & \textrm{N} & {\bf \overline{N}} & {\bf 1} & {\bf N_a} & 1 & -1 & 0 & 0 \\
\hdashline
\eta & & {\bf 1} & {\bf 1} & {\bf 1} & 0 & N & 1 & 0 \\
 \hline
 \textrm{VM} & \mathcal{D} & {\bf 1} & {\bf 1} & {\bf 1} & 0 & 0 & {\bf Adj} & 0 \\
 v_{\pm} & \mathcal{D} & {\bf 1} & {\bf 1} & {\bf 1} & -N & 0 & \pm 1 & 1 \\
M_{I \alpha} & \textrm{N} & {\bf 1} & {\bf N_f} & {\bf N_a} & 2 & 0 & 0 & 0 \\
\hline
M_{I \alpha} & \textrm{N} & {\bf 1} & {\bf N_f} & {\bf N_a} & 2 & 0 & 0 & 0 \\
B & \textrm{N} & {\bf 1} & {\bf 1} & {\bf 1} & N & N & 0 & 0 \\
\overline{B} & \textrm{N} & {\bf 1} & {\bf 1} & {\bf 1} & N & -N & 0 & 0 \\
V & \textrm{D} & {\bf 1} & {\bf 1} & {\bf 1} & -2N & 0 & 0 & 2 \\
\hdashline
\eta & & {\bf 1} & {\bf 1} & {\bf 1} & 0 & N & 1 & 0
\end{array}
\end{align}

With these boundary conditions the anomalies match for all three theories provided we include a background mixed CS coupling for $U(1)_G \times U(1)_y$ in theory B, contributing $2NGy$ to the boundary 't Hooft anomaly. In particular the anomalies match for these theories and boundary conditions since
\begin{align}
\label{bdy_SUN_N_anom}
\Acal = &  \underbrace{N\Tr(s^2) + \frac{N^2 - 1}{2}r^2}_{SU(N) \; \textrm{VM}, \; \Ncal}
 - \underbrace{\left( \frac{N}{2} \Tr(s^2) + \frac{N}{2} \Tr(x^2) + \frac{N^2}{2}(A+G-r)^2 \right)}_{Q_I, \; N}
   \nonumber \\
 & - \underbrace{\left( \frac{N}{2} \Tr(s^2) + \frac{N}{2} \Tr(\tilde{x}^2) + \frac{N^2}{2}(A-G-r)^2 \right)}_{\overline{Q}_{\alpha}, \; N}
 + \underbrace{(NG + y)^2}_{\eta}
 \nonumber \\
  = & -\underbrace{\frac{1}{2}r^2}_{U(1) \; \textrm{VM}, \; \Dcal}
   + \underbrace{\frac{1}{2}(-NA + y)^2}_{v_{+}, \; D} + \underbrace{\frac{1}{2}(-NA - y)^2}_{v_{-}, \; D}
   \nonumber \\
 &  - \underbrace{\left( \frac{N}{2} \Tr(x^2) + \frac{N}{2} \Tr(\tilde{x}^2) + \frac{N^2}{2}(2A-r)^2 \right)}_{M_{I \alpha}, \; N}
   + \underbrace{2NGy}_{\textrm{FI}}
   \nonumber \\
  = & - \underbrace{\left( \frac{N}{2} \Tr(x^2) + \frac{N}{2} \Tr(\tilde{x}^2) + \frac{N^2}{2}(2A-r)^2 \right)}_{M_{I \alpha}, \; N}
   - \underbrace{\frac{1}{2}(NA + NG - r)^2}_{B, \; N}
   - \underbrace{\frac{1}{2}(NA - NG - r)^2}_{\overline{B}, \; N}
  \nonumber \\
  & + \underbrace{\frac{1}{2}(-2NA + r)^2}_{V, \; D} + \underbrace{(NG + y)^2}_{\eta}
 \nonumber \\
  = & - \frac{N}{2} \Tr(x^2) - \frac{N}{2} \Tr(\tilde{x}^2) - N^2 A^2 + 2N^2 Ar + y^2 + 2NGy - \frac{N^2 + 1}{2}r^2. 
\end{align}
To compare to the previous table~(\ref{SUN_N_charges}), note that we can map $U(1)_A \times U(1)_G$ to $U(1)_a \times U(1)_b$ with the following linear combinations of charges $Q_a = \frac{1}{2}(Q_A + Q_G)$ and $Q_b = \frac{1}{2}(Q_A - Q_G)$.

To summarize, we propose the following confining duality of $\mathcal{N}=(0,2)$ boundary conditions: 
\begin{align}
\label{bcdual_suNnfNnaN}
&\textrm{$SU(N)$ $+$ $N$ fund. chirals $Q_I$ and $N$ antifund. chirals $\overline{Q}_{\alpha}$ with b.c. $(\mathcal{N},N,N)$}
\nonumber\\
\Leftrightarrow &
\textrm{
$SU(N) \times SU(N)$ bifundamental chiral $M_{I \alpha}$ $+$ singlet chirals $B$
and $\overline{B}$
}
\nonumber\\
&\textrm{with Neumann b.c.\ and a singlet chiral $V$ with b.c. $(N,N,N,D)$}. 
\end{align}
The operators can be identified as $M_{I \alpha} \sim Q_I \overline{Q}_{\alpha}$ (mesons in theory A), $B \sim \epsilon Q_{1} \cdots Q_{N} = \det Q$ and
$\overline{B} \sim \epsilon \overline{Q}_1 \cdots \overline{Q}_N = \det \overline{Q}$ (Barons in theory A) and $V$ corresponds to the minimal monopole in theory A. Theory B has superpotential $W = -V (\det M - B\overline{B})$.

We also note that there is a bulk duality and the matching of full indices can be checked as all monopoles in theory A have positive dimension (R-charge). Specifically the dimensions of the monopoles are given by
\begin{align}
\label{Mono_Ch_suN_N}
    N \left( 1 - \frac{r_a + r_b}{2} \right) \sum_{i=1}^N |m_i| - \sum_{i < j}^N |m_i - m_j| \ge \left( 1 - \frac{N(r_a + r_b)}{2} \right) \sum_{i=1}^N |m_i| \ge 2
\end{align}
for $r_a \ge 0$, $r_b \ge 0$, $m_i \in \Zb$ with the constraint $\sum_{i=1}^N m_i = 0$ (and at least one $m_i \ne 0$). The lower bound is given for $r_a = r_b = 0$ and saturated for a single $m_i = \pm 1$. Choosing positive values for $r_a$ and $r_b$ will decrease the R-charge of the monopoles but there will still be a range of values $0 < r_a + r_b < 2/N$ so that all monopoles and chirals have positive R-charge.

\subsection{$SU(N)$ with $N_f = N + 1$, $N_a = N (+ 1)$}
\label{sec_GNR_integrals}
Consider a higher-rank generalization of the $SU(N)$ for the theory studied in section \ref{sec_su2nf3na3} 
by adding to theory A in section \ref{sec_GAW_SUN_integrals} a pair of chiral multiplets, one 
in the fundamental representation of $SU(N)$ with Neumann boundary conditions,
and the other in the antifundamental representation of $SU(N)$ with
Dirichlet boundary conditions. This theory has a $SU(N_f = N+1) \times SU(N_a = N)$
flavor symmetry (and some $U(1)$ global symmetries). 

The half-index of theory A is given by
\begin{align}
\label{SUN_Np1_halfindex}
&\mathbb{II}^A_{(\mathcal{N},N,N,D)}
=\frac{(q)_{\infty}^{N-1}}{N!} \prod_{i=1}^N \oint \frac{ds_i}{2\pi i s_i}
\nonumber\\
&\times 
\prod_{i \ne j}^N (s_i s_j^{-1}; q)_{\infty}
\prod_{i = 1}^N 
\frac{(q^{((N+1)r_a + Nr_b)/2} a^{N+1} b^N s_i; q)_{\infty}}{\left( \prod_{I = 1}^{N+1} (q^{r_a/2} a s_i x_{I}; q)_{\infty} \right) \left( \prod_{\alpha = 1}^N (q^{r_b/2} b s_i^{-1} \tilde{x}_{\alpha}; q)_{\infty} \right)}. 
\end{align}
It is shown in \cite{MR1266569} (see eq.(3.3)) that the integral is equivalent to
\begin{align}
\label{SUN_Np1_halfindexB}
\frac{\left( \prod_{I = 1}^{N+1} \left( q^{(Nr_a + Nr_b)/2} a^N b^N x_I^{-1}; q \right)_{\infty} \right) \prod_{\alpha = 1}^N (q^{((N+1)r_a + r_b)/2} a^{N+1} b \tilde{x}_{\alpha}; q)_{\infty}}{\left( \prod_{I = 1}^{N+1} \left( q^{Nr_a/2} a^N x_I^{-1}; q \right)_{\infty} \right) \left( q^{Nr_b/2} b^N; q \right)_{\infty} \prod_{I = 1}^{N+1} \prod_{\alpha = 1}^{N} (q^{(r_a + r_b)/2} ab x_I \tilde{x}_{\alpha}; q)_{\infty}}. 
\end{align}
Again we can regard the expression (\ref{SUN_Np1_halfindexB}) as the half-index of the dual boundary conditions in theory B. 
In theory B we again have no gauge group. We have various chirals in representations of
the flavor symmetry group $SU(N_f) \times SU(N_a)$. Those with
Neumann boundary conditions are
\begin{itemize}
\item Bifundamental $M_{I\alpha}$ of $SU(N_f) \times SU(N_a)$.
\item Antifundamental $B^I$ of $SU(N_f)$.
\item a singlet $\overline{B}$.
\end{itemize}
Those with Dirichlet boundary conditions are
\begin{itemize}
\item Fundamental $\widetilde{B}^{\alpha}$ of $SU(N_f)$.
\item Antifundamental $\widetilde{M}_I$ of $SU(N_a)$.
\end{itemize}

The full description of the theories including global charges and boundary conditions is 
\begin{align}
\label{SUN_Np1_charges}
\begin{array}{c|c|c|c|c|c|c|c}
& \textrm{bc} & SU(N) & SU(N_f = N+1) & SU(N_a = N) & U(1)_a & U(1)_b & U(1)_R \\ \hline
\textrm{VM} & \mathcal{N} & {\bf Adj} & {\bf 1} & {\bf 1} & 0 & 0 & 0 \\
Q_I & \textrm{N} & {\bf N} & {\bf N_f} & {\bf 1} & 1 & 0 & 0 \\
\overline{Q}_{\alpha} & \textrm{N} & {\bf \overline{N}} & {\bf 1} & {\bf N_a} & 0 & 1 & 0 \\
\widetilde{Q} & \textrm{D} & {\bf \overline{N}} & {\bf 1} & {\bf 1} & -(N + 1) & -N & 2 \\
 \hline
M_{I \alpha} & \textrm{N} & {\bf 1} & {\bf N_f} & {\bf N_a} & 1 & 1 & 0 \\
B^I & \textrm{N} & {\bf 1} & {\bf \overline{N_f}} & {\bf 1} & N & 0 & 0 \\
\overline{B} & \textrm{N} & {\bf 1} & {\bf 1} & {\bf 1} & 0 & N & 0 \\
\widetilde{B}^{\alpha} & \textrm{D} & {\bf 1} & {\bf 1} & {\bf \overline{N_a}} & -(N + 1) & -1 & 2 \\
\widetilde{M}_I & \textrm{D} & {\bf 1} & {\bf N_f} & {\bf 1} & -N & -N & 2 \\
\end{array}
\end{align}

The anomalies match for these theories and boundary conditions since
\begin{align}
\label{bdy_SUN_Np1_anom}
\Acal & = \underbrace{N\Tr(s^2) + \frac{N^2 - 1}{2}r^2}_{\textrm{VM}, \; \Ncal}
 - \underbrace{\left( \frac{N+1}{2} \Tr(s^2) + \frac{N}{2} \Tr(x^2) + \frac{N(N+1)}{2}(a-r)^2 \right)}_{Q_I, \; N}
  \nonumber \\
 & - \underbrace{\left( \frac{N}{2} \Tr(s^2) + \frac{N}{2} \Tr(\tilde{x}^2) + \frac{N^2}{2}(b-r)^2 \right)}_{\overline{Q}_{\alpha}, \; N}
 + \underbrace{\frac{1}{2} \Tr(s^2) + \frac{N}{2}\big( -(N+1)a-Nb+r \big)^2}_{\widetilde{Q}, \; D}
  \nonumber \\
  = & - \underbrace{\left( \frac{N}{2} \Tr(x^2) + \frac{N+1}{2} \Tr(\tilde{x}^2) + \frac{N(N+1)}{2}(a+b-r)^2 \right)}_{M_{I \alpha}, \; N}
  \nonumber \\
 &  - \underbrace{\left( \frac{1}{2} \Tr(x^2) + \frac{N+1}{2}(Na-r)^2 \right)}_{B^I, \; N}
   - \underbrace{\frac{1}{2}(Nb-r)^2}_{\overline{B}, \; N}
   \nonumber \\
 & + \underbrace{\left( \frac{1}{2} \Tr(\tilde{x}^2) + \frac{N}{2}\big( -(N+1)a-b+r \big)^2 \right)}_{\widetilde{B}^{\alpha}, \; D}
 + \underbrace{\left( \frac{1}{2} \Tr(x^2) + \frac{N+1}{2}(-Na-Nb+r)^2 \right)}_{\widetilde{M}_I, \; D}
  \nonumber \\
  = & - \frac{N}{2} \Tr(x^2) - \frac{N}{2} \Tr(\tilde{x}^2) + \frac{N^2(N+1)}{2}a^2 + \frac{N^2(N-1)}{2}b^2 + N^2(N+1)ab - \frac{N^2 + 1}{2}r^2.
\end{align}

The mapping of operators between theory A and theory B is given by:
$M_{I \alpha} \sim Q_I \overline{Q}_{\alpha}$, $B^I \sim \epsilon Q_{J_1} \cdots Q_{J_N} \epsilon^{I J_1 \cdots J_N}$, $\overline{B} \sim \epsilon \overline{Q}_1 \cdots \overline{Q}_N$, $\widetilde{B}^{\alpha} \sim \epsilon \overline{Q}_{\beta_1} \cdots \overline{Q}_{\beta_{N-1}} \widetilde{Q} \epsilon^{\alpha \beta_1 \cdots \beta_{N-1}}$ and $\widetilde{M}_I \sim Q_I \widetilde{Q}$.

So we propose the boundary confining duality: 
\begin{align}
\label{bcdual_suN_N+1_N+1}
&\textrm{$SU(N)$ $+$ $(N+1)$ fund. $+N$ antifund. $+1$ antifund. with b.c. $(\mathcal{N},N;N,D)$}
\nonumber\\
&\Leftrightarrow 
\textrm{
$SU(N+1)\times SU(N)$ bifund. chiral $M_{I\alpha}$ $+$ an $SU(N+1)$ antifund. chiral $B^I$}
\nonumber\\
&
\textrm{$+$ a singlet $\overline{B}$ 
$+$ an $SU(N)$ antifund. chiral $\widetilde{B}_{\alpha}$ 
$+$ an $SU(N+1)$ fund. chiral $\widetilde{M}_I$ 
}
\nonumber\\
&\textrm{
with b.c. $(N,N,N,D,D)$}. 
\end{align}

This can be viewed as the deformation of the boundary confining duality (\ref{bcdual_suNnfNnaN}) 
by adding extra fundamental chirals and in pairs with Neumann and Dirichlet boundary conditions in theory A which we have seen in section~\ref{sec_GAW_SUN_integrals}. 
In theory B this leads to an extension of the rank-$2$ flavor symmetry chiral 
with Neumann boundary condition and an additional chiral in the vector of the flavor symmetry group having Dirichlet boundary condition.

This boundary confining duality corresponds to a bulk confining duality which arises as a direct dimensional reduction of the 4d Seiberg duality between $SU(N)$ with $N_f = N_a = N + 1$ flavors and an $SU(N_f - N) = SU(1)$ theory. This compactification where theory A has a linear monopole superpotential is explained in \cite{Aharony:2013dha}. 
The full indices can be computed as
\begin{align}
\label{suN_N+1_N+1_fullA}
I^A&=\frac{1}{N!}
\sum_{m_1,\cdots,m_N\in \mathbb{Z}}
\prod_{i=1}^N
\oint \frac{ds_i}{2\pi is_i}
(1-q^{\frac{|m_i-m_j|}{2}} s_i^{\pm}s_j^{\mp})
\nonumber\\
&\times 
\prod_{i=1}^N 
\prod_{I=1}^{N+1}
\frac{
(q^{1-\frac{r_a}{2}+\frac{|m_i|}{2}} a^{-1} s_i^{-1} x_I^{-1};q)_{\infty}
}{
(q^{\frac{r_a}{2}+\frac{|m_i|}{2}} a s_i x_I;q)_{\infty}
}
\prod_{\alpha=1}^{N}
\frac{
(q^{1-\frac{r_b}{2}+\frac{|m_i|}{2}} b^{-1} s_i \tilde{x}_{\alpha}^{-1};q)_{\infty}
}{
(q^{\frac{r_b}{2}+\frac{|m_i|}{2}} b s_i^{-1} \tilde{x}_\alpha;q)_{\infty}
}
\nonumber\\
&\times 
\prod_{i=1}^N
\frac{
(q^{\frac{(N+1)r_a+Nr_a}{2}+\frac{|m_i|}{2}} a^{N+1} b^Ns_i;q)_{\infty}
}
{
(q^{1-\frac{(N+1)r_a+Nr_a}{2}+\frac{|m_i|}{2}} a^{-N-1} b^{-N}s_i^{-1};q)_{\infty}
}q^{ 2\sum_{i=1}^N |m_i|-\frac12\sum_{i<j}|m_i\pm m_j|}
\end{align}
where $\sum_{i=1}^N m_i=0$ and 
\begin{align}
\label{suN_N+1_N+1_fullB}
I^B&=
\prod_{I=1}^{N+1}
\frac{
(q^{\frac{Nr_a+Nr_b}{2}}a^Nb^Nx_I^{-1};q)_{\infty}
}
{
(q^{1-\frac{Nr_a+Nr_b}{2}}a^{-N}b^{-N}x_I;q)_{\infty}
}
\prod_{\alpha=1}^{N}
\frac{
(q^{\frac{(N+1)r_a+r_b}{2}}a^{N+1} b \tilde{x}_{\alpha};q)_{\infty}
}
{
(q^{1-\frac{(N+1)r_a+r_b}{2}}a^{-N-1} b^{-1} \tilde{x}_{\alpha}^{-1};q)_{\infty}
}
\nonumber\\
&\times 
\prod_{I=1}^{N+1}
\frac{
(q^{1-\frac{Nr_a}{2}} a^{-N} x_I;q )_{\infty}
}
{
(q^{\frac{Nr_a}{2}} a^{N} x_I^{-1};q )_{\infty}
}
\frac{
(q^{1-\frac{Nr_b}{2}} b^{-N};q )_{\infty}
}
{
(q^{\frac{Nr_b}{2}} b^N;q )_{\infty}
}
\prod_{I=1}^{N+1}
\prod_{\alpha=1}^{N}
\frac{
(q^{1-\frac{(N+1)r_a+r_b}{2}} a^{-1}b^{-1} x_I^{-1} \tilde{x}_{\alpha}^{-1};q )_{\infty}
}
{
(q^{\frac{(N+1)r_a+r_b}{2}} ab x_I \tilde{x}_{\alpha};q )_{\infty}
}.
\end{align}
In particular the theory A index is well-defined as all monopoles have positive dimension. In fact, given the charges of $\tilde{Q}$, the monopole dimensions are given by
\begin{align}
\label{Mono_Ch_suN_Np1}
    N \sum_{i=1}^N |m_i| - \sum_{i < j}^N |m_i - m_j| \ge \sum_{i=1}^N |m_i| \ge 2
\end{align}
with $\sum_{i=1}^N m_i = 0$ and no dependence on the values of $r_a$ and $r_b$.
We have confirmed that the full-indices 
(\ref{suN_N+1_N+1_fullA}) and (\ref{suN_N+1_N+1_fullB}) match. 
For example, for $N=4$ with $r_a=r_b=\frac18$, $x_I=\tilde{x}_{\alpha}=1$, 
we have
\begin{align}
I^A&=I^B
\nonumber\\
&=1+20abq^{1/8}
+(5a^4+210a^2b^2+b^4)q^{1/4}
+4(24a^5b+385a^3b^3+5ab^5)q^{3/8}\nonumber\\
&+(15a^8+5a^{-4}b^{-4}+970a^6b^2+8855a^4b^4+210a^2b^6+b^8)q^{1/2}
+\cdots.
\end{align}

As for the $SU(2)$ example in section~\ref{sec_su2nf3na3}, both theories have non-zero superpotentials
\begin{align}
    \Wcal_A = & V \\
    \Wcal_B = & B^I M_{I \alpha} \widetilde{B}^{\alpha} + \overline{B} \widetilde{M}_I B^I
\end{align}
where $V$ is the minimal monopole operator in theory A having R-charge $2$ and no other charges. This again explains the global charges in theory A which do not include a potential $U(1)_c$ symmetry under which $\widetilde{Q}$ would have charge $1$. Including such a symmetry would result in $V$ having non-zero $U(1)_c$ charge, hence the monopole superpotential excludes such a $U(1)_c$. Another way to view this is to shift the R-charges but a linear combination of the $U(1)_a$ and $U(1)_b$ charges so that all $Q_I$, $\overline{Q}_{\alpha}$ and $\widetilde{Q}$ have the same R-charge. Then, compared to the case of vanishing superpotential, the monopole superpotential reduces the flavor symmetry from $U(N+1) \times U(N+1)$ to $S(U(N+1) \times U(N+1))$ and the dual theory is becomes the theory B described here rather than a $U(1)$ gauge theory. In this description the chirals $M_{I \alpha}$ and $\widetilde{M}_I$ are combined, as are $\overline{B}$ and $\widetilde{B}^{\alpha}$.

Of course, this leads to the obvious conjecture that the general case where theory A has gauge group $SU(N)$ and $N_f = N_a$ fundamental and antifundamental chirals, all having Neumann boundary conditions except for one of the fundamental or antifundamental chirals, will have a boundary dual theory B. In this case theory B will have the same chirals as described above along with an $SU(N_f - N)$ vector multiplet with Dirichlet boundary conditions. The chirals $B^I$ will be in the fundamental representation of $SU(N_f - N)$ while the chirals $\overline{B}$ and $\widetilde{B}^{\alpha}$ will be in the antifundamental representation of $SU(N_f - N)$ while $M_{I \alpha}$ and $\widetilde{M}_I$ will be singlets of $SU(N_f - N)$. We will also have Fermis in theory A to cancel the gauge anomaly.

\subsection{$USp(2n)$ with $2n+2$ fundamental chirals}
\label{sec_USp2n_2np2_integrals}
Consider an alternative higher-rank generalization of the $SU(2)$ gauge theory in section \ref{sec_Askey_Wilson} to $USp(2n)$ gauge group. 
Theory A has gauge group $USp(2n)$ with $2n + 2$ fundamental chirals
with R-charge $0$. These all have Neumann boundary conditions.

The half-index takes the form
\begin{align}
\label{USp2n_2n+2_hindexA}
\mathbb{II}_{(\mathcal{N},N,N)}^A
&=\frac{(q)_{\infty}^n}{n! 2^n} \prod_{i=1}^n \oint \frac{ds_i}{2\pi i s_i}
\prod_{i \ne j}^n (s_i s_j^{-1}; q)_{\infty} \prod_{i \le j}^n (s_i^{\pm} s_j^{\pm}; q)_{\infty}
\nonumber\\
&\times 
\frac{1}{\prod_{\alpha = 1}^{2n + 2} \prod_{i = 1}^n (q^{r/2} a s_i^{\pm} x_{\alpha}; q)_{\infty}}. 
\end{align}
According to Theorem 7.1 in \cite{MR1139492}, 
the half-index (\ref{USp2n_2n+2_hindexA}) can be evaluated as
\begin{align}
\label{USp2n_2n+2_hindexB}
\frac{\left( q^{(n+1)r} a^{2n+2}; q \right)_{\infty}}{\prod_{\alpha < \beta}^{2n + 2} (q^r a^2 x_{\alpha} x_{\beta}; q)_{\infty}} \; .
\end{align}
The expression (\ref{USp2n_2n+2_hindexA}) has a simple implication as a half-index of theory B. 
Theory B  is the Seiberg-like dual of theory A \cite{Karch:1997ux}. It has an  $SU(2n + 2)$ antisymmetric rank-$2$ chiral $M_{\alpha\beta}$ with R-charge $0$
with Neumann boundary conditions and
a singlet chiral $V$ with R-charge $2$ with Dirichlet boundary condition. 
The field content and the boundary conditions are shown in the following: 
\begin{align}
\label{USp2n_2np2_charges}
\begin{array}{c|c|c|c|c|c}
& \textrm{bc} & USp(2n) & SU(N_f = 2n + 2) & U(1)_a & U(1)_R \\ \hline
\textrm{VM} & \mathcal{N} & {\bf Adj} & {\bf 1} & 0 & 0 \\
Q_{\alpha} & \textrm{N} & {\bf 2n} & {\bf N_f} & 1 & 0 \\
 \hline
M_{\alpha \beta} & \textrm{N} & {\bf 1} & {\bf N_f(N_f - 1)/2} & 2 & 0 \\
V & \textrm{D} & {\bf 1} & {\bf 1} & -N_f & 2
\end{array}
\end{align}

The mapping of operators between theory A and theory B is given by
$M_{\alpha \beta} \sim Q_{\alpha} Q_{\beta}$ where the gauge indices are contracted with the $USp(2n)$-invariant rank-$2$ antisymmetric tensor, and $V$ is dual to the minimal monopole operator in theory A in the bulk. 

The anomalies match for these theories and boundary conditions since
\begin{align}
\label{bdy_USp2n_2np2_anom}
\Acal & = \underbrace{(n+1)\Tr(s^2) + \frac{n(2n+1)}{2}r^2}_{\textrm{VM}, \; \Ncal}
 - \underbrace{\left( (n+1)\Tr(s^2) + n\Tr(x^2) + 2n(n+1)(a-r)^2 \right)}_{Q_{\alpha}, \; N}
  \nonumber \\
  = & - \underbrace{\left( n \Tr(x^2) + \frac{(n+1)(2n+1)}{2}(2a-r)^2 \right)}_{M_{\alpha \beta}, \; N}
   + \underbrace{\frac{1}{2}\big( -2(n+1)a+r \big)^2}_{V, \; D}
  \nonumber \\
  = & - n \Tr(x^2) - 2n(n+1)a^2 + 4n(n+1)ar - \frac{n(2n+3)}{2}r^2.
\end{align}

The matching of the half-indices and boundary anomalies demonstrate that 
the $USp(2n)$ higher-rank generalization of the boundary confining duality (\ref{bcdual_su2nf2na2}) is given by
\begin{align}
\label{bcdual_USp2n_2n}
&\textrm{$USp(2n)$ $+$ $2n+2$ fund. chirals $Q_{\alpha}$ with b.c. $(\mathcal{N},N)$}
\nonumber\\
&\Leftrightarrow 
\textrm{
$SU(2n+2)$ antisym. chiral $M_{\alpha \beta}$ $+$ a single chiral $V$ with b.c. $(N,D)$}. 
\end{align}

Again the boundary confining duality can be generalized to a bulk duality. The matching of full indices 
\begin{align}
I^A&=\frac{1}{n!2^n}
\sum_{m_1,\cdots,m_n\in \mathbb{Z}}
\prod_{i=1}^n
\oint \frac{ds_i}{2\pi i s_i}
\prod_{i\neq j}^n (1-q^{\frac{|m_i-m_j|}{2}}s_i s_j^{-1})
\prod_{i\le j}^{n} (1-q^{\frac{|m_i+m_j|}{2}}s_i^{\pm} s_j^{\pm})
\nonumber\\
&\times 
\prod_{\alpha=1}^{2n+2}
\prod_{i=1}^{n}
\frac{(q^{1-\frac{r}{2}+\frac{|m_i|}{2}} a^{-1} s^{\mp} x_{\alpha}^{-1};q )_{\infty}}
{(q^{\frac{r}{2}+\frac{|m_i|}{2}}as^{\pm}x_{\alpha};q)_{\infty}}
q^{(n+1)(1-r)\sum_{i=1}^n |m_i|-\sum_{i=1}^n |m_i|-\frac12\sum_{i<j}|m_i\pm m_j|}
a^{-6\sum_{i=1}^n|m_i| }
\end{align}
and
\begin{align}
I^B&=\frac{
(q^{(n+1)r} a^{2n+2};q)_{\infty}
}{
(q^{1-(n+1)r} a^{-2n-2};q)_{\infty}
}
\prod_{\alpha<\beta}^{2n+2}
\frac{
(q^{1-r} a^{-2}x_{\alpha}^{-1}x_{\beta}^{-1};q)_{\infty}
}{
(q^{r} a^{2} x_{\alpha}x_{\beta};q)_{\infty}
}
\end{align}
can be checked as all monopoles in theory A have positive dimension
\begin{align}
\label{Mono_Ch_usp2n_2np2}
    &2(n - (n+1)r_a) \sum_{i=1}^n |m_i| - \sum_{i < j}^n |m_i - m_j| - \sum_{i < j}^n |m_i + m_j| 
    \nonumber\\
    &= \sum_{i=1}^n 2\left( i - (n+1)r_a \right) |m_{\sigma(i)}| \ge 2
\end{align}
for $r_a \ge 0$, $m_i \in \Zb$ (with at least one $m_i \ne 0$) and $\sigma$ is a permutation giving the ordering $|m_{\sigma(i)}| \ge |m_{\sigma(j)}|$ for $i < j$. The lower bound is given for $r_a = 0$ and saturated for a single $m_i = \pm 1$. Choosing positive values for $r_a$ will decrease the R-charge of the monopoles but there is still be a range of values $0 < r_a < 1/(n+1)$ so that all monopoles and chirals have positive R-charge.
Theory B has superpotential \cite{Karch:1997ux}
\begin{align}
    W = & V \Pf M \; .
\end{align}

For example, for $n=2$ with $r=\frac18$, $x_{\alpha}=1$ we have 
\begin{align}
I^A&=I^B
\nonumber\\
&=1+15a^2q^{1/8}+120a^4q^{1/4}+679a^6q^{3/8}
+3045a^8q^{1/2}+(a^{-6}+11508a^{10})q^{5/8}
\nonumber\\
&+5(3a^{-4}+7616a^{12})q^{3/4}
+15(7a^{-2}+7548a^{14})q^{7/8}
+(452+308142a^{16})q+\cdots.
\end{align}

\subsection{$USp(2n)$ with $2n+3(+1)$ fundamental chirals}
\label{sec_USp2n_2np4_integrals}
Theory A has gauge group $USp(2n)$ with $N_f = 2n + 3$ fundamental chirals
with R-charge $0$. These all have Neumann boundary
conditions. There is another fundamental chiral with R-charge $2$ which has Dirichlet boundary condition. 
This can be thought of as a higher-rank $USp(2n)$ generalization of the $SU(2)$ gauge theory discussed in section \ref{sec_NR_integral}.

The half-index reads
\begin{align}
\label{USp2n_2np4_hindexA}
\mathbb{II}_{(\mathcal{N},N,D)}^{A}&=
\frac{(q)_{\infty}^n}{n! 2^n} \prod_{i=1}^n \oint \frac{ds_i}{2\pi i s_i}
\prod_{i \ne j}^n (s_i s_j^{-1}; q)_{\infty} \prod_{i \le j}^n (s_i^{\pm} s_j^{\pm}; q)_{\infty}
\nonumber\\
&\times 
\frac{\prod_{i = 1}^n (q^{N_f r/2} a^{N_f} s_i^{\pm}; q)_{\infty}}{\prod_{\alpha = 1}^{2n + 3} \prod_{i = 1}^n (q^{r/2} s_i^{\pm} a x_{\alpha}; q)_{\infty}}. 
\end{align}
It follows from Theorem 4.1 in \cite{MR1147876} that 
the half-index (\ref{USp2n_2np4_hindexA}) is equal to 
\begin{align}
\label{USp2n_2np4_hindexB}
\frac{\prod_{\alpha=1}^{2n + 3} (q^{(n+1)r} a^{N_f - 1} x_{\alpha}^{-1}; q)_{\infty}}{\prod_{\alpha < \beta} (q^r a^2 x_{\alpha} x_{\beta}; q)_{\infty}}
\end{align}
where $\prod_{\alpha = 1}^{2n + 3} x_{\alpha} = 1$. 
The equality of (\ref{USp2n_2np4_hindexA}) and (\ref{USp2n_2np4_hindexB}) can be simply understood as the boundary confining duality where 
the dual theory B has an antisymmetric rank-2 $SU(N_f)$ chiral with R-charge $0$
with Neumann boundary conditions and a chiral in the fundamental representation of $SU(N_f)$
with R-charge $2$ with Dirichlet boundary condition. 

The charges of fields are summarized as follows: 
\begin{align}
\label{USp2n_2np4_charges}
\begin{array}{c|c|c|c|c|c}
& \textrm{bc} & USp(2n) & SU(N_f = 2n + 3) & U(1)_a & U(1)_R \\ \hline
\textrm{VM} & \mathcal{N} & {\bf Adj} & {\bf 1} & 0 & 0 \\
Q_{\alpha} & \textrm{N} & {\bf 2n} & {\bf N_f} & 1 & 0 \\
\widetilde{Q} & \textrm{D} & {\bf 2n} & {\bf 1} & -N_f & 2 \\
 \hline
M_{\alpha \beta} & \textrm{N} & {\bf 1} & {\bf N_f(N_f - 1)/2} & 2 & 0 \\
\widetilde{M}_{\alpha} & \textrm{D} & {\bf 1} & {\bf N_f} & 1 - N_f & 2
\end{array}
\end{align}

The mapping of operators between theory A and theory B is given by
$M_{\alpha \beta} \sim Q_{\alpha} Q_{\beta}$ and $M_{\alpha} \sim Q_{\alpha} \widetilde{Q}$ where the gauge indices are contracted with the $USp(2n)$-invariant rank-$2$ antisymmetric tensor.

The anomalies match for these theories and boundary conditions since
\begin{align}
\label{bdy_USp2n_2np4_anom}
\Acal & = \underbrace{(n+1)\Tr(s^2) + \frac{n(2n+1)}{2}r^2}_{\textrm{VM}, \; \Ncal}
 - \underbrace{\left( \frac{2n+3}{2}\Tr(s^2) + n\Tr(x^2) + n(2n+3)(a-r)^2 \right)}_{Q_{\alpha}, \; N}
  \nonumber \\
 & + \underbrace{\left( \frac{1}{2}\Tr(s^2) + n\big( -(2n+3) a + r \big)^2 \right)}_{\widetilde{Q}, \; D}
   \nonumber \\
  = & - \underbrace{\left( \frac{2n+1}{2} \Tr(x^2) + \frac{(n+1)(2n+3)}{2}(2a-r)^2 \right)}_{M_{\alpha \beta}, \; N}
   + \underbrace{\frac{1}{2} \Tr(x^2) + \frac{2n+3}{2}\big( -2(n+1)a+r \big)^2}_{\widetilde{M}_{\alpha}, \; D}
  \nonumber \\
  = & - n \Tr(x^2) + 2n(n+1)(2n+3)a^2 - \frac{n(2n+3)}{2}r^2.
\end{align}

To summarize, we propose the $USp(2n)$ higher-rank generalization of the boundary confining duality (\ref{bcdual_su2nf3na3})
\begin{align}
\label{bdy_USp2n_2np4_dual}
&\textrm{$USp(2n)$ $+$ $(2n+3)$ fund. chirals $Q_{\alpha}$ $+$ $1$ fund. chiral $\widetilde{Q}$ with b.c. $(\mathcal{N},N,D)$}
\nonumber\\
&\Leftrightarrow 
\textrm{
$SU(2n+3)$ antisym. chiral $M_{\alpha \beta}$ $+$ a fund. chiral $\widetilde{M}_{\alpha}$ with b.c. $(N,D)$}. 
\end{align}

This case also corresponds to a bulk confining duality which
arises as a direct dimensional reduction of the 4d Seiberg duality between $USp(2n)$
with $2n + 4$ flavors and a $USp(2n + 4 - 2n - 4) = $USp(0) theory which is the confining duality example of the dualities presented in \cite{Intriligator:1995ne}. This dimensional
reduction is explained in \cite{Aharony:2013dha}.
The full indices can be computed. In particular the theory A index is well-defined as all monopoles have positive dimension. In fact, given the charges of $\tilde{Q}$, the monopole dimensions are given by
\begin{align}
\label{Mono_Ch_usp2n_2np3p1}
    2n \sum_{i=1}^n |m_i| - \sum_{i < j}^n |m_i - m_j| - \sum_{i < j}^n |m_i + m_j| = \sum_{i=1}^n 2i |m_{\sigma(i)}| \ge 2
\end{align}
for $m_i \in \Zb$ (with at least one $m_i \ne 0$) and $\sigma$ is a permutation giving the ordering $|m_{\sigma(i)}| \ge |m_{\sigma(j)}|$ for $i < j$. The lower bound is saturated for a single $m_i = \pm 1$.
The full-indices are given by
\begin{align}
\label{usp2n_2np3p1_fullA}
I^A&=\frac{1}{n!2^n}
\sum_{m_1,\cdots,m_n\in \mathbb{Z}}
\prod_{i=1}^n
\oint \frac{ds_i}{2\pi i s_i}
\prod_{i\neq j}^n (1-q^{\frac{|m_i-m_j|}{2}}s_i s_j^{-1})
\prod_{i\le j}^{n} (1-q^{\frac{|m_i+m_j|}{2}}s_i^{\pm} s_j^{\pm})
\nonumber\\
&\times 
\prod_{\alpha=1}^{2n+3}
\prod_{i=1}^{n}
\frac{(q^{1-\frac{r}{2}+\frac{|m_i|}{2}} a^{-1} s^{\mp} x_{\alpha}^{-1};q )_{\infty}}
{(q^{\frac{r}{2}+\frac{|m_i|}{2}}as^{\pm}x_{\alpha};q)_{\infty}}
\prod_{i=1}^n 
\frac{
(q^{\frac{N_f r}{2}+\frac{|m_i|}{2}}a^{N_f}s_i^{\mp};q)_{\infty}
}
{
(q^{1-\frac{N_f r}{2}+\frac{|m_i|}{2}}a^{-N_f}s_i^{\pm};q)_{\infty}
}
\nonumber\\
&\times 
q^{(n+\frac32)(1-r)\sum_{i=1}^n |m_i|+(-1+N_f r)\sum_{i=1}^n |m_i|
-\sum_{i=1}^n |m_i|-\frac12\sum_{i<j}|m_i\pm m_j|}
\end{align}
and
\begin{align}
\label{usp2n_2np3p1_fullB}
I^B&=
\prod_{\alpha=1}^{2n+3}
\frac{
(q^{(n+1)r}a^{N_f-1}x_{\alpha}^{-1};q)_{\infty}
}
{
(q^{1-(n+1)r}a^{-N_f+1}x_{\alpha};q)_{\infty}
}
\prod_{\alpha<\beta}^{2n+2}
\frac{
(q^{1-r} a^{-2}x_{\alpha}^{-1}x_{\beta}^{-1};q)_{\infty}
}{
(q^{r} a^{2} x_{\alpha}x_{\beta};q)_{\infty}
}
\end{align}
For example, for $n=2$ with $r=\frac18$, $x_{\alpha}=1$ we have
\begin{align}
I^A&=I^B
\nonumber\\
&=1+21a^2q^{1/8}+231a^4q^{1/4}
+1764a^6q^{3/8}+10479a^{8}q^{1/2}
+(7a^{-6}+51513a^{10})q^{5/8}
\nonumber\\
&+147(a^{-4}+1482a^{12})q^{3/4}
+(1596a^{-2}+814089a^{14})q^{7/8}
+21(567+130526a^{16})q+\cdots.
\end{align}

Both theories have non-zero superpotentials
\begin{align}
    \Wcal_A = & V \\
    \Wcal_b = & \epsilon^{\alpha \alpha_1 \beta_1 \cdots \alpha_{n+1} \beta_{n+1}} \widetilde{M}_{\alpha} M_{\alpha_1 \beta_1} \cdots M_{\alpha_{n+1} \beta_{n+1}}
\end{align}
where $V$ is the minimal monopole in theory A and, through the operator mapping, the theory B superpotential corresponds to imposing the fact that the antisymmetric product of more than $2n$ $Q_{\alpha}$ in theory A must vanish. Similarly to the case of $SU(N)$ discussed in section~\ref{sec_GNR_integrals}, the theory A monopole superpotential, after shifting the R-charge by an appropriate multiple of the $U(1)_a$ charge, the $Q_{\alpha}$ and $\widetilde{Q}$ will have a $U(2n + 4)$ flavor symmetry broken to $SU(2n + 4)$ by the linear monopole superpotential.

Of course, this leads to the obvious conjecture that the general case where theory
A has gauge group $USp(N = 2n)$ and $N_f$ fundamental chirals,
all having Neumann boundary conditions except for one of the fundamental chirals, will have a boundary dual theory B with Dirichlet boundary condition for the $USp(N_f - N - 4)$ vector multiplet. In this case theory B will
have the same chirals as described above along with a $USp(N_f - N - 4)$ vector multiplet
with Dirichlet boundary conditions and $N_f$ fundamental chirals which will have Dirichlet boundary condition except for one with Neumann boundary conditions. The chirals $M_{\alpha \beta}$ and $\widetilde{M}_{\alpha}$ will combine to the $SU(N_f)$ rank-$2$ antisymmetric representation. We will also have Fermis in theory A to cancel the gauge anomaly.

\subsection{$USp(4)$ with $2$ fundamental and $2$ antisymmetric chirals}
\label{sec_GNR_USp4_2_2AS_integral}
Now consider theory A which has gauge group $USp(4)$ with $2$ fundamental and $2$ rank-$2$ antisymmetric chirals \footnote{Here and in other examples in this paper, rank-$2$ antisymmetric representations of $USp(2n)$ groups are reducible representations consisting of a $(n(2n-1) - 1)$-dimensional irrep.\ and a singlet.}, all with Neumann boundary conditions. 
The boundary anomaly is evaluated as
\begin{align}
\label{bdy_USp4_2_2AS_anomA}
\Acal &=\underbrace{3\Tr(s^2) + 5r^2}_{\textrm{VM}, \; \Ncal}
 - \underbrace{\left( \Tr(s^2) + 2\Tr(x^2) + 4(a-r)^2 \right)}_{Q_{\alpha}, \; N}
 \nonumber\\
& - \underbrace{\left( 2\Tr(s^2) + 3\Tr(\tilde{x}^2) + 6(A-r)^2 \right)}_{\Phi_I, \; N}
\nonumber\\
&=- 2\Tr(x^2) - 3\Tr(\tilde{x}^2) - 4a^2 + 8ar - 6A^2 + 12Ar - 5r^2. 
\end{align}
As it is free from the gauge anomaly, this Neumann boundary condition for gauge field is quantum mechanically consistent. 

We can express the half-index as
\begin{align}
\label{USp4_2_2AS_hindexA}
\mathbb{II}_{(\mathcal{N},N,N)}^{A}
&=\frac{(q)_{\infty}^2}{8} \prod_{i=1}^2 \oint \frac{ds_i}{2\pi i s_i}
(s_1^{\pm} s_2^{\mp}; q)_{\infty} \prod_{i \le j}^2 (s_i^{\pm} s_j^{\pm}; q)_{\infty}
\frac{1}{\prod_{\alpha = 1}^2 \prod_{i = 1}^2 (q^{r_a/2} a s_i^{\pm} x_{\alpha}; q)_{\infty}}
\nonumber \\
& \times
\frac{1}{\prod_{I = 1}^2 (q^{r_A/2} A s_1^{\pm} s_2^{\mp} \tilde{x}_I; q)_{\infty} (q^{r_A/2} A s_1^{\pm} s_2^{\pm} \tilde{x}_I; q)_{\infty} (q^{r_A/2} A \tilde{x}_I; q)_{\infty}^2}, 
\end{align}
which again takes the form of the Askey-Wilson type $q$-beta integral. 
By means of Theorem 4.1 in \cite{MR1266569}, 
we can rewrite the half-index (\ref{USp4_2_2AS_hindexA}) as
\begin{align}
\label{USp4_2_2AS_hindexB}
&
\frac{(q^{2r_a + 2r_A} a^4 A^4; q)_{\infty}}
{(q^{r_a} a^2; q)_{\infty} \left( \prod_{I \le J}^{2} (q^{r_A} A^2 \tilde{x}_I \tilde{x}_J; q)_{\infty} \right) \left( \prod_{I=1}^{2} (q^{r_A/2} A \tilde{x}_I; q)_{\infty} (q^{r_a + r_A/2} a^2 A \tilde{x}_I; q)_{\infty} \right)}
\nonumber\\
&\times 
\frac{1}{
\prod_{\alpha \le \beta}^{2} (q^{r_a + r_A} a^2 A^2 x_{\alpha} x_{\beta}; q)_{\infty}
}
\end{align}
where $\prod_{\alpha = 1}^2 x_{\alpha} = 1 = \prod_{I = 1}^2 \tilde{x}_I$. 
The expression (\ref{USp4_2_2AS_hindexB}) enables us to find the dual boundary condition in theory B 
which consists of several chiral multiplets; 
$2$ singlet chirals $M$, $V$, 
two $SU(2)_A$ fundamental chirals $\phi_I$, $B_{I}$, 
an $SU(2)_A$ symmetric chiral $\phi_{IJ}$, 
and an $SU(2)_a$ symmetric chiral $B_{\alpha\beta}$. 
The content and boundary conditions of both theories are summarized as
\begin{align}
\label{USp4_2_2AS_charges}
\begin{array}{c|c|c|c|c|c|c|c}
& \textrm{bc} & USp(4) & SU(2)_a & SU(2)_A & U(1)_a & U(1)_A & U(1)_R \\ \hline
\textrm{VM} & \mathcal{N} & {\bf Adj} & {\bf 1} & {\bf 1} & 0 & 0 & 0 \\
Q_{\alpha} & \textrm{N} & {\bf 4} & {\bf 2} & {\bf 1} & 1 & 0 & 0 \\
\Phi_I & \textrm{N} & {\bf 6} & {\bf 1} & {\bf 2} & 0 & 1 & 0 \\
 \hline
M & \textrm{N} & {\bf 1} & {\bf 1} & {\bf 1} & 2 & 0 & 0 \\
\phi_I & \textrm{N} & {\bf 1} & {\bf 1} & {\bf 2} & 0 & 1 & 0 \\
\phi_{IJ} & \textrm{N} & {\bf 1} & {\bf 1} & {\bf 3} & 0 & 2 & 0 \\
B_{\alpha \beta} & \textrm{N} & {\bf 1} & {\bf 3} & {\bf 1} & 2 & 2 & 0 \\
B_I & \textrm{N} & {\bf 1} & {\bf 1} & {\bf 2} & 2 & 1 & 0 \\
V & \textrm{D} & {\bf 1} & {\bf 1} & {\bf 1} & -4 & -4 & 2
\end{array}
\end{align}

The mapping of operators between theory A and theory B is given by
$M \sim Q_1 Q_2$, $\phi_I \sim \Tr \Phi_I$, $\phi_{IJ} \sim \Tr(\Phi_I \Phi_J)$, $B_{\alpha \beta} \sim Q_{\alpha} \Phi_1 \Phi_2 Q_{\beta}$, $B_I \sim Q_1 \Phi_I  Q_2$ with suitable contractions of the gauge indices using a convention that expressions with two $Q_{\alpha}$ include full contraction of indices, and $V$ is dual to the minimal monopole operator in theory A in the bulk.

The anomalies of boundary condition for theory B is computed as
\begin{align}
\label{bdy_USp4_2_2AS_anom}
&\Acal = - \underbrace{\frac{1}{2}(2a-r)^2}_{M, \; N}
  - \underbrace{\left( \frac{1}{2}\Tr(\tilde{x}^2) + (A-r)^2 \right)}_{\phi_I, \; N}
  - \underbrace{\left( 2\Tr(\tilde{x}^2) + \frac{3}{2}(2A-r)^2 \right)}_{\phi_{IJ}, \; N}
  \nonumber \\
  & - \underbrace{\left( 2\Tr(x^2) + \frac{3}{2}(2a+2A-r)^2 \right)}_{B_{\alpha \beta}, \; N}
  - \underbrace{\left( \frac{1}{2}\Tr(\tilde{x}^2) + (2a+A-r)^2 \right)}_{B_I, \; N}
   + \underbrace{\frac{1}{2}(-4a-4A+r)^2}_{V, \; D}. 
\end{align}
This precisely agrees with the boundary anomaly (\ref{bdy_USp4_2_2AS_anomA}) for theory A. 

Therefore we propose the following boundary confining duality: 
\begin{align}
\label{bcdual_USp4_2_2AS}
&\textrm{$USp(4)$ $+$ $2$ fund. chirals $Q_{\alpha}$ $+$ $2$ antisym. $\Phi_I$ with b.c. $(\mathcal{N},N,N)$}
\nonumber\\
&\Leftrightarrow 
\textrm{
a singlet $M$ $+$ an $SU(2)_A$ fund. $\phi_I$ $+$ 
an $SU(2)_A$ sym. chiral $\phi_{IJ}$}
\nonumber\\
&\textrm{an $SU(2)_a$ sym. chiral $B_{\alpha\beta}$ $+$ and $SU(2)_A$ fund. chiral $B_I$}
\nonumber\\
&\textrm{$+$ a singlet chiral $V$ with b.c. $(N,N,N,N,N,D)$}. 
\end{align}

We remark that all monopoles in theory A have positive dimension
\begin{align}
\label{Mono_Ch_usp4_2_2}
   -2r_a(|m_1| + |m_2|) + (1 - 2r_A)|m_1 - m_2| + (1 - 2r_A)|m_1 + m_2| \ge 2\max \{|m_1|, |m_2| \} \ge 2
\end{align}
for $m_i \in \Zb$ (with at least one $m_i \ne 0$) where the inequalities are given for $r_a = r_A = 0$ and the bound is saturated for a single $m_i = \pm 1$ and for $|m_1| = |m_2| = 1$. Choosing positive values for $r_a$ or $r_A$ will decrease the R-charge of the monopoles but there will still be a range of values so that all monopoles have positive R-charge. For $r_a > 0$, the minimal monopole has $|m_1| = |m_2| = 1$ and $U(1)_A$ and $U(1)_A$ charges equal to $-4$.
Thus there is a bulk confining duality, which we believe has not been considered before, but appears to be similar to the symplectic dualities considered in \cite{Amariti:2022iaz}. \footnote{There are however some differences in the operator mapping and global symmetries so this should be investigated further. It would also be interesting to explore the possibility of generalizing the example in this section to other $USp(2n)$ confining dualities with two antisymmetric chirals.} Theory B has a superpotential which we don't write explicitly where $V$ plays the role of a Lagrange multiplier imposing an algebraic constraint on a linear combination of $\det B_{\alpha \beta}$, $B_I B_J \Phi$, $B_I M \Phi$ and $M^2 \Phi$ where $\Phi$ is used to schematically indicate various algebraic combinations of the $\phi_I$ and $\phi_{IJ}$ chirals. The precise details can be deduced from the operator mapping to theory A.
The matching of full indices 
\begin{align}
I^A&=
\frac{1}{8}\sum_{m_1,m_2}
\prod_{i=1}^2 
\oint  \frac{ds_i}{2\pi is_i}
(1-q^{|m_i|}s_i^{\pm 2})
(1-q^{\frac{|m_1-m2|}{2}}s_1^{\pm}s_2^{\mp})
(1-q^{\frac{|m_1+m2|}{2}}s_1^{\pm}s_2^{\pm})
\nonumber\\
&\times 
\prod_{I=1}
\frac{
(q^{1-\frac{r_A}{2}+\frac{|m_1-m_2|}{2}}As_1^{\mp}s_2^{\pm} \tilde{x}_{I}^{-1};q)_{\infty}
}
{
(q^{\frac{r_A+|m_1-m_2|}{2}}A^{-1}s_1^{\pm}s_2^{\mp} \tilde{x}_{I};q)_{\infty}
}
\frac{
(q^{1-\frac{r_A}{2}+\frac{|m_1+m_2|}{2}}As_1^{\mp}s_2^{\mp} \tilde{x}_{I}^{-1};q)_{\infty}
}
{
(q^{\frac{r_A+|m_1+m_2|}{2}}A^{-1}s_1^{\pm}s_2^{\pm} \tilde{x}_{I};q)_{\infty}
}
\nonumber\\
&\times 
\frac{
(q^{1-\frac{r_A}{2}}A \tilde{x}_{I}^{-1};q)_{\infty}^2
}
{
(q^{\frac{r_A}{2}}A^{-1} \tilde{x}_{I};q)_{\infty}^2
}
q^{(1-r_a)\sum_{i=1}^2|m_i|+(1-r_A)|m_1\pm m_2|-\frac12|m_1\pm m_2|-\sum_{i=1}^2|m_i|}
\nonumber\\
&\times 
a^{-2\sum_{i=1}^2|m_i|} A^{-2|m_1\pm m_2|}
\end{align}
and 
\begin{align}
I^B&=
\frac{(q^{2r_a+2r_A}a^4 A^4;q)_{\infty}}
{(q^{1-2r_a-2r_A}a^{-4}A^{-4};q)_{\infty}}
\frac{(q^{1-r_a}a^{-2};q)}{(q^{r_a}a^2;q)_{\infty}}
\prod_{I\le J}^{2}
\frac{(q^{1-r_A}A^{-2}\tilde{x}_I^{-1} \tilde{x}_J^{-1};q)_{\infty}}
{(q^{r_A}A^2\tilde{x}_I \tilde{x}_J;q)_{\infty}}
\nonumber\\
&\times 
\frac{
(q^{1-\frac{r_A}{2}}A^{-1}\tilde{x}_{I}^{-1};q)_{\infty}
}
{
(q^{\frac{r_A}{2}}A\tilde{x}_{I};q)_{\infty}
}
\frac{
(q^{1-\frac{r_a+r_A}{2}}a^{-2}A^{-1}\tilde{x}_{I}^{-1};q)_{\infty}
}
{
(q^{\frac{r_a+r_A}{2}}a^2A\tilde{x}_{I};q)_{\infty}
}
\prod_{\alpha\le \beta}^2
\frac{
(q^{1-r_a+r_A}a^{-2}A^{-2}x_{\alpha}^{-1}x_{\beta}^{-1};q)_{\infty}
}
{
(q^{r_a+r_A}a^2A^2 x_{\alpha}x_{\beta};q)_{\infty}
}
\end{align}
can be checked. 
For example, by setting $x_{\alpha}=\tilde{x}_{I}=1$ and $r_a=r_A=\frac18$, 
we find that both full-indices can be expanded as
\begin{align}
I^A&=I^B
\nonumber\\
 = & 1 + 2Aq^{1/16} + (a^2 + 6A^2)q^{1/8} + (4a^2A + 10A^3)q^{3/16} 
 + (a^4 + 13a^2A^2 + 20A^4)q^{1/4}
\nonumber\\
&+(4a^4A+28a^2A^3+30A^5)q^{5/16}+
(a^6+16a^4A^2+58a^2A^4+50A^6)q^{3/8}+\cdots. 
\end{align}

From the above expansion, we can study several types of  gauge invariant local operators in theory A and the composite operators in theory B. 
For example, the terms involving only the fugacity $A$ enumerate the gauge invariant operators consisting of antisymmetric chirals in theory A. They can be extracted for the full index by defining $t = Aq^{1/16}$ and taking the limit $q \rightarrow 0$ while keeping $a$ and $t$ fixed. The resulting expression is given by 
\begin{align}
\frac{1}{(1-t)^2(1-t^2)^3}
&=1+2t+6t^2+10t^3+20t^4 + 30t^5 + 50t^6+70t^7+\cdots, 
\end{align}
which is essentially the Poincar\'{e} series $P(C_{2,2};t)$ of the pure trace algebra $C_{2,2}$  of $2$ generic $2\times 2$ matrices \cite{MR2303199}.

\subsection{$USp(4)$ with $3$ antisymmetric chirals}
\label{sec_GNR_USp4_3AS_integral}
It is intriguing to add more rank-$2$ chiral multiplets to theory A. 
Let us consider theory A with gauge group $USp(4)$ and $3$ rank-$2$ antisymmetric chirals, all with Neumann boundary conditions. 
These boundary conditions lead to the boundary 't Hoof anomaly
\begin{align}
\label{bdy_USp4_3AS_anomA}
\Acal&= \underbrace{3\Tr(s^2) + 5r^2}_{\textrm{VM}, \; \Ncal}
- \underbrace{\left( 3\Tr(s^2) + 3\Tr(\tilde{x}^2) + 9(A-r)^2 \right)}_{\Phi_I, \; N}
\nonumber\\
&=- 3\Tr(\tilde{x}^2) - 9A^2 + 18Ar - 4r^2. 
\end{align}
As there is no gauge anomaly, these boundary conditions are consistent with Neumann boundary conditions preserving $USp(4)$ gauge group. 

The Neumann half-index of theory A is given by
\begin{align}
\label{USp4_3AS_hindexA}
&
\mathbb{II}_{(\mathcal{N},N)}^A
\nonumber\\
&=\frac{(q)_{\infty}^2}{8} \prod_{i=1}^2 \oint \frac{ds_i}{2\pi i s_i}
\frac{(s_1^{\pm} s_2^{\mp}; q)_{\infty} \prod_{i \le j}^2 (s_i^{\pm} s_j^{\pm}; q)_{\infty}}
{\prod_{I = 1}^3 (q^{r_A/2} A s_1^{\pm} s_2^{\mp} \tilde{x}_I; q)_{\infty} (q^{r_A/2} A s_1^{\pm} s_2^{\pm} \tilde{x}_I; q)_{\infty} (q^{r_A/2} A \tilde{x}_I; q)_{\infty}^2}
\end{align}
where $\prod_{I = 1}^3 \tilde{x}_I = 1$.
According to Theorem 4.4 in \cite{MR1266569}, 
we can write the half-index (\ref{USp4_3AS_hindexA}) as
\begin{align}
\label{USp4_3AS_hindexB}
\frac{(q^{1 + 3r_A/2} A^3; q)_{\infty}}{\prod_{I=1}^{3} (q^{r_A/2} A \tilde{x}_I; q)_{\infty} \prod_{J \ge I}^{3} (q^{r_A} A^2 \tilde{x}_I \tilde{x}_J; q)_{\infty}}. 
\end{align}
Again this immediately tells the dual boundary condition in theory B. 
We show the content and boundary conditions of both theories in the following table: 
\begin{align}
\label{USp4_3AS_charges}
\begin{array}{c|c|c|c|c|c}
& \textrm{bc} & USp(4) & SU(3) & U(1)_A & U(1)_R \\ \hline
\textrm{VM} & \mathcal{N} & {\bf Adj} & {\bf 1} & 0 & 0 \\
\Phi_I & \textrm{N} & {\bf 6} & {\bf 3} & 1 & 0 \\
 \hline
\phi_I & \textrm{N} & {\bf 1} & {\bf 3} & 1 & 0 \\
\phi_{IJ} & \textrm{N} & {\bf 1} & {\bf 6} & 2 & 0 \\
V & \textrm{D} & {\bf 1} & {\bf 1} & -3 & 0
\end{array}
\end{align}
The 't Hooft anomaly for the boundary conditions in theory B is given by
\begin{align}
\label{bdy_USp4_3AS_anomB}
\Acal= & - \underbrace{\left( \frac{1}{2}\Tr(\tilde{x}^2) + \frac{3}{2}(A-r)^2 \right)}_{\phi_I, \; N}
  - \underbrace{\left( \frac{5}{2}\Tr(\tilde{x}^2) + 3(2A-r)^2 \right)}_{\phi_{IJ}, \; N}
   + \underbrace{\frac{1}{2}(-3A-r)^2}_{V, \; D}. 
\end{align}
This is equal to the boundary anomaly (\ref{bdy_USp4_3AS_anomA}) for theory A. 

The mapping of operators between theory A and theory B is given by
$\phi_I \sim \omega \Phi_I$, $\phi_{IJ} \sim \Phi_I \Phi_J$ and $V$ is dual to the monopole operator with flux $|m_1| = |m_2| = 1$ in theory A in the bulk. 

We therefore find the boundary confining duality 
\begin{align}
\label{bcdual_USp4_3AS_dual}
&\textrm{$USp(4)$ $+$ $3$ antisym. chirals $\Phi_{I}$ with b.c. $(\mathcal{N},N)$}
\nonumber\\
&\Leftrightarrow 
\textrm{
an $SU(3)$ fund. $\phi_I$ $+$ 
an $SU(3)$ rank3-sym. chiral $\phi_{IJ}$ $+$ 
a singlet chiral $V$
}
\nonumber\\
&\textrm{with b.c. $(N,N,D)$}. 
\end{align}

In this case the bulk theory A has monopoles of zero (or negative) dimension since the dimensions are given by
\begin{align}
\label{Mono_Ch_usp4_0_3}
    (2 - 3r_A)|m_1 - m_2| + (2 - 3r_A)|m_1 + m_2| - 2|m_1| - 2|m_2| \ge 2\big| |m_1| - |m_2| \big| \ge 0
\end{align}
for $m_i \in \Zb$ (with at least one $m_i \ne 0$) where the inequalities are given for $r_A = 0$.
We can see that the lower bound is indeed saturated if $r_A = 0$ when $|m_1| = |m_2|$. Choosing positive values for $r_A$ will decrease the R-charge of the monopoles so in that case there will be negative dimension monopoles. Similarly, in theory B any choice of $r_A$ results in at least one chiral having non-positive dimension.
Therefore this is an example of a boundary duality which does not have a corresponding standard bulk Seiberg-like bulk duality of IR unitary CFTs.

\subsection{$USp(6)$ with $2$ antisymmetric chirals}
\label{sec_GNR_USp6_2AS_integral}
Consider theory A with gauge group $USp(6)$ and $2$ rank-$2$ antisymmetric chirals, all with Neumann boundary conditions. 
The boundary 't Hooft anomaly is given by
\begin{align}
\label{bdy_USp6_2AS_anomA}
\Acal = & \underbrace{4\Tr(s^2) + \frac{21}{2}r^2}_{\textrm{VM}, \; \Ncal}
 - \underbrace{\left( 4\Tr(s^2) + \frac{15}{2}\Tr(\tilde{x}^2) + 15(A-r)^2 \right)}_{\Phi_I, \; N}
  \nonumber \\
  = & - \frac{15}{2}\Tr(\tilde{x}^2) - 15A^2 + 30Ar - \frac{9}{2}r^2. 
\end{align}

The Neumann half-index reads
\begin{align}
\label{bdy_USp6_2AS_hindexA}
&\mathbb{II}_{(\mathcal{N},N)}^A
\nonumber\\
&=\frac{(q)_{\infty}^3}{48} \prod_{i=1}^3 \oint \frac{ds_i}{2\pi i s_i}
\frac{\left( \prod_{i < j}^3 (s_i^{\pm} s_j^{\mp}; q)_{\infty} \right) \prod_{i \le j}^3 (s_i^{\pm} s_j^{\pm}; q)_{\infty}}
{\prod_{I = 1}^2 (q^{r_A/2} A \tilde{x}_I; q)_{\infty}^3 \prod_{i < j}^3 (q^{r_A/2} A s_i^{\pm} s_j^{\mp} \tilde{x}_I; q)_{\infty} (q^{r_A/2} A s_i^{\pm} s_j^{\pm} \tilde{x}_I; q)_{\infty}}
\end{align}
where $\prod_{I = 1}^2 \tilde{x}_I = 1$.
According to Theorem 4.6 in \cite{MR1266569}, it follows that 
the integral (\ref{bdy_USp6_2AS_hindexA}) is equal to 
\begin{align}
\label{bdy_USp6_2AS_hindexB}
 \frac{(q^{1 + 3r_A} A^6; q)_{\infty}}{(q^{2r_A} A^4; q)_{\infty} \prod_{I=1}^{2} (q^{r_A/2} A \tilde{x}_I; q)_{\infty} \prod_{J \ge I}^{2} (q^{r_A} A^2 \tilde{x}_I \tilde{x}_J; q)_{\infty} \prod_{K \ge J \ge I}^2 (q^{3r_A/2} A^3 \tilde{x}_I \tilde{x}_J \tilde{x}_K; q)_{\infty}}. 
\end{align}
From the infinite product (\ref{bdy_USp6_2AS_hindexB}) we can read the dual boundary condition in theory B. 
Theory B involves 
an $SU(2)$ fundamental chiral $\phi_I$, 
an $SU(2)$ adjoint chiral $\phi_{IJ}$, 
an $SU(2)$ symmetric chiral $\phi_{IJK}$, 
two singlet chirals $\phi$, $V$. 
They obey the boundary conditions $(N,N,N,N,D)$. 
The anomalies for theory B is given by
\begin{align}
\label{bdy_USp6_2AS_anom}
\Acal = 
 & - \underbrace{\left( 5\Tr(\tilde{x}^2) + 2(3A-r)^2 \right)}_{\phi_{IJK}, \; N}
  - \underbrace{\frac{1}{2}(4A-r)^2}_{\phi, \; N}
   + \underbrace{\frac{1}{2}(-6A-r)^2}_{V, \; D}
\end{align}
This precisely agrees with the anomaly (\ref{bdy_USp6_2AS_anomA}) in theory A. 

The content and boundary conditions of both theories are summarized as
\begin{align}
\label{USp6_2AS_charges}
\begin{array}{c|c|c|c|c|c}
& \textrm{bc} & USp(6) & SU(2) & U(1)_A & U(1)_R \\ \hline
\textrm{VM} & \mathcal{N} & {\bf Adj} & {\bf 1} & 0 & 0 \\
\Phi_I & \textrm{N} & {\bf 15} & {\bf 2} & 1 & 0 \\
 \hline
\phi_{I} & \textrm{N} & {\bf 1} & {\bf 2} & 1 & 0 \\
\phi_{IJ} & \textrm{N} & {\bf 1} & {\bf 3} & 2 & 0 \\
\phi_{IJK} & \textrm{N} & {\bf 1} & {\bf 4} & 3 & 0 \\
\phi & \textrm{N} & {\bf 1} & {\bf 1} & 4 & 0 \\
V & \textrm{D} & {\bf 1} & {\bf 1} & -6 & 0
\end{array}
\end{align}

We propose a confining boundary duality 
\begin{align}
\label{bcdual_USp6_2AS_dual}
&\textrm{$USp(6)$ $+$ $2$ antisym. chirals $\Phi_{I}$ with b.c. $(\mathcal{N},N)$}
\nonumber\\
&\Leftrightarrow 
\textrm{
an $SU(2)$ fund. $\phi_I$ $+$ 
an $SU(2)$ adj. chiral $\phi_{IJ}$ $+$ 
}
\nonumber\\
&\textrm{
$+$ an $SU(2)$ sym. chiral $\phi_{IJK}$
}
\nonumber\\
&
\textrm{
$+$ two singlet chirals $\phi$, $V$ with b.c. $(N,N,N,N,D)$
}. 
\end{align}

The mapping of operators between theory A and theory B is given by
$\phi_I \sim \omega \Phi_I$, $\phi_{IJ} \sim \Phi_I \Phi_J$, $\phi_{IJK} \sim \Phi_I \Phi_J \Phi_K$, $\phi \sim \Phi_1^2 \Phi_2^2$ and $V$ is dual to the monopole operator with $|m_1| = |m_2| = 1$ and $|m_3| = 0$ in theory A in the bulk. 

In this case in the bulk theory A has monopoles of zero (or negative) dimension since the dimensions are given by
\begin{align}
\label{Mono_Ch_usp6_0_2}
    (1 - 2r_A)\sum_{i < j}^3 \left( |m_i - m_j| + |m_i + m_j| \right) - 2 \sum_{i = 1}^3 |m_i| \ge 2\left( \max \{ |m_i| \} - \min \{ |m_i| \} \right) \ge 0
\end{align}
for $m_i \in \Zb$ (with at least one $m_i \ne 0$) and the inequalities are given for $r_A = 0$. We can see that the lower bound is indeed saturated when $|m_1| = |m_2| = |m_3|$. Choosing positive values for $r_A$ will decrease the R-charge of the monopoles so in that case there will be negative dimension monopoles. Similarly, in theory B any choice of $r_A$ results in at least one chiral having non-positive dimension. 
Therefore this is another example of a boundary duality which does not seem to have a corresponding standard bulk Seiberg-like bulk duality of IR unitary CFTs.

\subsection{$SO(N)$ with $N-2$ fundamental chirals}
\label{sec_GAW_SO_integrals}
Let us consider a case with orthogonal gauge group. 
We take theory A with gauge group $SO(N)$ and $N - 2$ fundamental chirals $Q_{\alpha}$
with R-charge $0$. We choose the $\mathcal{N}=(0,2)$ boundary condition 
so that these all have Neumann boundary conditions.
Then the boundary anomaly is given by
\begin{align}
\label{bdy_SON_Nm2_anomA}
\Acal & = \underbrace{(N-2)\Tr(s^2) + \frac{N(N-1)}{4}r^2}_{\textrm{VM}, \; \Ncal}
 - \underbrace{\left( (N-2)\Tr(s^2) + \frac{N}{2}\Tr(x^2) + \frac{N(N-2)}{2}(a-r)^2 \right)}_{Q_{\alpha}, \; N}
    \nonumber \\
  = & - \frac{N}{2} \Tr(x^2) - \frac{N(N-2)}{2}a^2 + N(N-2)ar - \frac{N(N-3)}{4}r^2. 
\end{align}
As it is free from the gauge anomaly, 
the set of Neumann boundary conditions is quantum mechanically consistent. 

We can write the half-index as
\begin{align}
\label{bdy_SON_Nm2_hindexA}
\mathbb{II}_{(\mathcal{N},N)}^{A}
&=
\frac{(q)_{\infty}^n}{n! 2^{n - 1 + \epsilon}} \prod_{i=1}^n \oint \frac{ds_i}{2\pi i s_i}
(s_i^{\pm}; q)_{\infty}^{\epsilon}
\prod_{i < j}^n (s_i^{\pm} s_j^{\mp}; q)_{\infty} (s_i^{\pm} s_j^{\pm}; q)_{\infty}
 \nonumber \\
& \times
\frac{1}{\prod_{\alpha = 1}^{N - 2} \left( (q^{r/2} a x_\alpha; q)_{\infty}^{\epsilon} \prod_{i = 1}^n (q^{r/2} s_i^{\pm} a x_\alpha; q)_{\infty} \right)}
\end{align}
where $N = 2n + \epsilon$ with $n \in \Zb$ and $\epsilon \in \{0, 1\}$. 
By means of Theorems 7.14 and 7.16 in \cite{MR1139492}, 
the half-index (\ref{bdy_SON_Nm2_hindexA}) can be simply expressed as
\begin{align}
\label{bdy_SON_Nm2_hindexB}
 \frac{\left( q^{1 + (N/2 - 1)r} a^{N-2}; q \right)_{\infty}}{\prod_{\alpha \le \beta}^{N-2} (q^r a^2 x_\alpha x_\beta; q)_{\infty}}. 
\end{align}

Note that this identity can be derived as a limit of the symplectic identity discussed in section~\ref{sec_USp2n_2np2_integrals} \cite{MR1139492}. Specifically, combining the axial and flavor fugacities as $X_{\alpha} = a x_{\alpha}$ ($1 \le \alpha \le 2n+3$), then taking $X_{2n+1} = -1$, $X_{2n+3} = q^{1/2}$ and $X_{2n+3} = -q^{1/2}$, before redefining $X_{\alpha} = a x_{\alpha}$ ($1 \le \alpha \le 2n$) with $\prod_{\alpha = 1}^{2n} x_{\alpha} = 1$ gives the above identity for $SO(2n+1)$ after some simple manipulations including using the $q$-Pochhammer identity $(\pm z; q)_{\infty} (\pm q^{1/2}z; q)_{\infty} = (z^2; q)_{\infty}$. Taking also $X_{2n} = 1$ gives the identity for $SO(2n)$. The same limits can be taken for other symplectic identities with at least $3$ fundamental chirals\footnote{The crucial point is that this limit specialising the $3$ or $4$ flavor fugacities for fundamental chirals results in part of the fundamental chiral contribution to the half-index combining with the $USp(2n)$ vector multiplet contribution to give the contribution of an $SO(2n+1)$ or $SO(2n)$ vector multiplet.} and this is considered further in section~\ref{sec_SON_N_integrals} for the identities in section~\ref{sec_USp2n_2np4_integrals}.

We can also generalize these identities by including fugacities for discrete
$\Zb_2$ symmetries $\mathcal{C}$ and $\mathcal{M}$. We present the results in section~\ref{sec_SON_Nm2_discrete_integrals}.

The equation (\ref{bdy_SON_Nm2_hindexB}) is identified with the half-index of the dual boundary condition in theory B 
that has a symmetric rank-$2$ $SU(N_f = N - 2)$ chiral $M_{\alpha\beta}$ with R-charge $0$
with Neumann boundary conditions and
a singlet chiral $V$ with $U(1)_a$ charge $-N_f$
and R-charge $0$ with Dirichlet boundary condition. Note that in comparison to the case of symplectic or unitary gauge groups with only fundamental (and antifundamental) chirals, here the singlet $V$ has R-charge $0$ rather than $2$. 

The field content and boundary conditions are summarized as
\begin{align}
\label{SON_Nm2_charges}
\begin{array}{c|c|c|c|c|c}
& \textrm{bc} & SO(N) & SU(N_f = N - 2) & U(1)_a & U(1)_R \\ \hline
\textrm{VM} & \mathcal{N} & {\bf Adj} & {\bf 1} & 0 & 0 \\
Q_{\alpha} & \textrm{N} & {\bf N} & {\bf N_f} & 1 & 0 \\
 \hline
M_{\alpha \beta} & \textrm{N} & {\bf 1} & {\bf N_f(N_f + 1)/2} & 2 & 0 \\
V & \textrm{D} & {\bf 1} & {\bf 1} & -N_f & 0
\end{array}
\end{align}

The anomaly for theory B is computed as
\begin{align}
\label{bdy_SON_Nm2_anomB}
\Acal  = & - \underbrace{\left( \frac{N}{2} \Tr(x^2) + \frac{(N-1)(N-2)}{4}(2a-r)^2 \right)}_{M_{\alpha \beta}, \; N}
   + \underbrace{\frac{1}{2}\big( -(N-2)a - r \big)^2}_{V, \; D}, 
\end{align}
which precisely matches with the anomaly (\ref{bdy_SON_Nm2_anomA}) for theory A. 

The mapping of operators between theory A and theory B is given by
$M_{\alpha \beta} \sim Q_{\alpha} Q_{\beta}$ where the gauge indices are contracted with the $SO(N)$-invariant rank-$2$ symmetric tensor and $V$ is dual to the minimal monopole operator in theory A.

The boundary confining duality is summarized as
\begin{align}
\label{bdy_SON_Nm2_dual}
&\textrm{$SO(N)$ $+$ $(N-2)$ fund. chirals $Q_{\alpha}$ with b.c. $(\mathcal{N},N)$}
\nonumber\\
&\Leftrightarrow 
\textrm{
an $SU(N-2)$ sym. $M_{\alpha\beta}$ $+$ 
a singlet chiral $V$ with b.c. $(N,D)$}. 
\end{align}

In this case in the bulk theory A has monopoles of zero (or negative) dimension since the dimensions are given by
\begin{align}
\label{Mono_Ch_soN_Nm2}
    (2(n-1) - (N-2)r_a) \sum_{i = 1}^n |m_i| - \sum_{i < j}^n \left( |m_i - m_j| + |m_i + m_j| \right)
    \ge \sum_{i=1}^n 2(i-1)|m_{\sigma(i)}| \ge 0
\end{align}
for $m_i \in \Zb$ (with at least one $m_i \ne 0$) and $\sigma$ is a permutation giving the ordering $|m_{\sigma(i)}| \ge |m_{\sigma(j)}|$ for $i < j$. The inequalities are given for $r_a = 0$ and we can see that the lower bound is indeed saturated when a single $|m_i| = 1$. Choosing positive values for $r_a$ will (for $N > 2$) decrease the R-charge of the monopoles so in that case there will be negative dimension monopoles.
Therefore this is an another example of a boundary duality which does not originate from the standard Seiberg-like bulk duality of IR unitary CFTs.

\section{Gustafson-Rakha integrals}
\label{sec_GR_integral}
Here we consider the case where theory A has general $USp(2n)$ or $SU(N)$ gauge group with rank-$2$ antisymmetric chirals as well as fundamental chirals. We begin with the simplest example involving a symplectic gauge group. The Neumann half-indices can be identified with the Gustafson-Rakha integrals \cite{MR1266569}.

\subsection{$USp(2n)$ with rank-$2$ antisymmetric and $5(+1)$ fundamental chirals}
\label{sec_GNR_USp_AS_integrals}
Consider theory A with a $USp(2n)$ vector multiplet obeying Neumann boundary condition, 
a $USp(2n)$ rank-2 antisymmetric chiral $\Phi$ with Neumann boundary condition, 
5 fundamental chirals $Q_{\alpha}$ with Neumann boundary conditions, 
and one fundamental chiral $\widetilde{Q}$ with Dirichlet boundary conditions. 
The boundary 't Hooft anomaly is 
\begin{align}
\label{bdy_USp2n_AS_6_anomA}
\Acal & = \underbrace{(n+1)\Tr(s^2) + \frac{n(2n+1)}{2}r^2}_{\textrm{VM}, \; \Ncal}
 - \underbrace{\left( \frac{5}{2}\Tr(s^2) + n\Tr(x^2) + 5n(a-r)^2 \right)}_{Q_{\alpha}, \; N}
  \nonumber \\
 & - \underbrace{\left( (n-1)\Tr(s^2) + \frac{n(2n-1)}{2}(A - r)^2 \right)}_{\Phi, \; \Ncal}
  + \underbrace{\left( \frac{1}{2}\Tr(s^2) + n\big( (2 - 2n)A - 5a + r \big)^2 \right)}_{\widetilde{Q}, \; D}
  \nonumber\\
  = & - n \Tr(x^2) + 20n a^2 + 20n(n-1)Aa + \frac{n(2n-3)(4n-3)}{2}A^2 -(2n-3)n Ar - 3n r^2. 
\end{align}
Again this is a consistent $\mathcal{N}=(0,2)$ Neumann boundary condition as it has no gauge anomaly. 

The Neumann half-index takes the form: 
\begin{align}
\label{bdy_USp2n_AS_6_hindexA}
&
\mathbb{II}_{(\mathcal{N,N,N,D})}^{A}
\nonumber\\
&=\frac{(q)_{\infty}^n}{n! 2^n} \prod_{i=1}^n \oint \frac{ds_i}{2\pi i s_i}
\prod_{i \ne j}^n (s_i s_j^{-1}; q)_{\infty} \prod_{i \le j}^n (s_i^{\pm} s_j^{\pm}; q)_{\infty}
\frac{\prod_{i = 1}^n (q^{(n-1)r_A+5r_a/2} A^{2n - 2} a^5 s_i^{\pm}; q)_{\infty}}{\prod_{\alpha = 1}^{5} \prod_{i = 1}^n (q^{r_a/2} s_i^{\pm} a x_{\alpha}; q)_{\infty}}
\nonumber \\
& \times \frac{1}{(q^{r_A/2} A; q)_{\infty}^n \prod_{i < j}^n (q^{r_A/2} A s_i^{\pm} s_j^{\mp}; q)_{\infty} (q^{r_A/2} A s_i^{\pm} s_j^{\pm}; q)_{\infty}}
\end{align}
where $\prod_{\alpha = 1}^{5} x_{\alpha} = 1$.
By making use of Theorem 2.1 in \cite{MR1266569}, 
we can write the Neumann half-index (\ref{bdy_USp2n_AS_6_hindexA}) as
\begin{align}
\label{bdy_USp2n_AS_6_hindexB}
\prod_{\lambda = 1}^n \frac{\prod_{\alpha=1}^{5} (q^{(2n - 1 - \lambda)r_A/2 + 2r_a} A^{2n - 1 - \lambda} a^4 x_{\alpha}^{-1}; q)_{\infty}}{(q^{\lambda r_A/2} A^\lambda; q)_{\infty} \prod_{\alpha < \beta} (q^{(\lambda-1)r_A/2 + r_a} A^{\lambda-1} a^2 x_{\alpha} x_{\beta}; q)_{\infty}} 
\end{align}
We can understand the infinite product (\ref{bdy_USp2n_AS_6_hindexB}) as the half-index of theory B. 
Theory B has $n$ copies of the following chirals (with different global Abelian charges) in representations of the $SU(5)$ global flavor symmetry group:
rank-2 antisymmetric with Neumann boundary conditions, singlet with
Neumann boundary conditions and fundamental with Dirichlet boundary conditions. 
The field content and the charges are given by

\begin{align}
\label{USp2n_AS_6_charges}
\begin{array}{c|c|c|c|c|c|c|c|c}
& \textrm{bc} & USp(2n) & SU(N_f = 5) & U(1)_A & U(1)_a & U(1)_R \\ \hline
\textrm{VM} & \mathcal{N} & {\bf Adj} & {\bf 1} & 0 & 0 & 0 \\
\Phi &\textrm{N}& {\bf n(2n-1)} & {\bf 1} & 1 & 0 & 0 \\
Q_{\alpha} & \textrm{N} & {\bf 2n} & {\bf 5} & 0 & 1 & 0 \\
\widetilde{Q} & \textrm{D} & {\bf 2n} & {\bf 1} & 2 - 2n & -5 & 2 \\
 \hline
M_{\alpha \beta}^{(\lambda)} & \textrm{N} & {\bf 1} & {\bf 10} & \lambda-1 & 2 & 0 \\
\phi^{(\lambda)} & \textrm{N} & {\bf 1} & {\bf 1} & \lambda & 0 & 0 \\
\widetilde{M}_{\alpha}^{(\lambda)} & \textrm{D} & {\bf 1} & {\bf 5} & \lambda + 1 - 2n & -4 & 2
\end{array}
\end{align}
where $\lambda \in \{1, 2, \ldots , n\}$.

The 't Hooft anomaly for theory B is calculated as
\begin{align}
\label{bdy_USp2n_AS_6_anomB}
\Acal 
  &=  - \underbrace{\left( \frac{3n}{2} \Tr(x^2) + 5 \sum_{\lambda = 1}^n \big( (\lambda-1)A + 2a - r\big)^2 \right)}_{M_{\alpha \beta}^{(\lambda)}, \; N}
  - \underbrace{\frac{1}{2} \sum_{\lambda = 1}^n \big( \lambda A - r\big)^2}_{\phi^{(\lambda)}, \; N}
  \nonumber \\
  & + \underbrace{\frac{n}{2} \Tr(x^2) + \frac{5}{2} \sum_{\lambda = 1}^n \big( (\lambda + 1 -2n)A - 4a + r \big)^2}_{\widetilde{M}_{\alpha}^{(\lambda)}, \; D}. 
\end{align}
This agrees with the anomaly (\ref{bdy_USp2n_AS_6_anomA}) for theory A. 

The mapping of operators between theory A and theory B is given by
$M_{\alpha \beta}^{(\lambda)} \sim Q_{\alpha} \Phi^{\lambda-1} Q_{\beta}$, $\phi^{(\lambda)} \sim \Phi^\lambda$ and $\widetilde{M}_{\alpha}^{(\lambda)} \sim Q_{\alpha} \Phi^{\lambda-1} \widetilde{Q}$ where the gauge indices are contracted with the $USp(2n)$-invariant rank-$2$ antisymmetric tensor. 

We propose the following boundary confining duality: 
\begin{align}
\label{bdy_USp2n_AS_6_dual}
&\textrm{$USp(2n)$ $+$ antisym. chiral $\Phi$ $+$ $5$ fund. chirals $Q_{\alpha}$}
\nonumber\\
&\textrm{$+$ $1$ fund. chiral $\widetilde{Q}$ with b.c. $(\mathcal{N},N,N,D)$}
\nonumber\\
&\Leftrightarrow 
\textrm{
an $SU(5)$ antisym. $M_{\alpha\beta}^{(\lambda)}$ $+$ a singlet $\phi^{(\lambda)}$}
\nonumber\\
&\textrm{$+$ a $SU(5)$ fund. chiral $\widetilde{M}_{\alpha}^{(\lambda)}$ with b.c. $(N,N,D)$}. 
\end{align}

All monopoles in theory A have positive dimension
\begin{align}
\label{Mono_Ch_usp2n_6_1}
    2(1 + (n-1)r_A) \sum_{i = 1}^n|m_i| - r_A \sum_{i<j}^n |m_i + m_j| - r_A \sum_{i<j}^n |m_i - m_j| \ge 2
\end{align}
for $m_i \in \Zb$ (with at least one $m_i \ne 0$)and the inequality is given for $r_A = 0$. Choosing positive values for $r_A$ may decrease the R-charge of the monopoles but there will still be a range of values so that all monopoles have positive R-charge. 
Hence the boundary duality can be extended to a bulk duality where theory A has superpotential $\Wcal_A = V + \Tr \Phi^{n+1}$ and theory B has a superpotential which is a linear combination of terms of the form $\epsilon^{\alpha \alpha_1 \beta_1 \alpha_2 \beta_2} \widetilde{M}_{\alpha}^{(\lambda)} M_{\alpha_1 \beta_1}^{(\lambda_1)} M_{\alpha_1 \beta_2}^{(\lambda_2)} \widehat{\phi}$ where $\widehat{\phi}$ schematically indicates some algebraic combination of the $\phi^{(\lambda)}$ chirals. The precise details of the theory B superpotential can be deduced from the operator mapping to theory A. This is the confining duality case of a 4d Seiberg-like duality derived in \cite{Intriligator:1995ff, Intriligator:1995ax} and with the inclusion of the linear monopole term in the theory A superpotential it reduces to a 3d duality \cite{Aharony:2013dha}. \footnote{We haven't checked the full details of the reduction from 4d to 3d including the counting of fermionic zero modes to check that the linear monopole superpotential is indeed generated from the circle compactification. However, we believe this is the case as the spectrum of the 3d theories matches expectations from such a reduction and the absence of a chiral multiplet in theory B dual to the minimal monopole in theory A is consistent with a linear monopole superpotential. This applies also to other cases described later where we interpret the 3d duality as arising from a 4d duality compactified on a circle.}
In fact, the matching of full indices 
\begin{align}
&
I^A=
\frac{1}{n! 2^n}\sum_{m_1,\cdots, m_n}
\prod_{i=1}^n 
\oint \frac{ds_i}{2\pi is_i}
\prod_{i\neq j}^{n}
(1-q^{\frac{|m_i-m_j|}{2}}s_i s_j^{-1})
\prod_{i\le j}^n
(1-q^{\frac{|m_i+m_j|}{2}}s_i^{\pm} s_j^{\pm})
\nonumber\\
&\times 
\prod_{i=1}^{n}
\prod_{\alpha=1}^5
\frac{
(q^{1-\frac{r_a}{2}+\frac{|m_i|}{2}} s_i^{\mp}a^{-1}x_{\alpha}^{-1};q)_{\infty}
}
{
(q^{\frac{r_a+|m_i|}{2}} s_i^{\pm}ax_{\alpha};q)_{\infty}
}
\nonumber\\
&\times 
\prod_{i<j}^n 
\frac{
(q^{1-\frac{r_A}{2}+\frac{|m_i-m_j|}{2}} s_i^{\pm}s_j^{\mp}A^{-1};q)_{\infty}
}
{
(q^{\frac{r_A+|m_i-m_j|}{2}} s_i^{\pm}s_j^{\mp}A;q)_{\infty}
}
\frac{
(q^{1-\frac{r_A}{2}+\frac{|m_i+m_j|}{2}} s_i^{\mp}s_j^{\mp}A^{-1};q)_{\infty}
}
{
(q^{\frac{r_A+|m_i+m_j|}{2}} s_i^{\pm}s_j^{\pm}A;q)_{\infty}
}
\nonumber\\
&\times 
\frac{
(q^{1-\frac{r_A}{2}}A^{-1};q)_{\infty}^n
}
{
(q^{\frac{r_A}{2}}A;q)_{\infty}^n
}
\prod_{i=1}^n
\frac{
(q^{(n-1)r_A+\frac52 r_a+\frac{|m_i|}{2}}A^{2n-2}a^5s_i^{\pm};q)_{\infty}
}
{
(q^{1-(n-1)r_A-\frac52 r_a+\frac{|m_i|}{2}}A^{-2n+2}a^{-5}s_i^{\mp};q)_{\infty}
}
\nonumber\\
&\times 
q^{\frac{5}{2}(1-r_a)\sum_i|m_i|+\frac{1}{2}(1-r_A)\sum_{i<j}|m_i\pm m_j|+\frac{1}{2}[-1+2(n-1)r_A+5r_a]\sum_{i}|m_i|
-\frac12 \sum_{i<j}|m_i\pm m_j|
-\sum_i|m_i|}
\nonumber\\
&\times 
A^{-\sum_{i<j}|m_i\pm m_j|+2(n-1)\sum_{i}|m_i|}
\end{align}
and 
\begin{align}
I^B&=
\prod_{\lambda=1}^n 
\prod_{\alpha<\beta}
\frac{
(q^{1-(\lambda-1)\frac{r_A}{2}-r_a}A^{-\lambda+1}a^{-2}x_{\alpha}^{-1}x_{\beta}^{-1};q)_{\infty}
}
{
(q^{(\lambda-1)\frac{r_A}{2}+r_a}A^{\lambda-1}a^{2}x_{\alpha}x_{\beta};q)_{\infty}
}
\nonumber\\
&\times 
\prod_{\alpha=1}^5 
\frac{
(q^{(2n-1-\lambda)\frac{r_A}{2}+2r_a}A^{2n-1-\lambda}a^4 x_{\alpha}^{-1};q)_{\infty}
}
{
(q^{1-(2n-1-\lambda)\frac{r_A}{2}-2r_a}A^{-2n+1+\lambda}a^{-4} x_{\alpha};q)_{\infty}
}
\frac{
(q^{1-\lambda\frac{r_A}{2}}A^{-\lambda};q)_{\infty}
}
{
(q^{\lambda\frac{r_A}{2}}A^{\lambda};q)_{\infty}
}
\end{align}
can be checked. 

For example, 
for $n=2$ with $r_a=\frac16$, $r_A=\frac14$ we have 
\begin{align}
I^A&=I^B
\nonumber\\&=1+Aq^{1/8}+10a^2 q^{1/6}+10a^2q^{1/6}
+2A^2q^{1/4}+20a^2Aq^{7/24}+55a^4q^{1/3}
\nonumber\\
&+2A^3q^{3/8}+(5a^{-4}A^{-2}+30a^2A^2)q^{5/12}
+150a^4Aq^{11/24}+(220a^6+3A^4)q^{1/2}+\cdots.
\end{align}

\subsection{$SU(N)$ with rank-$2$ antisymmetric, $3(+1)$ fundamental and $N$ antifundamental chirals}
\label{sec_GR_integral_AS}
Let us consider theory A with gauge group $SU(N)$, an antisymmetric chiral $\Phi$, $N_f = 3$
fundamental chirals $Q_{I}$ and $N_a = N$ antifundamental chirals $\overline{Q}_{\alpha}$ with R-charge $0$. 
These all obey Neumann boundary conditions. 
Also we introduce an additional fundamental chiral $\widetilde{Q}$ with Dirichlet boundary conditions.  
One finds the boundary 't Hooft anomaly 
\begin{align}
\label{bdy_SUN_AS_3_N_anomA}
\Acal^A & = \underbrace{N\Tr(s^2) + \frac{N^2 - 1}{2}r^2}_{\textrm{VM}, \; \Ncal}
 - \underbrace{\left( \frac{3}{2}\Tr(s^2) + \frac{N}{2}\Tr(x^2) + \frac{3N}{2}(a-r)^2 \right)}_{Q_I, \; N}
  \nonumber \\
 & - \underbrace{\left( \frac{N}{2}\Tr(s^2) + \frac{N}{2}\Tr(\tilde{x}^2) + \frac{N^2}{2}(b-r)^2 \right)}_{\overline{Q}_{\alpha}, \; N}
  \nonumber \\
 & - \underbrace{\left( \frac{N-2}{2}\Tr(s^2) + \frac{N(N-1)}{4}(A - r)^2 \right)}_{\Phi, \; \Ncal}
 \nonumber\\
 & + \underbrace{\left( \frac{1}{2}\Tr(s^2) + \frac{N}{2}\big( (2 - N)A - 3a - Nb + r \big)^2 \right)}_{\widetilde{Q}, \; D}
 \nonumber\\
 &=- N \Tr(x^2) - N \Tr(\tilde{x}^2) + 3Na^2 + 3N^2 ab + \frac{N^2(N-1)}{2}b^2 + \frac{N(N-3)(2N-3)}{4}A^2
  \nonumber \\
  & + 3N(N-2)Aa + N^2(N-2)Ab - \frac{N(N-3)}{2}Ar - \frac{(N+1)(N+2)}{4}r^2.
\end{align}
The absence of gauge anomaly ensures that the Neumann boundary condition is consistent. 

We can evaluate the Neumann-half-index as 
\begin{align}
\label{bdy_SUN_AS_3_N_hindexA}
&\mathbb{II}_{(\mathcal{N},N,N,N,D)}^{A}
\nonumber\\
&=\frac{(q)_{\infty}^{N-1}}{N!} \prod_{i=1}^{N-1} \oint \frac{ds_i}{2\pi i s_i}
\prod_{i \ne j}^N (s_i s_j^{-1}; q)_{\infty} 
\nonumber \\
&\times \prod_{i = 1}^N \frac{(q^{((N-2)r_A + 3r_a + Nr_b)/2} A^{N-2} a^3 b^N s_i^{-1}; q)_{\infty}}{\left( \prod_{\alpha = 1}^N (q^{r_b/2} b s_i^{-1} \tilde{x}_{\alpha}; q)_{\infty} \right) \left( \prod_{I = 1}^3 (q^{r_a/2} a s_i x_I; q)_{\infty} \right) \left( \prod_{i < j}^N (q^{r_A/2} A s_i s_j; q)_{\infty} \right)}
\end{align}
where $\prod_{i=1}^N s_i = \prod_{I = 1}^3 x_I = \prod_{\alpha = 1}^N \tilde{x}_{\alpha} = 1$. 
The half-index (\ref{bdy_SUN_AS_3_N_hindexA}) is identified with the integral studied in \cite{MR1811060}. 
For $N=2n$ it can be expresses as
\begin{align}
\label{bdy_SUN_AS_3_N_hindexB1}
&
\frac{\left( q^{(nr_A + Nr_b)/2} A^n b^N; q \right)_{\infty} \left( \prod_{\alpha = 1}^N \left( q^{((N-2)r_A + 3r_a + (N-1)r_b)/2} A^{N-2} a^3 b^{N-1} \tilde{x}_{\alpha}^{-1};q \right)_{\infty} \right)}
 {\left( q^{nr_A/2} A^n; q \right)_{\infty} \left( q^{Nr_b/2} b^N; q \right)_{\infty} \prod_{I = 1}^3 \prod_{\alpha = 1}^N (q^{(r_a + r_b)/2} ab x_I \tilde{x}_{\alpha}; q)_{\infty}}
\nonumber \\
&\times \frac{\left( \prod_{I = 1}^3 \left( q^{((n-1)r_A + 2r_a + Nr_b)/2} A^{n-1} a^2 b^N x_I^{-1};q \right)_{\infty} \right)}
 {\left( \prod_{I = 1}^3 \left( q^{((n-1)r_A + 2r_a)/2} A^{n-1} a^2 x_I^{-1}; q \right)_{\infty} \right) \prod_{\alpha < \beta}^N (q^{(r_A + 2r_b)/2} A b^2 \tilde{x}_{\alpha} \tilde{x}_{\beta}; q)_{\infty}}. 
\end{align}
For $N=2n+1$ the Neumann half-index (\ref{bdy_SUN_AS_3_N_hindexA}) can be written as
\begin{align}
\label{bdy_SUN_AS_3_N_hindexB2}
&
\frac{\left( q^{((n-1)r_A + 3r_a + Nr_b)/2} A^{n-1} a^3 b^N; q \right)_{\infty} \left( \prod_{\alpha = 1}^N \left( q^{((N-2)r_A + 3r_a + (N-1)r_b)/2} A^{N-2} a^3 b^{N-1} \tilde{x}_{\alpha}^{-1};q \right)_{\infty} \right)}
 {\left( q^{((n-1)r_A + 3r_a)/2} A^{n-1} a^3; q \right)_{\infty} \left( q^{Nr_b/2} b^N; q \right)_{\infty} \prod_{I = 1}^3 \prod_{\alpha = 1}^N (q^{(r_a + r_b)/2} ab x_I \tilde{x}_{\alpha}; q)_{\infty}} \nonumber \\
&\times \frac{\left( \prod_{I = 1}^3 \left( q^{(nr_A + r_a + Nr_b)/2} A^n a b^N x_I;q \right)_{\infty} \right)}
 {\left( \prod_{I = 1}^3 \left( q^{((n+1)r_A + r_a)/2} A^{n+1} a x_I; q \right)_{\infty} \right) \prod_{\alpha < \beta}^N (q^{(r_A + 2r_b)/2} A b^2 \tilde{x}_{\alpha} \tilde{x}_{\beta}; q)_{\infty}}. 
\end{align}
These half-indices should be convergent for any positive values for
$r_A$, $r_a$ and $r_b$.

For example if we have $SU(4)$ the identity is
\begin{align}
&\mathbb{II}_{(\mathcal{N},N,N,N,D)}^{A}
\nonumber\\
&=
\frac{(q)_{\infty}^{3}}{4!} \prod_{i=1}^{3} \oint \frac{ds_i}{2\pi i s_i}
\prod_{i \ne j}^4 (s_i s_j^{-1}; q)_{\infty} \nonumber \\
&\times \prod_{i = 1}^4 \frac{(q^{(2r_A + 3r_a + 4r_b)/2} A^{2} a^3 b^4 s_i^{-1}; q)_{\infty}}{\left( \prod_{\alpha = 1}^4 (q^{r_b/2} b s_i^{-1} \tilde{x}_{\alpha}; q)_{\infty} \right) \left( \prod_{I = 1}^3 (q^{r_a/2} a s_i x_I; q)_{\infty} \right) \left( \prod_{i < j}^4 (q^{r_A/2} A s_i s_j; q)_{\infty} \right)}
\nonumber \\
&=\frac{\left( q^{r_A + 2r_b} A^2 b^4; q \right)_{\infty} \left( \prod_{\alpha = 1}^4 \left( q^{(2r_A + 3r_a + 3r_b)/2} A^{2} a^3 b^{3} \tilde{x}_{\alpha}^{-1};q \right)_{\infty} \right)}
 {\left( q^{r_A} A^2; q \right)_{\infty} \left( q^{2r_b} b^4; q \right)_{\infty} \prod_{I = 1}^3 \prod_{\alpha = 1}^4 (q^{(r_a + r_b)/2} ab x_I \tilde{x}_{\alpha}; q)_{\infty}}
\nonumber \\
&\times \frac{\left( \prod_{I = 1}^3 \left( q^{(r_A + 2r_a + 4r_b)/2} A a^2 b^4 x_I^{-1};q \right)_{\infty} \right)}
 {\left( \prod_{I = 1}^3 \left( q^{(r_A + 2r_a)/2} A a^2 x_I^{-1}; q \right)_{\infty} \right) \prod_{\alpha < \beta}^4 (q^{(r_A + 2r_b)/2} A b^2 \tilde{x}_{\alpha} \tilde{x}_{\beta}; q)_{\infty}}
\label{GR_su4}. 
\end{align}
When the R-charges are chosen as $r_a=r_b=r_A=\frac14$ and fugacities $x_I$ and $\tilde{x}_{\alpha}$ are switched off, 
it can be expanded as
\begin{align}
&
\mathbb{II}_{(\mathcal{N},N,N,N,D)}^{A}
\nonumber\\
&=1+(A^2+12ab)q^{1/4}
+(3a^2A+6Ab^2)q^{8/3}
\nonumber\\
&+(A^4+12aA^2b+78a^2b^2+b^4)q^{1/2}
+(3a^2A^3+36a^3Ab+6A^3b^2+72aAb^3)q^{5/8}+\cdots.
\end{align}

The expressions (\ref{bdy_SUN_AS_3_N_hindexB1}) and (\ref{bdy_SUN_AS_3_N_hindexB2}) allow us to read the confined descriptions of the Neumann boundary condition of theory A as the boundary condition for dual theory B consisting of chiral multiplets. 
The field content of theory B depends on whether $N$ is even or odd, i.e.\ for $n \in \Nb$, $N = 2n$ or $N = 2n+1$, and is summarized in the following tables.

For $N = 2n$ with $n \ge 2$ we have
\begin{align}
\hspace*{-1cm}
\label{SUN_even_AS_charges}
\begin{array}{c|c|c|c|c|c|c|c|c}
& \textrm{bc} & SU(N = 2n) & SU(N_f = 3) & SU(N_a = N) & U(1)_A & U(1)_a & U(1)_b & U(1)_R \\ \hline
\textrm{VM} & \mathcal{N} & {\bf Adj} & {\bf 1} & {\bf 1} & 0 & 0 & 0 & 0 \\
\Phi & \textrm{N} & {\bf N(N-1)/2} & {\bf 1} & {\bf 1} & 1 & 0 & 0 & 0 \\
Q_I & \textrm{N} & {\bf N} & {\bf 3} & {\bf 1} & 0 & 1 & 0 & 0 \\
\overline{Q}_{\alpha} & \textrm{N} & {\bf \overline{N}} & {\bf 1} & {\bf N} & 0 & 0 & 1 & 0 \\
\widetilde{Q} & \textrm{D} & {\bf N} & {\bf 1} & {\bf 1} & 2 - N & -3 & -N & 2 \\
 \hline
M_{_I \alpha} & \textrm{N} & {\bf 1} & {\bf 3} & {\bf N} & 0 & 1 & 1 & 0 \\
\overline{B} & \textrm{N} & {\bf 1} & {\bf 1} & {\bf 1} & 0 & 0 & N & 0 \\
\phi & \textrm{N} & {\bf 1} & {\bf 1} & {\bf 1} & n & 0 & 0 & 0 \\
\overline{M}_{\alpha \beta} & \textrm{N} & {\bf 1} & {\bf 1} & {\bf N(N-1)/2} & 1 & 0 & 2 & 0 \\
B^I & \textrm{N} & {\bf 1} & {\bf \overline{3}} & {\bf 1} & n - 1 & 2 & 0 & 0 \\
\widetilde{M}_{\alpha} & \textrm{D} & {\bf 1} & {\bf 1} & {\bf N} & 2 - N & -3 & 1-N & 2 \\
\widehat{B} & \textrm{D} & {\bf 1} & {\bf 1} & {\bf 1} & -n & 0 & -N & 2 \\
\widetilde{B}_I & \textrm{D} & {\bf 1} & {\bf 3} & {\bf 1} & 1 - n & -2 & -N & 2
\end{array}
\end{align}
with the operator mapping $M_{_I \alpha} \sim Q_I \overline{Q}_{\alpha}$, $\overline{B} \sim \epsilon \overline{Q}_1 \cdots \overline{Q}_N$, $\phi \sim \epsilon \Phi^n$, $\overline{M}_{\alpha \beta} \sim \Phi \overline{Q}_{\alpha} \overline{Q}_{\beta}$, $B^I \sim \epsilon \Phi^{n-1} Q_J Q_K \epsilon^{IJK}$, $\widetilde{M}_{\alpha} \sim \overline{Q}_{\alpha} \widetilde{Q}$, $\widehat{B} \sim \epsilon \Phi^{n-2} Q_1 Q_2 Q_3 \widetilde{Q}$ and $\widetilde{B}_I \sim \epsilon \Phi^{n-1} Q_I \widetilde{Q}$. 

In the case where $N = 2$ the chirals $\Phi$ in theory A and $\phi$ in theory B are both singlets with the same charges so can be removed from both theories. We then see that theory A is the same as theory A in section~\ref{sec_su2nf3na3}. Theory B also matches once we notice that in this case the chirals $\overline{M}_{\alpha \beta}$ and $\widehat{B}$ can be ignored as they cancel in the half-index.

For $N=2n+1$ we have
\begin{align}
\hspace*{-1cm}
\label{SUN_odd_AS_charges}
\begin{array}{c|c|c|c|c|c|c|c|c}
& \textrm{bc} & SU(N = 2n+1) & SU(N_f = 3) & SU(N_a = N) & U(1)_A & U(1)_a & U(1)_b & U(1)_R \\ \hline
\textrm{VM} & \mathcal{N} & {\bf Adj} & {\bf 1} & {\bf 1} & 0 & 0 & 0 & 0 \\
\Phi & \textrm{N} & {\bf N(N-1)/2} & {\bf 1} & {\bf 1} & 1 & 0 & 0 & 0 \\
Q_I & \textrm{N} & {\bf N} & {\bf 3} & {\bf 1} & 0 & 1 & 0 & 0 \\
\overline{Q}_{\alpha} & \textrm{N} & {\bf \overline{N}} & {\bf 1} & {\bf N} & 0 & 0 & 1 & 0 \\
\widetilde{Q} & \textrm{D} & {\bf N} & {\bf 1} & {\bf 1} & 2 - N & -3 & -N & 2 \\
 \hline
M_{_I \alpha} & \textrm{N} & {\bf 1} & {\bf 3} & {\bf N} & 0 & 1 & 1 & 0 \\
\overline{B} & \textrm{N} & {\bf 1} & {\bf 1} & {\bf 1} & 0 & 0 & N & 0 \\
B_I & \textrm{N} & {\bf 1} & {\bf 3} & {\bf 1} & n & 1 & 0 & 0 \\
\overline{M}_{\alpha \beta} & \textrm{N} & {\bf 1} & {\bf 1} & {\bf N(N-1)/2} & 1 & 0 & 2 & 0 \\
B & \textrm{N} & {\bf 1} & {\bf 1} & {\bf 1} & n - 1 & 3 & 0 & 0 \\
\widetilde{M}_{\alpha} & \textrm{D} & {\bf 1} & {\bf 1} & {\bf N} & 2 - N & -3 & 1-N & 2 \\
\widehat{B}^I & \textrm{D} & {\bf 1} & {\bf \overline{3}} & {\bf 1} & -n & -1 & -N & 2 \\
\widetilde{B} & \textrm{D} & {\bf 1} & {\bf 1} & {\bf 1} & 1 - n & -3 & -N & 2
\end{array}
\end{align}
with the operator mapping $M_{_I \alpha} \sim Q_I \overline{Q}_{\alpha}$, $\overline{B} \sim \epsilon \overline{Q}_1 \cdots \overline{Q}_N$, $B_I \sim \epsilon \Phi^n Q_I$, $\overline{M}_{\alpha \beta} \sim \Phi \overline{Q}_{\alpha} \overline{Q}_{\beta}$, $B \sim \epsilon \Phi^{n-1} Q_1 Q_2 Q_3$, $\widetilde{M}_{\alpha} \sim \overline{Q}_{\alpha} \widetilde{Q}$, $\widehat{B}^I \sim \epsilon \Phi^{n-1} Q_J Q_K \widetilde{Q} \epsilon^{IJK}$ and $\widetilde{B} \sim \epsilon \Phi^n \widetilde{Q}$.

In the case where $N = 3$, the antisymmetric representation is the same as the antifundamental representation of $SU(3)$ so this duality is the same as discussed in section~\ref{sec_GNR_integrals} up to relabelling of fields and exchanging gauge group representations ${\bf 3} \leftrightarrow {\bf \overline{3}}$.

For $N=2n$ the boundary anomaly for theory B is 
\begin{align}
\label{bdy_SUN_AS_3_N_anomB1}
&
\Acal^{B, N=2n} 
\nonumber\\
&=  - \underbrace{\left( \frac{N}{2} \Tr(x^2) + \frac{3}{2} \Tr(\tilde{x}^2) + \frac{3N}{2}(a+b-r)^2 \right)}_{M_{I \alpha}, \; N}
  - \underbrace{\frac{1}{2}(Nb-r)^2}_{\overline{B}, \; N}
  - \underbrace{\frac{1}{2}(nA-r)^2}_{\phi, \; N}
  \nonumber \\
 & - \underbrace{\left( \frac{N-2}{2} \Tr(\tilde{x}^2) + \frac{N(N-1)}{4}(A + 2b - r)^2 \right)}_{\overline{M}_{\alpha \beta}, \; N}
  - \underbrace{\left( \frac{1}{2} \Tr(x^2) + \frac{3}{2}\big( (n-1)A + 2a-r \big)^2 \right)}_{B^I, \; N}
   \nonumber \\
 & + \underbrace{\frac{1}{2}(-nA - Nb + r)^2}_{\widehat{B}, \; D}
 + \underbrace{\left( \frac{1}{2} \Tr(\tilde{x}^2) + \frac{N}{2}\big( (2-N)A - 3a +(1-N)b + r)^2 \right)}_{\widetilde{M}_{\alpha}, \; D}
  \nonumber \\
  & + \underbrace{\left( \frac{1}{2} \Tr(x^2) + \frac{3}{2}\big( (1-n)A - 2a - Nb + r \big)^2 \right)}_{\widetilde{B}^I, \; D}
\end{align}
and for $N=2n+1$ it is
\begin{align}
\label{bdy_SUN_AS_3_N_anomB2}
&\Acal^{B, N=2n+1} 
\nonumber\\
&= - \underbrace{\left( \frac{N}{2} \Tr(x^2) + \frac{3}{2} \Tr(\tilde{x}^2) + \frac{3N}{2}(a+b-r)^2 \right)}_{M_{I \alpha}, \; N}
  - \underbrace{\frac{1}{2}(Nb-r)^2}_{\overline{B}, \; N}
  - \underbrace{\frac{1}{2} \big( (n-1)A + 3a - r \big)^2}_{B, \; N}
  \nonumber \\
 & - \underbrace{\left( \frac{N-2}{2} \Tr(\tilde{x}^2) + \frac{N(N-1)}{4}(A + 2b - r)^2 \right)}_{\overline{M}_{\alpha \beta}, \; N}
  - \underbrace{\left( \frac{1}{2} \Tr(x^2) + \frac{3}{2}(nA + a-r)^2 \right)}_{B_I, \; N}
   \nonumber \\
 & + \underbrace{\left( \frac{1}{2} \Tr(x^2) + \frac{3}{2}\big( -nA - a - Nb + r \big)^2 \right)}_{\widehat{B}^I, \; D}
 + \underbrace{\frac{1}{2}\big( (1-n)A - 3a - Nb + r)^2}_{\widetilde{B}, \; D}
  \nonumber \\
 & + \underbrace{\left( \frac{1}{2} \Tr(\tilde{x}^2) + \frac{N}{2}\big( (2-N)A - 3a +(1-N)b + r)^2 \right)}_{\widetilde{M}_{\alpha}, \; D}. 
\end{align}
The anomalies (\ref{bdy_SUN_AS_3_N_anomB1}) and (\ref{bdy_SUN_AS_3_N_anomB2}) exactly coincide with the anomaly (\ref{bdy_SUN_AS_3_N_anomA}) for theory A. 

The matching of the half-indices and the anomalies support the following boundary confining dualities: 
\begin{align}
\label{bdy_SUN_AS_3_N_dual1}
&\textrm{$SU(N)$ $+$ antisym. chiral $\Phi$ $+$ $3$ fund. chirals $Q_{I}$}
\nonumber\\
&\textrm{$+$ $N$ antifund. chiral $\overline{Q}_{\alpha}$ $+$ $1$ antifund. chiral $\widetilde{Q}$ with b.c. $(\mathcal{N},N,N,N,D)$}
\nonumber\\
&\Leftrightarrow 
\textrm{
an $SU(3)\times SU(N)$ bifund. $M_{I \alpha}$ $+$ a singlet $\overline{B}$ $+$ a singlet $\phi$}
\nonumber\\
&\textrm{$+$ an $SU(N)$ antisym. chiral $\overline{M}_{\alpha\beta}$ $+$ an $SU(3)$ antifund. $B^I$ 
$+$ an $SU(N)$ fund. chiral $\widetilde{M}_{\alpha}$}
\nonumber\\
&\textrm{
$+$ a singlet $\widehat{B}$ $+$ an $SU(3)$ fund. chiral $\widetilde{B}_I$ with b.c. $(N,N,N,N,N,D,D,D)$}. 
\end{align}
for $N=2n$ and
\begin{align}
\label{bdy_SUN_AS_3_N_dual2}
&\textrm{$SU(N)$ $+$ antisym. chiral $\Phi$ $+$ $3$ fund. chirals $Q_{I}$}
\nonumber\\
&\textrm{$+$ $N$ antifund. chiral $\overline{Q}_{\alpha}$ $+$ $1$ antifund. chiral $\widetilde{Q}$ with b.c. $(\mathcal{N},N,N,N,D)$}
\nonumber\\
&\Leftrightarrow 
\textrm{
an $SU(3)\times SU(N)$ bifund. $M_{I \alpha}$ $+$ a singlet $\overline{B}$ $+$ an $SU(3)$ fund. $B_I$}
\nonumber\\
&\textrm{$+$ an $SU(N)$ antisym. chiral $\overline{M}_{\alpha\beta}$ $+$ a singlet $B$ 
$+$ an $SU(N)$ fund. chiral $\widetilde{M}_{\alpha}$}
\nonumber\\
&\textrm{
$+$ an $SU(3)$ antifund. $\widehat{B}^I$ $+$ a singlet $\widetilde{B}$ with b.c. $(N,N,N,N,N,D,D,D)$}. 
\end{align}
for $N=2n+1$. 

These boundary confining dualities (\ref{bdy_SUN_AS_3_N_dual1}) and (\ref{bdy_SUN_AS_3_N_dual2}) are extended to bulk confining dualities with superpotentials
\begin{align}
    \Wcal_A = & V \\
    \Wcal_B^{N = 2n} = & \widetilde{M}_{\alpha_1} \left( M_{I \beta_1} B^I \overline{M}_{\alpha_2 \beta_2} - \epsilon^{IJK} M_{I \beta_1} M_{J \alpha_2} M_{K \beta_2} \right) \overline{M}_{\alpha_3 \beta_3} \cdots \overline{M}_{\alpha_n \beta_n} \epsilon^{\alpha_1 \beta_1 \cdots \alpha_n \beta_n}
    \nonumber \\
    & + \widetilde{B}_I \left( B^I \overline{B} - \epsilon^{IJK} M_{J \alpha_1} M_{K \beta_1} \overline{M}_{\alpha_2 \beta_2} \cdots \overline{M}_{\alpha_n \beta_n} \epsilon^{\alpha_1 \beta_1 \cdots \alpha_n \beta_n} \right)
    \nonumber \\
    & + \widehat{B} \left( \overline{B} \phi - \Pf \overline{M}_{\alpha \beta} \right)
    \\
    \Wcal_B^{N = 2n+1} = & \widetilde{M}_{\alpha} \left( B \overline{M}_{\alpha_1 \beta_1} - \epsilon^{IJK} B_I M_{J \alpha_1} M_{K \beta_1} \right) \overline{M}_{\alpha_2 \beta_2} \cdots \overline{M}_{\alpha_n \beta_n} \epsilon^{\alpha \alpha_1 \beta_1 \cdots \alpha_n \beta_n},
    \nonumber \\
    & + \widetilde{B} \left( B \overline{B} - \epsilon^{IJK} M_{I \alpha} M_{J \alpha_1} M_{K \beta_1} \overline{M}_{\alpha_2 \beta_2} \cdots \overline{M}_{\alpha_n \beta_n} \epsilon^{\alpha \alpha_1 \beta_1 \cdots \alpha_n \beta_n} \right)
    \nonumber \\
    & + \widehat{B}^I \left( B_I \overline{B} - M_{I \alpha} \overline{M}_{\alpha_1 \beta_1} \cdots \overline{M}_{\alpha_n \beta_n} \epsilon^{\alpha \alpha_1 \beta_1 \cdots \alpha_n \beta_n} \right)
\end{align}
where $V$ is the dimension $2$ monopole. This is the confining duality case of the 4d dualities presented in \cite{Berkooz:1995km, Intriligator:1995ax, Pouliot:1995me} compactified to 3d with the addition of the linear monopole superpotential \cite{Aharony:2013dha}. 

For example, for $N=4$ with $r_a=\frac16$, $r_b=\frac16$ and $r_A=\frac14$ we have checked the precise agreement of the full-indices for theory A and B
\begin{align}
I^A&=I^B
\nonumber\\
&=1+12abq^{1/6}+(A^2+4a^{-3}A^{-2}b^{-3})q^{1/4}
+3A(a^2+2b^2)q^{7/24}+(78a^2b^2+b^4)q^{1/3}
\nonumber\\
&+3a^{-2}A^{-1}b^{-4}q^{3/8}
+(A^{-2}b^{-4}+48a^{-2}A^{-2}b^{-2}+12aA^2b)q^{5/12}+\cdots
\end{align}
where 
\begin{align}
I^A&=
\frac{1}{4!}\sum_{m_1,m_2,m_3\in \mathbb{Z}}
\prod_{i=1}^3 
\oint 
\frac{ds_i}{2\pi is_i}
\prod_{i \ne j}^4 (1-q^{\frac{|m_i-m_j|}{2}}s_i^{\pm}s_j^{\mp})
\prod_{i=1}^4
\prod_{\alpha=1}^{4}
\frac{
(q^{1-\frac{r_b}{2}+\frac{|m_i|}{2}}b^{-1}s_i\tilde{x}_{\alpha}^{-1};q)_{\infty}
}{
(q^{\frac{r_b}{2}+\frac{|m_i|}{2}}bs_i^{-1}\tilde{x}_{\alpha};q)_{\infty}
}
\nonumber\\
&\times 
\prod_{i=1}^4 \prod_{I=1}^3 
\frac{
(q^{1-\frac{r_a}{2}+\frac{|m_i|}{2}}a^{-1}s_i\tilde{x}_{I}^{-1};q)_{\infty}
}
{
(q^{\frac{r_a}{2}+\frac{|m_i|}{2}}as_i^{-1}\tilde{x}_{I};q)_{\infty}
}
\prod_{i<j}
\frac{
(q^{1-\frac{r_A}{2}+\frac{|m_i+m_j|}{2}}A^{-1}s_i^{-1}s_j^{-1};q)_{\infty}
}
{
(q^{\frac{r_A}{2}+\frac{|m_i+m_j|}{2}}As_is_j;q)_{\infty}
}
\nonumber\\
&\times 
\frac{
(q^{\frac{2r_A+3r_a+4r_b+|m_i|}{2}}A^{2}a^3 b^4 s_i^{-1};q)_{\infty}
}
{
(q^{1-\frac{2r_A+3r_a+4r_b}{2}+\frac{|m_i|}{2}}A^{-2}a^{-3}b^{-4}s_i;q)_{\infty}
}
\nonumber\\
&\times 
q^{
\Delta
}
A^{-\sum_{i<j}|m_i-m_j|+2\sum_{i=1}^4|m_i|}
\end{align}
and 
\begin{align}
I^B&=
\frac{
(q^{r_A+2r_b}A^2b^4;q)_{\infty}
}
{
(q^{1-r_A-2r_b}A^{-2}b^{-4};q)_{\infty}
}
\frac{
(q^{1-r_A}A^{-2};q)_{\infty}
}
{
(q^{r_A}A^{2};q)_{\infty}
}
\frac{
(q^{1-2r_b}b^{-4};q)_{\infty}
}
{
(q^{r_b}b^{4};q)_{\infty}
}
\nonumber\\
&\times 
\prod_{\alpha=1}^4
\frac{
(q^{\frac{2r_A+3r_a+3r_b}{2}}Aa^3b^3\tilde{x}_{\alpha}^{-1};q)_{\infty}
}
{
(q^{1-\frac{2r_A+3r_a+3r_b}{2}}A^{-1}a^{-3}b^{-3}\tilde{x}_{\alpha};q)_{\infty}
}
\nonumber\\
&\times 
\prod_{i=1}^{3}
\prod_{\alpha=1}^{4}
\frac{
(q^{1-\frac{r_a+r_b}{2}}a^{-1}b^{-1}x_I^{-1}\tilde{x}_{\alpha}^{-1}\;q)_{\infty}
}
{
(q^{\frac{r_a+r_b}{2}}abx_I\tilde{x}_{\alpha};q)_{\infty}
}
\prod_{I=1}^3 
\frac{
(q^{\frac{r_A+2r_a+4r_b}{2}}Aa^2b^{4}x_I^{-1};q)_{\infty}
}
{
(q^{1-\frac{r_A+2r_a+4r_b}{2}}A^{-1}a^{-2}b^{-4}x_I;;q)_{\infty}
}
\nonumber\\
&\times 
\prod_{I=1}^3
\frac{
(q^{1-\frac{r_A+2r_a}{2}}A^{-1}a^{-2}x_I;q)_{\infty}
}
{
(q^{\frac{r_A+2r_a}{2}}Aa^{2}x_I^{-1};;q)_{\infty}
}
\prod_{\alpha<\beta}
\frac{
(q^{1-\frac{r_A+2r_b}{2}}A^{-1}b^{-2}\tilde{x}_{\alpha}^{-1}\tilde{x}_{\beta}^{-1};q)_{\infty}
}
{
(q^{\frac{r_A+2r_b}{2}}Ab^{2}\tilde{x}_{\alpha}x_{\beta};;q)_{\infty}
}
\end{align}
with $m_4 = - (m_1 + m_2 + m_3)$ and
\begin{align}
\Delta&=
\frac34 (1-r_a)\sum_{i=1}^4 |m_i|
+(1-r_b)\sum_{i=1}^4|m_i|
+\frac12 (1-r_A)\sum_{i<j}|m_i+m_j|
\nonumber\\
&+\frac14(-1+2r_A+3r_a+4r_b)\sum_{i=1}^4|m_i|
-\frac12 \sum_{i<j}|m_i-m_j| \; .
\end{align}

For general $N$ the monopoles have dimensions given, with $\sum_{i=1}^N m_i = 0$, by
\begin{align}
\label{Mono_Ch_suN_Np4_1}
    \frac{N+2+(N-2)r_A}{2}\sum_{i = 1}^N |m_i| + (1 - r_A) \sum_{i < j}^N |m_i + m_j| - \sum_{i < j}^N |m_i - m_j|
\end{align}
which we believe are bounded from below by $2$ for $r_A = 0$ (and $N \ge 2$) but we haven't proven this.
For $r_A > 0$ the monopole dimensions may decrease but there will still be a range of values of $r_A$ for which all monopoles will have positive dimension.

We will see in section~\ref{sec_New_SUN_integral_AS} that there is a similar duality for $SU(N)$ with an antisymmetric rank-$2$ chiral, four fundamental and $N$ antifundamental chirals where one of the antifundamental (rather than fundamental) chirals has Dirichlet boundary condition.
These both correspond to the same 3d bulk duality described above.

\subsection{$SU(N)$ with $2$ rank-$2$ antisymmetric, $3$ fundamental and $2(+1)$ antifundamental chirals}
\label{sec_G_integral_ASAS}
Next consider theory A that has an $SU(N)$ vector multiplet with Neumann boundary condition, 
two chirals $\Phi$, $\overline{\Phi}$ with Neumann boundary conditions in the antisymmetric 
and conjugate antisymmetric rank-2 representations of $SU(N)$, 3 fundamental $Q_{I}$ and
2 antifundamental chirals $\overline{Q}_{\alpha}$ also with Neumann boundary conditions, 
and one antifundamental $\widetilde{Q}$ with Dirichlet boundary condition. 
This setup can be obtained by adding one more rank-$2$ antisymmetric chiral multiplet $\overline{\Phi}$ with Neumann boundary condition 
to theory A in subsection \ref{sec_GR_integral_AS}.  

We obtain the boundary 't Hooft anomaly 
\begin{align}
\label{bdy_SUN_2AS_3_3_anomA}
\Acal^A & = \underbrace{N\Tr(s^2) + \frac{N^2 - 1}{2}r^2}_{\textrm{VM}, \; \Ncal}
 - \underbrace{\left( \frac{3}{2}\Tr(s^2) + \frac{N}{2}\Tr(x^2) + \frac{3N}{2}(a-r)^2 \right)}_{Q_I, \; N}
  \nonumber \\
 & - \underbrace{\left( \Tr(s^2) + \frac{N}{2}\Tr(\tilde{x}^2) + N(b-r)^2 \right)}_{\overline{Q}_{\alpha}, \; N}
 - \underbrace{\left( \frac{N-2}{2}\Tr(s^2) + \frac{N(N-1)}{4}(A - r)^2 \right)}_{\Phi, \; \Ncal}
  \nonumber \\
 & - \underbrace{\left( \frac{N-2}{2}\Tr(s^2) + \frac{N(N-1)}{4}(B - r)^2 \right)}_{\overline{\Phi}, \; \Ncal}
 \nonumber \\
 & + \underbrace{\left( \frac{1}{2}\Tr(s^2) + \frac{N}{2}\big( (2 - N)A + (2 - N)B - 3a - 2b + r \big)^2 \right)}_{\widetilde{Q}, \; D}
   \nonumber \\
  &= - \frac{N}{2} \Tr(x^2) - \frac{N}{2} \Tr(\tilde{x}^2) +
  3 a^2 N + 3N (N-2) Aa + 2 N (N-2) Ab
  \nonumber \\ 
  & + N (N-2)^2 AB - \frac{1}{2} (N-3) N Ar + 3 (N-2) N Ba + 2 (N-2) N Bb
  \nonumber \\
  & - \frac{1}{2} (N-3) N Br + 6N a b + \frac{1}{4} N (N-3) (2N-3) A^2 + b^2 N
  \nonumber \\
  & + \frac{1}{4} N (N-3) (2N-3) B^2 - \frac{1}{2} (3N + 1) r^2. 
\end{align}

The Neumann half-index is evaluated as
\begin{align}
\label{bdy_SUN_2AS_3_3_hindexA}
\II^A = & \frac{(q)_{\infty}^{N-1}}{N!} \prod_{i=1}^{N-1} \oint \frac{ds_i}{2\pi i s_i}
\prod_{i \ne j}^N (s_i s_j^{-1}; q)_{\infty}
\frac{1}{\prod_{i < j}^N  (q^{r_A/2} A s_i s_j; q)_{\infty} (q^{r_B/2} B s_i^{-1} s_j^{-1}; q)_{\infty}}
\nonumber \\
& \times \prod_{i = 1}^N \frac{(q^{((N-2)r_A + (N-2)r_B + 3r_a + 2r_b)/2} A^{N-2} B^{N-2} a^3 b^2 s_i; q)_{\infty}}{\left( \prod_{\alpha = 1}^2 (q^{r_b/2} b s_i^{-1} \tilde{x}_{\alpha}; q)_{\infty} \right) \left( \prod_{I = 1}^3 (q^{r_a/2} a s_i x_I; q)_{\infty} \right)}. 
\end{align}
The Neumann half-index (\ref{bdy_SUN_2AS_3_3_hindexA}) is identified with the integral studied in \cite{MR1266569}. 
According to the equation (3.2a) in \cite{MR1266569}, 
the half-index (\ref{bdy_SUN_2AS_3_3_hindexA}) for $SU(2n)$ can be written as
\begin{align}
\label{bdy_SUN_2AS_3_3_hindexB1}
&
\frac{ \prod_{\alpha = 1}^2 \left( q^{((N-2)r_A + (n-1)r_B + 3r_a + r_b)/2} A^{N-2} B^{n-1} a^3 b \tilde{x}_{\alpha};q \right)_{\infty} }
 {\left( q^{nr_A/2} A^n; q \right)_{\infty} \left( q^{nr_B/2} B^n; q \right)_{\infty} \left( q^{(n-1)r_B/2 + r_b} B^{n-1} b^2; q \right)_{\infty} \prod_{I = 1}^3 (q^{(n-1)r_A/2 + r_a} A^{n-1} a^2 x_I^{-1}; q)_{\infty}}
 \nonumber \\
& \times \prod_{\lambda = 0}^{n-1} \prod_{I = 1}^3 \frac{ \left( q^{((n-1-\lambda/2)r_A + (n-1-\lambda/2)r_B + r_a + r_b)} A^{N-2-\lambda} B^{N-2-\lambda} a^2 b^2 x_I^{-1};q \right)_{\infty} }
 { \prod_{\alpha = 1}^2 \left( q^{(\lambda r_A + \lambda r_B + r_a + r_b)/2} A^{\lambda} B^{\lambda} a b x_I \tilde{x}_{\alpha}; q \right)_{\infty} }
 \nonumber \\
& \times \frac{\prod_{\lambda = 0}^{n-2} \prod_{\alpha = 1}^2 \left( q^{((n-3/2-\lambda/2)r_A + (n-1-\lambda/2)r_B + 3r_a/2 + r_b/2)} A^{N-3-\lambda} B^{N-2-\lambda} a^3 b x_{\alpha}^{-1};q \right)_{\infty} }
 { \prod_{\lambda = 1}^{n-1} \left( q^{\lambda r_A/2 + \lambda r_B/2} A^{\lambda} B^{\lambda}; q \right)_{\infty} \left( q^{\lambda r_A/2 + (\lambda - 1)r_B/2 + r_b} A^{\lambda} B^{\lambda - 1} b^2; q \right)_{\infty} }
 \nonumber \\
& \times \prod_{\lambda = 1}^{n-1} \prod_{I = 1}^3 \frac{1}{\left( q^{(\lambda - 1)r_A/2 + \lambda r_B/2 + r_a} A^{\lambda - 1} B^{\lambda} a^2 x_I^{-1}; q \right)_{\infty}}. 
\end{align}
On the other hand, for $SU(2n+1)$ it follows 
from equation (3.2b) in \cite{MR1266569} 
that it can be expressed as
\begin{align}
\label{bdy_SUN_2AS_3_3_hindexB2}
&
\frac{ \left( q^{((N-2)r_A + (n-1)r_B/2 + 3r_a + 2r_b)/2} A^{N-2} B^{n-1} a^3 b^2;q \right)_{\infty} }
 {\left( \prod_{\alpha = 1}^2 \left( q^{nr_B/2} B^n b \tilde{x}_{\alpha}; q \right)_{\infty} \right) \left( q^{(n-1)r_A/2 + 3r_a/2} A^{n-1} a^3; q \right)_{\infty} }
 \nonumber \\
& \times \frac{ \left( q^{((N-2)r_A + nr_B + 3r_a)/2} A^{N-2} B^n a^3;q \right)_{\infty} }
{\prod_{I = 1}^3 (q^{nr_A/2 + r_a/2} A^n a x_I; q)_{\infty}}
 \nonumber \\
& \times \prod_{\lambda = 0}^{n-1} \prod_{I = 1}^3 \frac{ \left( q^{((n-1/2-\lambda/2)r_A + (n-1/2-\lambda/2)r_B + r_a + r_b)} A^{N-2-\lambda} B^{N-2-\lambda} a^2 b^2 x_I^{-1};q \right)_{\infty} }
 { \prod_{\alpha = 1}^2 \left( q^{(\lambda r_A + \lambda r_B + r_a + r_b)/2} A^{\lambda} B^{\lambda} a b x_I \tilde{x}_{\alpha}; q \right)_{\infty} }
 \nonumber \\
& \times \prod_{\lambda = 0}^{n-1} \prod_{\alpha = 1}^2 \left( q^{((n-1-\lambda/2)r_A + (n-1/2-\lambda/2)r_B + 3r_a/2 + r_b/2)} A^{N-3-\lambda} B^{N-2-\lambda} a^3 b x_{\alpha};q \right)_{\infty}
 \nonumber \\
& \times \prod_{\lambda = 1}^{n} \frac{1}
 { \left( q^{\lambda r_A/2 + \lambda r_B/2} A^{\lambda} B^{\lambda}; q \right)_{\infty} \left( q^{\lambda r_A/2 + (\lambda - 1)r_B/2 + r_b} A^{\lambda} B^{\lambda - 1} b^2; q \right)_{\infty} }
 \nonumber \\
& \times \prod_{\lambda = 1}^{n} \prod_{I = 1}^3 \frac{1}{\left( q^{(\lambda - 1)r_A/2 + \lambda r_B/2 + r_a} A^{\lambda - 1} B^{\lambda} a^2 x_I^{-1}; q \right)_{\infty}}. 
\end{align}
The equations  (\ref{bdy_SUN_2AS_3_3_hindexB1}) and (\ref{bdy_SUN_2AS_3_3_hindexB2}) can be understood as the half-indices of dual boundary conditions for theory B. 
Theory B has no gauge group and the chirals depend on whether $N$ is even, i.e.\ $N = 2n$, or odd, i.e.\ $N = 2n + 1$. Note that in both cases there is a
global $SU(3) \times SU(2)$ flavor symmetry in theory A, and we can label the
chirals by their representations of this group.

For $N = 2n$ we have
\begin{align}
\hspace*{-1.5cm}
\label{SU2n_AS_charges}
\begin{array}{c|c|c|c|c|c|c|c|c|c}
& \textrm{bc} & SU(N = 2n) & SU(3) & SU(2) & U(1)_A & U(1)_B & U(1)_a & U(1)_b & U(1)_R \\ \hline
\textrm{VM} & \mathcal{N} & {\bf Adj} & {\bf 1} & {\bf 1} & 0 & 0 & 0 & 0 & 0 \\
\Phi & \textrm{N} & {\bf N(N-1)/2} & {\bf 1} & {\bf 1} & 1 & 0 & 0 & 0 & 0 \\
\overline{\Phi} & \textrm{N} & {\bf \overline{N(N-1)/2}} & {\bf 1} & {\bf 1} & 0 & 1 & 0 & 0 & 0 \\
Q_I & \textrm{N} & {\bf N} & {\bf 3} & {\bf 1} & 0 & 0 & 1 & 0 & 0 \\
\overline{Q}_{\alpha} & \textrm{N} & {\bf \overline{N}} & {\bf 1} & {\bf 2} & 0 & 0 & 0 & 1 & 0 \\
\widetilde{Q} & \textrm{D} & {\bf \overline{N}} & {\bf 1} & {\bf 1} & 2-2n & 2-2n & -3 & -2 & 2 \\ \hline
B^I & \textrm{N} & {\bf 1} & {\bf \overline{3}} & {\bf 1} & n-1 & 0 & 2 & 0 & 0 \\
\overline{B} & \textrm{N} & {\bf 1} & {\bf 1} & {\bf 1} & 0 & n-1 & 0 & 2 & 0 \\
\phi & \textrm{N} & {\bf 1} & {\bf 1} & {\bf 1} & n & 0 & 0 & 0 & 0 \\
\overline{\phi} & \textrm{N} & {\bf 1} & {\bf 1} & {\bf 1} & 0 & n & 0 & 0 & 0 \\
M^{(0 \le \lambda \le n-1)}_{I \alpha} & \textrm{N} & {\bf 1} & {\bf 3} & {\bf 2} & \lambda&\lambda& 1 & 1 & 0 \\
\widehat{\phi}^{(1 \le \lambda \le n-1)} & \textrm{N} & {\bf 1} & {\bf 1} & {\bf 1} & \lambda & \lambda & 0 & 0 & 0 \\
M^{(1 \le \lambda \le n-1)} & \textrm{N} & {\bf 1} & {\bf 1} & {\bf 1} & \lambda & \lambda-1 & 0 & 2 & 0 \\
M^{I \; (1 \le \lambda \le n-1)} & \textrm{N} & {\bf 1} & {\bf \overline{3}} & {\bf 1} & \lambda-1 & \lambda & 2 & 0 & 0 \\
\widetilde{M}_{\alpha} & \textrm{D} & {\bf 1} & {\bf 1} & {\bf 2} & 2-2n & 1-n & -3 & -1 & 2 \\
\widetilde{M}^{(0 \le \lambda \le n-1)}_I & \textrm{D} & {\bf 1} & {\bf 3} & {\bf 1} & 2-2n+\lambda & 2-2n+\lambda & -2 & -2 & 2 \\
\widetilde{M}^{(0 \le \lambda \le n-2)}_{\alpha} & \textrm{D} & {\bf 1} & {\bf 1} & {\bf 2} & 3 - 2n + \lambda & 2 - 2n + \lambda & -3 & -1 & 2
\end{array}
\end{align}
with the operator mapping $B^I \sim \epsilon Q_J Q_K \epsilon^{IJK} \Phi^{n-1}$, $\overline{B} \sim \epsilon \overline{Q}_1 \overline{Q}_2 \overline{\Phi}^{n-1}$, $\phi \sim \epsilon \Phi^n$, $\overline{\phi} \sim \epsilon \overline{\Phi}^n$, $M^{(\lambda)}_{I \alpha} \sim Q_I \Phi^\lambda \overline{\Phi}^\lambda \overline{Q}_{\alpha}$, $\widehat{\phi}^{(\lambda)} \sim \Phi^\lambda \overline{\Phi}^\lambda$, $M^{(\lambda)} \sim \Phi^\lambda \overline{\Phi}^{\lambda-1} \overline{Q}_1 \overline{Q}_2$, $M^{(\lambda)}_{IJ} \sim Q_I Q_J \Phi^{\lambda-1} \overline{\Phi}^\lambda$, $\widetilde{M}_{\alpha} \sim \widetilde{Q} \overline{Q}_{\alpha} \overline{\Phi}^n$, $\widetilde{M}^{(\lambda)}_I \sim Q_I \Phi^\lambda \overline{\Phi}^\lambda \widetilde{Q}$ and $\widetilde{M}^{(\lambda)}_{\alpha} \sim \Phi^{\lambda+1} \overline{\Phi}^\lambda \overline{Q}_{\alpha} \widetilde{Q}$. 
In the table, to save space, we have indicated the range of $\lambda$ within the superscripts.

In the case of $N=2$ this is equivalent to the duality discussed in section~\ref{sec_su2nf3na3} after removing $\Phi$ and $\overline{\Phi}$ from theory A and the matching $\phi$ and $\overline{\phi}$ from theory B.

The boundary anomaly for theory B is given by
\begin{align}
\label{bdy_SUN_2AS_3_3_anomB1}
&
\Acal^{B, N=2n}
\nonumber\\
&= - \underbrace{\left( n\Tr(x^2) + \frac{3n}{2} \Tr(\tilde{x}^2) + 3\sum_{\lambda = 0}^{n-1} (\lambda A + \lambda B + a + b - r)^2 \right)}_{M_{I \alpha}^{(\lambda)}, \; N}
\nonumber \\
 & - \underbrace{\frac{1}{2}\big( (n-1)B + 2b - r \big)^2}_{\overline{B}, \; N}
  - \underbrace{\frac{1}{2}(nA-r)^2}_{\phi, \; N}
  - \underbrace{\frac{1}{2}(nB-r)^2}_{\overline{\phi}, \; N}
  \nonumber \\
 & - \underbrace{\left( \frac{1}{2} \sum_{\lambda = 1}^{n-1} (\lambda A + (\lambda - 1)B + 2b - r)^2 \right)}_{M_{\alpha \beta}^{(\lambda)}, \; N}
  - \underbrace{\left( \frac{1}{2} \Tr(x^2) + \frac{3}{2}\big( (n-1)A + 2a-r \big)^2 \right)}_{B^I, \; N}
  \nonumber \\
  & - \underbrace{\left( \frac{n-1}{2} \Tr(x^2) + \frac{3}{2} \sum_{\lambda = 1}^{n-1} \big( (\lambda - 1)A + \lambda B + 2a - r \big)^2 \right)}_{M^{I \; (\lambda)}, \; N}
  - \underbrace{\frac{1}{2} \sum_{\lambda = 1}^{n-1} (\lambda A + \lambda B - r)^2}_{\widehat{\phi}^{(\lambda)}, \; N}
  \nonumber \\
 & + \underbrace{\left( \frac{n}{2} \Tr(\tilde{x}^2) + \sum_{\lambda = -1}^{n-2} \big( (3-2n+\lambda)A + (2-2n+\lambda)B - 3a - b + r)^2 \right)}_{\widetilde{M}_{\alpha}, \; \widetilde{M}_{\alpha}^{(\lambda)}, \; D}
  \nonumber \\
  & + \underbrace{\left( \frac{n}{2} \Tr(x^2) + \frac{3}{2} \sum_{\lambda = 0}^{n-1} \big( (2-2n+\lambda)A + (2-2n+\lambda)B - 2a - 2b + r \big)^2 \right)}_{\widetilde{M}_I^{(\lambda)}, \; D},
\end{align}
which agrees with the anomaly (\ref{bdy_SUN_2AS_3_3_anomA}) for theory A. 

For $N = 2n+1$ we have
\begin{align}
\hspace*{-1.5cm}
\label{SU2n1_AS_charges}
\begin{array}{c|c|c|c|c|c|c|c|c|c}
& \textrm{bc} & SU(N = 2n+1) & SU(3) & SU(2) & U(1)_A & U(1)_B & U(1)_a & U(1)_b & U(1)_R \\ \hline
\textrm{VM} & \mathcal{N} & {\bf Adj} & {\bf 1} & {\bf 1} & 0 & 0 & 0 & 0 & 0 \\
\Phi & \textrm{N} & {\bf N(N-1)/2} & {\bf 1} & {\bf 1} & 1 & 0 & 0 & 0 & 0 \\
\overline{\Phi} & \textrm{N} & {\bf \overline{N(N-1)/2}} & {\bf 1} & {\bf 1} & 0 & 1 & 0 & 0 & 0 \\
Q_I & \textrm{N} & {\bf N} & {\bf 3} & {\bf 1} & 0 & 0 & 1 & 0 & 0 \\
\overline{Q}_{\alpha} & \textrm{N} & {\bf \overline{N}} & {\bf 1} & {\bf 2} & 0 & 0 & 0 & 1 & 0 \\
\widetilde{Q} & \textrm{D} & {\bf \overline{N}} & {\bf 1} & {\bf 1} & 1-2n & 1-2n & -3 & -2 & 2 \\ \hline
B_I & \textrm{N} & {\bf 1} & {\bf 3} & {\bf 1} & n & 0 & 1 & 0 & 0 \\
\overline{B}_{\alpha} & \textrm{N} & {\bf 1} & {\bf 1} & {\bf 2} & 0 & n & 0 & 1 & 0 \\
B & \textrm{N} & {\bf 1} & {\bf 1} & {\bf 1} & n-1 & 0 & 3 & 0 & 0 \\
M^{(0 \le \lambda \le n-1)}_{I \alpha} & \textrm{N} & {\bf 1} & {\bf 3} & {\bf 2} & \lambda & \lambda & 1 & 1 & 0 \\
\widehat{\phi}^{(1 \le \lambda \le n)} & \textrm{N} & {\bf 1} & {\bf 1} & {\bf 1} & \lambda & \lambda & 0 & 0 & 0 \\
\overline{M}^{(1 \le \lambda \le n)} & \textrm{N} & {\bf 1} & {\bf 1} & {\bf 1} & \lambda & \lambda-1 & 0 & 2 & 0 \\
M^{I \; (1 \le \lambda \le n)} & \textrm{N} & {\bf 1} & {\bf \overline{3}} & {\bf 1} & \lambda-1 & \lambda & 2 & 0 & 0 \\
\widehat{B} & \textrm{D} & {\bf 1} & {\bf 1} & {\bf 1} & 1-2n & 1-n & -3 & -2 & 2 \\
\widetilde{B} & \textrm{D} & {\bf 1} & {\bf 1} & {\bf 1} & 1-2n & -n & -3 & 0 & 2 \\
\widetilde{M}^{(0 \le \lambda \le n-1)}_I & \textrm{D} & {\bf 1} & {\bf 3} & {\bf 1} & 1-2n+\lambda & 1-2n+\lambda & -2 & -2 & 2 \\
\widetilde{M}^{(0 \le \lambda \le n-1)}_{\alpha} & \textrm{D} & {\bf 1} & {\bf 1} & {\bf 2} & 2 - 2n + \lambda & 1 - 2n + \lambda & -3 & -1 & 2
\end{array}
\end{align}
with the operator mapping $B_I \sim \epsilon Q_I \Phi^n$, $\overline{B}_{\alpha} \sim \epsilon \overline{Q}_{\alpha} \overline{\Phi}^n$, $B \sim \epsilon Q_1 Q_2 Q_3 \Phi^{n-1}$, $M^{(\lambda)}_{I \alpha} \sim Q_I \Phi^\lambda \overline{\Phi}^\lambda \overline{Q}_{\alpha}$, $\widehat{\phi}^{(\lambda)} \sim \Phi^\lambda \overline{\Phi}^\lambda$, $\overline{M}^{(\lambda)} \sim \Phi^\lambda \overline{\Phi}^{\lambda-1} \overline{Q}_1 \overline{Q}_2$, $M^{I \; (\lambda)} \sim Q_J Q_K \epsilon^{IJK} \Phi^{\lambda-1} \overline{\Phi}^\lambda$, $\widehat{B} \sim \epsilon \widetilde{Q} \overline{\Phi}^n$, $\widetilde{B} \sim \epsilon \widetilde{Q} \overline{Q}_1 \overline{Q}_2 \overline{\Phi}^{n-1}$, $\widetilde{M}^{(\lambda)}_I \sim Q_I \Phi^\lambda \overline{\Phi}^\lambda \widetilde{Q}$ and $\widetilde{M}^{(\lambda)}_{\alpha} \sim \Phi^\lambda \overline{\Phi}^{\lambda+1} \overline{Q}_{\alpha} \widetilde{Q}$.

In the case of $N=3$ the antisymmetric and conjugate antisymmetric chirals $\Phi$ and $\overline{\Phi}$ are antifundamental and fundamental chirals so this duality is equivalent to the $N = 3$ example of the duality discussed in section~\ref{sec_GNR_integrals}.

The 't Hooft anomaly (\ref{bdy_SUN_2AS_3_3_anomB1}) for theory A matches with that for theory B, 
which is given by
\begin{align}
\label{bdy_SUN_2AS_3_3_anomB2}
&
\Acal^{B, N=2n+1}
\nonumber\\
&= - \underbrace{\left( n\Tr(x^2) + \frac{3n}{2} \Tr(\tilde{x}^2) + 3\sum_{\lambda = 0}^{n-1} (\lambda A + \lambda B + a + b - r)^2 \right)}_{M_{I \alpha}^{(\lambda)}, \; N}
\nonumber \\
 & - \underbrace{\frac{1}{2}\big( (n-1)B + 3b - r \big)^2}_{B, \; N}
 - \underbrace{\left( \frac{1}{2} \Tr(\tilde{x}^2) + \frac{3}{2}\big( nB +  - r \big)^2 \right)}_{\overline{B}_{\alpha}, \; N}
  \nonumber \\
 & - \underbrace{\left( \frac{1}{2} \sum_{\lambda = 1}^{n} (\lambda A + (\lambda - 1)B + 2b - r)^2 \right)}_{\overline{M}^{(\lambda)}, \; N}
  - \underbrace{\left( \frac{1}{2} \Tr(x^2) + \frac{3}{2}\big( nA + a - r \big)^2 \right)}_{B_I, \; N}
  \nonumber \\
  & - \underbrace{\left( \frac{n}{2} \Tr(x^2) + \frac{3}{2} \sum_{\lambda = 1}^{n} \big( (\lambda - 1)A + \lambda B + 2a - r \big)^2 \right)}_{M^{I \; (\lambda)}, \; N}
  - \underbrace{\frac{1}{2} \sum_{\lambda = 1}^{n} (\lambda A + \lambda B - r)^2}_{\widehat{\phi}^{(\lambda)}, \; N}
  \nonumber \\
 & + \underbrace{\left( \big( (1-2n)A - nB - 3a + r \big)^2 \right)}_{\widetilde{B}, \; D}
  \nonumber \\
  & + \underbrace{\left( \big( (1-2n)A + (1-n)B - 3a - 2b + r \big)^2 \right)}_{\widehat{B}, \; D}
  \nonumber \\
 & + \underbrace{\left( \frac{n}{2} \Tr(\tilde{x}^2) + \sum_{\lambda = 0}^{n-1} \big( (2-2n+\lambda)A + (1-2n+\lambda)B - 3a - b + r)^2 \right)}_{\widetilde{M}_{\alpha}^{(\lambda)}, \; D}
  \nonumber \\
  & + \underbrace{\left( \frac{n}{2} \Tr(x^2) + \frac{3}{2} \sum_{\lambda = 0}^{n-1} \big( (2-2n+\lambda)A + (2-2n+\lambda)B - 2a - 2b + r \big)^2 \right)}_{\widetilde{M}_I^{(\lambda)}, \; D}. 
\end{align}
This precisely matches with the boundary 't Hooft anomaly (\ref{bdy_SUN_2AS_3_3_anomA}) for theory A. 

The precise agreement of the half-indices and the boundary anomalies lead to the following boundary confining dualities: 
\begin{align}
\label{bdy_SUN_2AS_3_3_dual1}
&\textrm{$SU(N)$ $+$ 2 antisym. chirals $\Phi, \overline{\Phi}$ $+$ $3$ fund. chirals $Q_{I}$}
\nonumber\\
&\textrm{$+$ $2$ antifund. chiral $\overline{Q}_{\alpha}$ $+$ $1$ antifund. chiral $\widetilde{Q}$ with b.c. $(\mathcal{N},N,N,N,N,D)$}
\nonumber\\
&\Leftrightarrow 
\textrm{
an $SU(3)$ antifund. $B^{I}$ $+$ a singlet $\overline{B}$ $+$ $2$ singlets $\phi, \overline{\phi}$}
\nonumber\\
&\textrm{$+$ an $SU(3)\times SU(2)$ bifund. $M_{I\alpha}^{(0\le \lambda\le n-1)}$ $+$ a singlet $\widehat{\phi}^{(0\le \lambda\le n)}$ }
\nonumber\\
&\textrm{
$+$ a singlet $M^{(1\le \lambda\le n-1)}$ $+$ an $SU(3)$ antifund. $M^{I (1\le \lambda\le n-1)}$}
\nonumber\\
&\textrm{
$+$ an $SU(2)$ fund. $\widetilde{M}_{\alpha}$ $+$ an $SU(3)$ fund. $\widetilde{M}_I^{(0\le \lambda\le n-1)}$ 
$+$ an $SU(2)$ fund. $\widetilde{M}_{\alpha}^{(0\le \lambda\le n-2)}$ 
}
\nonumber\\
&
\textrm{with b.c. $(N,N,N,N,N,N,N,N,D,D,D)$}. 
\end{align}
for $N=2n$ and
\begin{align}
\label{bdy_SUN_2AS_3_3_dual2}
&\textrm{$SU(N)$ $+$ 2 antisym. chirals $\Phi, \overline{\Phi}$ $+$ $3$ fund. chirals $Q_{I}$}
\nonumber\\
&\textrm{$+$ $2$ antifund. chiral $\overline{Q}_{\alpha}$ $+$ $1$ antifund. chiral $\widetilde{Q}$ with b.c. $(\mathcal{N},N,N,N,N,D)$}
\nonumber\\
&\Leftrightarrow 
\textrm{
an $SU(3)$ antifund. $B^{I}$ $+$ an $SU(2)$ fund. $\overline{B}_{\alpha}$ $+$ a singlet $B$ }
\nonumber\\
&\textrm{$+$ an $SU(3)\times SU(2)$ bifund. $M_{I\alpha}^{(0\le \lambda\le n-1)}$ $+$ a singlet $\widehat{\phi}^{(1\le \lambda\le n)}$ }
\nonumber\\
&\textrm{
$+$ a singlet $\overline{M}^{(1\le \lambda\le n-1)}$ $+$ an $SU(3)$ antifund. $M^{I (1\le \lambda\le n-1)}$}
\nonumber\\
&\textrm{
$+$ $2$ singlets $\widehat{B}, \widetilde{B}$ 
$+$ an $SU(3)$ fund. $\widetilde{M}_I^{(0\le \lambda\le n-1)}$ 
$+$ an $SU(2)$ fund. $\widetilde{M}_{\alpha}^{(0\le \lambda\le n-2)}$ 
}
\nonumber\\
&
\textrm{with b.c. $(N,N,N,N,N,N,N,D,D,D,D)$}. 
\end{align}
for $N=2n+1$.

Moreover, the boundary confining dualities (\ref{bdy_SUN_2AS_3_3_dual1}) and (\ref{bdy_SUN_2AS_3_3_dual2}) can be extended to the bulk where theory A has superpotential $\Wcal_A = V + \Tr (\Phi \overline{\Phi})^{n+1}$ and theory B has a complicated superpotential which we do not write. This corresponds to a 4d duality described in \cite{Intriligator:1995ax} which compactifies to the same duality in 3d with the addition of the linear monopole superpotential term $V$ in theory A. 
For example, for $N=4$ with $r_a=\frac16$, $r_b=\frac16$ and $r_A=\frac14$, $r_B=\frac14$, we have confirmed that the full-indices for theory A and B beautifully match
\begin{align}
I^A&=I^B
\nonumber\\
&=1+(6ab+3a^{-2}A^{-2}b^{-2}B^{-2})q^{1/6}
+(A^2+AB+B^2)q^{1/4}
\nonumber\\
&
+(3a^2+b^2+2a^{-3}A^{-2}b^{-1}B^{-2})(A+B)q^{7/24}
\nonumber\\
&+3(7a^2b^2+2a^{-4}A^{-4}b^{-4}B^{-4}+6a^{-1}A^{-2}b^{-1}B^{-2})q^{1/3}+\cdots
\end{align}
where 
\begin{align}
I^A&=
\frac{1}{4!}\sum_{m_1,m_2,m_3\in \mathbb{Z}}
\prod_{i=1}^3 
\oint 
\frac{ds_i}{2\pi is_i}
\prod_{i \ne j}^4 (1-q^{\frac{|m_i-m_j|}{2}}s_i^{\pm}s_j^{\mp})
\prod_{i=1}^4
\prod_{\alpha=1}^{2}
\frac{
(q^{1-\frac{r_b}{2}+\frac{|m_i|}{2}}b^{-1}s_i\tilde{x}_{\alpha}^{-1};q)_{\infty}
}{
(q^{\frac{r_b}{2}+\frac{|m_i|}{2}}bs_i^{-1}\tilde{x}_{\alpha};q)_{\infty}
}
\nonumber\\
&\times 
\prod_{i=1}^4 \prod_{I=1}^3 
\frac{
(q^{1-\frac{r_a}{2}+\frac{|m_i|}{2}}a^{-1}s_i\tilde{x}_{I}^{-1};q)_{\infty}
}
{
(q^{\frac{r_a}{2}+\frac{|m_i|}{2}}as_i^{-1}\tilde{x}_{I};q)_{\infty}
}
\prod_{i<j}
\frac{
(q^{1-\frac{r_A}{2}+\frac{|m_i+m_j|}{2}}A^{-1}s_i^{-1}s_j^{-1};q)_{\infty}
}
{
(q^{\frac{r_A}{2}+\frac{|m_i+m_j|}{2}}As_is_j;q)_{\infty}
}
\nonumber\\
&\times 
\prod_{i<j}
\frac{
(q^{1-\frac{r_B}{2}+\frac{|m_i+m_j|}{2}}B^{-1}s_is_j;q)_{\infty}
}
{
(q^{\frac{r_B}{2}+\frac{|m_i+m_j|}{2}}Bs_i^{-1}s_j^{-1};q)_{\infty}
}
\frac{
(q^{\frac{2r_A+3r_a+2r_b+|m_i|}{2}}A^{2}a^3 b^2 s_i^{-1};q)_{\infty}
}
{
(q^{1-\frac{2r_A+3r_a+2r_b}{2}+\frac{|m_i|}{2}}A^{-2}a^{-3}b^{-2}s_i;q)_{\infty}
}
\nonumber\\
&\times 
q^{
\Delta
}
A^{-\sum_{i<j}|m_i-m_j|+2\sum_{i=1}^4|m_i|}
B^{-\sum_{i<j}|m_i-m_j|+2\sum_{i=1}^4|m_i|}
\end{align}
and 
\begin{align}
I^B&=
\prod_{\alpha=1}^2
\frac{
(q^{\frac{2r_A+r_B+3r_a+r_b}{2}} A^2Ba^3b\tilde{x}_{\alpha};q)_{\infty}
}
{
(q^{1-\frac{2r_A+r_B+3r_a+r_b}{2}} A^{-2}B^{-1}a^{-3}b^{-1}\tilde{x}_{\alpha}^{-1};q)_{\infty}
}
\frac{
(q^{1-r_A}A^{-2};q)_{\infty}
}
{
(q^{r_A}A^2;q)_{\infty}
}
\frac{
(q^{1-r_B}B^{-2};q)_{\infty}
}
{
(q^{r_B}B^2;q)_{\infty}
}
\nonumber\\
&\times 
\frac{
(q^{1-\frac{r_B}{2}-r_b}B^{-1}b^{-2};q)_{\infty}
}
{
(q^{\frac{r_B}{2}+r_b}Bb^{2};q)_{\infty}
}
\prod_{I=1}^3
\frac{
(q^{1-\frac{r_A}{2}}A^{-1}a^{-2}x_I;q)_{\infty}
}
{
(q^{\frac{r_A}{2}}Aa^2x_I^{-1};q)_{\infty}
}
\nonumber\\
&\times 
\prod_{\lambda=0}^1
\prod_{I=1}^3
\frac{
(q^{(1-\frac{\lambda}{2})r_A+(1-\frac{\lambda}{2})r_B+r_a+r_b} A^{2-\lambda} B^{2-\lambda} a^2 b^2x_I^{-1};q)_{\infty}
}
{
(q^{1-(1-\frac{\lambda}{2})r_A-(1-\frac{\lambda}{2})r_B-r_a-r_b} A^{-2+\lambda} B^{-2+\lambda} a^{-2} b^{-2}x_I;q)_{\infty}
}
\nonumber\\
&\times 
\prod_{\lambda=0}^1
\prod_{I=1}^3
\prod_{\alpha=1}^2
\frac{
(q^{1-\frac{\lambda r_A+\lambda r_B+r_a+r_b}{2}} A^{-\lambda} B^{-\lambda} a^{-1} b^{-1}x_I^{-1}\tilde{x}_{\alpha}^{-1};q)_{\infty}
}
{
(q^{\frac{\lambda r_A+\lambda r_B+r_a+r_b}{2}} A^{\lambda} B^{\lambda} a bx_I\tilde{x}_{\alpha};q)_{\infty}
}
\nonumber\\
&\times 
\prod_{\alpha=1}^2
\frac{
(q^{\frac{r_A}{2}+r_B+\frac32 r_a+\frac{r_b}{2}} AB^2a^3bx_{\alpha}^{-1};q)_{\infty}
}
{
(q^{1-\frac{r_A}{2}-r_B-\frac32 r_a-\frac{r_b}{2}} A^{-1}B^{-2}a^{-3}b^{-1}x_{\alpha};q)_{\infty}
}
\frac{
(q^{1-\frac{r_A}{2}-\frac{r_B}{2}} A^{-1}B^{-1};q)_{\infty}
}
{
(q^{\frac{r_A}{2}+\frac{r_B}{2}} AB;q)_{\infty}
}
\nonumber\\
&\times 
\frac{
(q^{1-\frac{r_A}{2}-r_b} A^{-1}b^{-2};q)_{\infty}
}
{
(q^{\frac{r_A}{2}+r_b} Ab^2;q)_{\infty}
}
\prod_{I=1}^3
\frac{
(q^{1-\frac{r_B}{2}-r_a} B^{-1}a^{-2}x_I;q)_{\infty}
}
{
(q^{\frac{r_B}{2}+r_a} Ba^2x_I^{-1};q)_{\infty}
}
\end{align}
with $m_4 = - (m_1 + m_2 + m_3)$ and
\begin{align}
\Delta&=
\frac34 (1-r_a)\sum_{i=1}^4 |m_i|
+\frac12(1-r_b)\sum_{i=1}^4|m_i|
\nonumber\\
&
+\frac12 (1-r_A)\sum_{i<j}|m_i+m_j|
+\frac12 (1-r_B)\sum_{i<j}|m_i+m_j|
\nonumber\\
&+\frac14(-1+2r_A+2r_B+3r_a+2r_b)\sum_{i=1}^4|m_i|
-\frac12 \sum_{i<j}|m_i-m_j| \; .
\end{align} 

For general $N$ the monopoles have dimensions given, with $\sum_{i=1}^N m_i = 0$, by
\begin{align}
\label{Mono_Ch_suN_6_2}
    \left( 2 + \frac{(N-2)(r_A + r_B)}{2} \right) \sum_{i = 1}^N |m_i| + (2 - r_A - r_B) \sum_{i < j}^N |m_i + m_j| - \sum_{i < j}^N |m_i - m_j|
\end{align}
which we believe are bounded from below by $2$ for $r_A = 0$ and $r_B = 0$ but we haven't proven this. For $r_A \ge 0$ and $r_B \ge 0$ the monopole dimensions may decrease but there will still be a range of values of $r_A$ and $r_B$ for which all monopoles will have positive dimension.

\section{New integrals}
\label{sec_New_integral_identities}
In this section we consider different pairs of theory A and theory B which admit the boundary confining dualities. 
As a consequence of the conjectural boundary confining dualities, we present new identities of the Askey-Wilson type $q$-beta integrals. 
\subsection{$SO(N)$ with $N-2$ fundamental chirals and $\Zb_2$ fugacities}
\label{sec_SON_Nm2_discrete_integrals}
In section~\ref{sec_GAW_SO_integrals} we interpreted the identity of the Askey-Wilson type $q$-beta integral as a consequence of the boundary confining duality
for $SO(N)$ with $N-2$ chirals. 
In the case of orthogonal gauge groups we can have gauge groups with the same Lie algebra but different global structure, distinguished by $\Zb_2$ global symmetries $\Zb_2^{\Mcal}$ (magnetic) and $\Zb_2^{\Ccal}$ (charge conjugation) with discrete fugacities $\zeta$ and $\chi$ as discussed for these 3d theories in \cite{Aharony:2013kma}. As shown in \cite{Okazaki:2021pnc} these discrete fugacities can be included in the half-indices and the dualities hold also with appropriate boundary conditions. In the case here the dual theory has gauge group $SO(0)$ which we interpret as the absence of a gauge group. The expressions for the half-indices in theory B also depend on discrete fugacities which are mapped from theory A as $\tilde{\zeta} = \zeta$ and $\tilde{\chi} = \zeta \chi$. In the case of $SO(0)$ there is no magnetic symmetry which we interpret as a restriction $\zeta \chi = \tilde{\chi} = 1$. Indeed, the general half-index expression makes sense with simply no contribution from the gauge group if $\tilde{\chi} = +1$ but it is not well-defined for $\tilde{\chi} = -1$ as in that case we should fix one of the gauge fugacities (for Neumann boundary condition) or one of the monopole fluxes (for Dirichlet boundary condition), but there are none. \footnote{See \cite{Okazaki:2021pnc} for the explicit expressions but the issue can also be seen here in the expression for $SO(N = 2n)$ in theory A.} So, we must restrict to the cases of $\zeta = \chi = +1$ (corresponding to the case discussed in section~\ref{sec_GAW_SO_integrals}) or $\zeta = \chi = -1$ in theory A. The latter leads to the following identities where, since $\chi = -1$, the theory A half-index takes a different form for the even $N = 2n$ and odd $N = 2n+1$ cases.

For the case $N = 2n+1$ we have the half-index of theory A
\begin{align}
\mathbb{II}_{(\mathcal{N},N)}^A
&=
\frac{(q)_{\infty}^n}{n! 2^n} \prod_{i=1}^n \oint \frac{ds_i}{2\pi i s_i}
(-s_i^{\pm}; q)_{\infty}
\prod_{i < j}^n (s_i^{\pm} s_j^{\mp}; q)_{\infty} (s_i^{\pm} s_j^{\pm}; q)_{\infty}
 \nonumber \\
& \times
\frac{1}{\prod_{\alpha = 1}^{N - 2} \left( (-q^{r/2} a x_{\alpha}; q)_{\infty}^{\epsilon} \prod_{i = 1}^n (q^{r/2} s_i^{\pm} a x_{\alpha}; q)_{\infty} \right)}. 
\end{align}
This coincides with 
\begin{align}
\mathbb{II}_{(N,D)}^B = & \frac{\left( -q^{1 + (N/2 - 1)r} a^{N-2}; q \right)_{\infty}}{\prod_{\alpha \le \beta}^{N-2} (q^r a^2 x_{\alpha} x_{\beta}; q)_{\infty}}. 
\end{align}
But in fact this is simply derived from the $\zeta = \chi = +1$ case by replacing $a \rightarrow -a$ and in the theory A half-index integrand $s_i \rightarrow -s_i$.

For the case $N = 2n$ we have
\begin{align}
\mathbb{II}_{(\mathcal{N},N)}^A
&=
\frac{(q)_{\infty}^{n-1} (-q; q)_{\infty}}{(n-1)! 2^{n-1}} \prod_{i=1}^{n-1} \oint \frac{ds_i}{2\pi i s_i}
(s_i^{\pm}; q)_{\infty} (-s_i^{\pm}; q)_{\infty}
\prod_{i < j}^{n-1} (s_i^{\pm} s_j^{\mp}; q)_{\infty} (s_i^{\pm} s_j^{\pm}; q)_{\infty}
 \nonumber \\
 &\times
\frac{1}{\prod_{\alpha = 1}^{N - 2} \left( (\pm q^{r/2} a x_{\alpha}; q)_{\infty} \prod_{i = 1}^{n-1} (q^{r/2} s_i^{\pm} a x_{\alpha}; q)_{\infty} \right)}. 
\end{align}
This agrees with 
\begin{align}
\mathbb{II}_{(N,D)}^B = & \frac{\left( -q^{1 + (N/2 - 1)r} a^{N-2}; q \right)_{\infty}}{\prod_{\alpha \le \beta}^{N-2} (q^r a^2 x_{\alpha} x_{\beta}; q)_{\infty}}. 
\end{align}
which is not trivially related to the $\zeta = \chi = +1$ case.

For example in the case of $N = 4$, turning off the fugacity $x_{\alpha}$ and setting $r=\frac14$, 
the half-indices are expanded as 
\begin{align}
& \mathbb{II}_{(\mathcal{N},N)}^A
 = \mathbb{II}_{(N,D)}^B
\nonumber\\
 = &   1 + 3 a^2 q^{1/4} + 6 a^4 q^{1/2} + 10 a^6 q^{3/4} + 15 a^8 q + 
 a^2 (4 + 21 a^8) q^{5/4} + 4 a^4 (3 + 7 a^8) q^{3/2}
 \nonumber \\
 & + 
 12 a^6 (2 + 3 a^8) q^{7/4} + 5 a^8 (8 + 9 a^8) q^2 + 
 a^2 (4 + 60 a^8 + 55 a^16) q^{9/4}
 + \cdots
\end{align}

We also note that identities for confining dualities can be derived from theory A with $SO(N)$ gauge group and $N-1$ fundamental chirals (and discrete fugacities). The dual theory B has the trivial gauge group $SO(1)$. Alternatively, we can start with gauge group $SO(1)$ in theory A which gives a theory of free chirals dual to an interacting theory B. For full details of how to construct the half-indices and several examples see \cite{Okazaki:2021pnc}.

As for the case with $\zeta = \chi = +1$ discussed in section~\ref{sec_GAW_SO_integrals} the bulk theory A has non-positive dimension monopole operators as the dimensions are
\begin{align}
\label{Mono_Ch_soN_Nm2_disc}
    2(n-2) \sum_{i = 1}^n |m_i| - \sum_{i < j}^{n-1} \left( |m_i - m_j| + |m_i + m_j| \right)
    = \sum_{i=1}^{n-1} 2(i-1)|m_{\sigma(i)}| \ge 0
\end{align}
for $m_i \in \Zb$ (with at least one $m_i \ne 0$) and $\sigma$ is a permutation giving the ordering $|m_{\sigma(i)}| \ge |m_{\sigma(j)}|$ for $i < j$. We can see that the lower bound is indeed saturated when a single $m_i \ne 0$. This expression is for $r_a = 0$. Choosing positive values for $r_a$ will decrease the R-charge of the monopoles so in that case there will be negative dimension monopoles.
Therefore, again we have a boundary duality which does not correspond to a bulk IR duality of unitary theories. 

\subsection{$SO(N)$ with $N-1 (+1)$ fundamental chirals}
\label{sec_SON_N_integrals}
Consider theory A with $SO(N = 2n + \epsilon)$ gauge group coupled to $N_f = N - 1$ fundamental chirals $Q_{\alpha}$ 
with R-charge $0$ satisfying Neumann boundary conditions and one fundamental chiral $\widetilde{Q}$ with R-charge $0$ satisfying Dirichlet boundary conditions. The boundary 't Hooft anomaly is
\begin{align}
\label{bdy_SON_N_anomA}
&
\Acal = \underbrace{(N-2)\Tr(s^2) + \frac{N(N-1)}{4}r^2}_{\textrm{VM}, \; \Ncal}
 - \underbrace{\left( (N-1)\Tr(s^2) + \frac{N}{2}\Tr(x^2) + \frac{N(N-1)}{2}(a-r)^2 \right)}_{Q_{\alpha}, \; N}
  \nonumber \\
 & + \underbrace{\left( \Tr(s^2) + \frac{N}{2}\big( -(N-1) a - r \big)^2 \right)}_{\widetilde{Q}, \; D}
   \nonumber \\
  &= - \frac{N}{2} \Tr(x^2) + \frac{N(N-1)(N-2)}{2}a^2 + 2N(N-1)ar - \frac{N(N-3)}{4}r^2. 
\end{align}
The Neumann half-index realizes the Askey-Wilson type $q$-beta integral of the form
\begin{align}
\label{SON_N_hindexA}
\mathbb{II}_{(\mathcal{N},N,D)}^A
&=\frac{(q)_{\infty}^n}{n! 2^{n + \epsilon - 1}} \prod_{i=1}^n \oint \frac{ds_i}{2\pi i s_i}
(s_i^{\pm}; q)_{\infty}^{\epsilon}
\prod_{i < j}^n (s_i^{\pm} s_j^{\mp}; q)_{\infty} (s_i^{\pm} s_j^{\pm}; q)_{\infty}
 \nonumber \\
& \times
\frac{(q^{1+N_fr/2} a^{N_f}; q)_{\infty}^{\epsilon} \prod_{i = 1}^n (q^{1+N_fr/2} s_i^{\pm} a^{N_f}; q)_{\infty}}{\prod_{\alpha = 1}^{N_f} \left( (q^{r/2} a x_{\alpha}; q)_{\infty}^{\epsilon} \prod_{i = 1}^n (q^{r/2} s_i^{\pm} a x_{\alpha}; q)_{\infty} \right)}
\end{align}
where $\prod_{\alpha = 1}^{N} x_{\alpha} = 1$.
This is identical to 
\begin{align}
\label{SON_N_hindexB}
\mathbb{II}_{(N,D,D)}^B = &
 \frac{(q^{1+N_f r} a^{2N_f}; q)_{\infty} \prod_{\alpha=1}^{N_f} (q^{1+(N_f - 1)r/2} a^{N_f - 1} x_{\alpha}^{-1}; q)_{\infty}}{\prod_{\alpha \le \beta}^{N_f} (q^r a^2 x_{\alpha} x_{\beta}; q)_{\infty}}. 
\end{align}

For example in the case of $N = 4$, turning off the fugacity $x_{\alpha}$ and setting $r=\frac14$, 
the half-indices are expanded as 
\begin{align}
& \mathbb{II}_{(\mathcal{N},N,D)}^A
 = \mathbb{II}_{(N,D,D)}^B
\nonumber\\
 = &  1 + 6 a^2 q^{1/4} + 21 a^4 q^{1/2} + 56 a^6 q^{3/4} + 126 a^8 q + 
 3 (a^2 + 84 a^10) q^{5/4} + 6 a^4 (3 + 77 a^8) q^{3/2}
 \nonumber \\
 & +
 2 a^6 (31 + 396 a^8) q^{7/4} + 9 a^8 (18 + 143 a^8) q^2 + 
 a^2 (3 + 357 a^8 + 2002 a^16) q^{9/4}
 + \cdots
\end{align}

These identities can be derived by 
starting with the symplectic half-index identity in section~\ref{sec_USp2n_2np4_integrals} and applying the limits \cite{MR1139492} discussed in section~\ref{sec_GAW_SO_integrals}. 
Specifically, combining the axial and flavor fugacities as $X_{\alpha} = a x_{\alpha}$ ($1 \le \alpha \le 2n+3$), then taking $X_{2n+1} = -1$, $X_{2n+3} = q^{1/2}$ and $X_{2n+3} = -q^{1/2}$, before redefining $X_{\alpha} = a x_{\alpha}$ ($1 \le \alpha \le 2n$) with $\prod_{\alpha = 1}^{2n} x_{\alpha} = 1$ gives the identity for $SO(N = 2n+1)$ after some simple manipulations including using the $q$-Pochhammer identity $(\pm z; q)_{\infty} (\pm q^{1/2}z; q)_{\infty} = (z^2; q)_{\infty}$. Taking also $X_{2n} = 1$ gives the identity for $SO(N = 2n)$. 

Now the equation (\ref{SON_N_hindexA}) can be interpreted as the half-index of dual theory B. 
Theory B has an symmetric rank-2 $SU(N_f)$ chiral with R-charge $0$
with Neumann boundary conditions and
a chiral in the fundamental representation of $SU(N_f)$ as well as a singlet
with R-charge $0$ with Dirichlet boundary conditions. The full details of the charges and boundary conditions are summarized as
\begin{align}
\label{SON_N_charges}
\begin{array}{c|c|c|c|c|c}
& \textrm{bc} & SO(N) & SU(N_f = N - 1) & U(1)_a & U(1)_R \\ \hline
\textrm{VM} & \mathcal{N} & {\bf Adj} & {\bf 1} & 0 & 0 \\
Q_{\alpha} & \textrm{N} & {\bf N} & {\bf N_f} & 1 & 0 \\
\widetilde{Q} & \textrm{D} & {\bf N} & {\bf 1} & -N_f & 0 \\
 \hline
M_{\alpha \beta} & \textrm{N} & {\bf 1} & {\bf N_f(N_f + 1)/2} & 2 & 0 \\
\widetilde{M}_{\alpha} & \textrm{D} & {\bf 1} & {\bf N_f} & 1 - N_f & 0 \\
M & \textrm{D} & {\bf 1} & {\bf 1} & -2N_f & 0
\end{array}
\end{align}

The boundary anomaly for theory B is 
\begin{align}
\label{bdy_SON_N_anomB}
\Acal & = - \underbrace{\left( \frac{N+1}{2} \Tr(x^2) + \frac{N(N-1)}{4}(2a-r)^2 \right)}_{M_{\alpha \beta}, \; N}
  \nonumber \\
 & + \underbrace{\frac{1}{2} \Tr(x^2) + \frac{N-1}{2}\big( -(N-2)a - r \big)^2}_{\widetilde{M}_{\alpha}, \; D}
  + \underbrace{\frac{1}{2} \big( -2(N-1)a - r \big)^2}_{M, \; D}. 
\end{align}
This precisely agrees with the anomaly (\ref{bdy_SON_N_anomA}) for theory A. 

The mapping of operators between theory A and theory B is given by
$M_{\alpha \beta} \sim Q_{\alpha} Q_{\beta}$, $M_{\alpha} \sim Q_{\alpha} \widetilde{Q}$ and $M \sim \widetilde{Q} \widetilde{Q}$ where the gauge indices are contracted with the $SO(N)$-invariant rank-$2$ symmetric tensor. 

In this case, although the monopoles would have positive dimension, there is no corresponding bulk duality since for any choice of $r_a \ne 0$ some of the chirals will have negative dimension (and all have zero dimension for $r_a = 0$). However, we note that this is closely related to the bulk duality between $SO(N)$ gauge theory with $N_f > N + 1$ fundamental chirals and an $SO(N_f - N)$ gauge theory with linear monopole superpotentials \cite{Aharony:2013kma, Amariti:2018gdc}. The bulk confining duality cases arising from this, including $N_f = N$, have been discussed in \cite{Benvenuti:2021nwt} and are seemingly very similar to the boundary dualities described above.

As in section~\ref{sec_SON_Nm2_discrete_integrals} we can include a $\Zb_2^{\Ccal}$ fugacity $\chi$, with the above discussion corresponding to $\chi = 1$. Considering the case $\chi = -1$ modifies the contributions of both the vector multiplet and fundamental chirals to the half-index for theory A. We also take $\widetilde{Q}$ to be charged under this symmetry. In theory B only $\widetilde{M}_{\alpha}$ are charged under this symmetry.

For the case $N = 2n+1$ we have the half-index of theory A
\begin{align}
\mathbb{II}_{(\mathcal{N},N,D)}^{A, \; \chi = -1}
 = &
\frac{(q)_{\infty}^n}{n! 2^n} \prod_{i=1}^n \oint \frac{ds_i}{2\pi i s_i}
(-s_i^{\pm}; q)_{\infty}
\prod_{i < j}^n (s_i^{\pm} s_j^{\mp}; q)_{\infty} (s_i^{\pm} s_j^{\pm}; q)_{\infty}
 \nonumber \\
& \times
\frac{(q^{1+N_fr/2} a^{N_f}; q)_{\infty} \prod_{i = 1}^n (-q^{1+N_fr/2} s_i^{\pm} a^{N_f}; q)_{\infty}}{\prod_{\alpha = 1}^{N_f} \left( (-q^{r/2} a x_{\alpha}; q)_{\infty} \prod_{i = 1}^n (q^{r/2} s_i^{\pm} a x_{\alpha}; q)_{\infty} \right)} \; . 
\end{align}
This coincides with 
\begin{align}
\mathbb{II}_{(N,D,D)}^{B, \; \chi = -1} = &
\frac{(q^{1+N_f r} a^{2N_f}; q)_{\infty} \prod_{\alpha=1}^{N_f} (-q^{1+(N_f - 1)r/2} a^{N_f - 1} x_{\alpha}^{-1}; q)_{\infty}}{\prod_{\alpha \le \beta}^{N_f} (q^r a^2 x_{\alpha} x_{\beta}; q)_{\infty}} \; . 
\end{align}
But in fact this is simply derived from the $\chi = +1$ case by replacing $a \rightarrow -a$, noting for theory A that $N_f = N - 1$ is even and for theory B that $N_f - 1 = N - 2$ is odd, and in the theory A half-index integrand also replacing $s_i \rightarrow -s_i$.

For the case $N = 2n$ we have
\begin{align}
\mathbb{II}_{(\mathcal{N},N,D)}^{A, \; \chi = -1}
 = &
\frac{(q)_{\infty}^{n-1} (-q; q)_{\infty}}{(n-1)! 2^{n-1}} \prod_{i=1}^{n-1} \oint \frac{ds_i}{2\pi i s_i}
(s_i^{\pm}; q)_{\infty} (-s_i^{\pm}; q)_{\infty}
\prod_{i < j}^{n-1} (s_i^{\pm} s_j^{\mp}; q)_{\infty} (s_i^{\pm} s_j^{\pm}; q)_{\infty}
 \nonumber \\
 &\times
\frac{(\pm q^{1+N_fr/2} a^{N_f}; q)_{\infty} \prod_{i = 1}^{n-1} (-q^{1+N_fr/2} s_i^{\pm} a^{N_f}; q)_{\infty}}{\prod_{\alpha = 1}^{N_f} (\pm q^{r/2} a x_{\alpha}; q)_{\infty} \prod_{i = 1}^{n-1} (q^{r/2} s_i^{\pm} a x_{\alpha}; q)_{\infty} } \; . 
\end{align}
This agrees with 
\begin{align}
\mathbb{II}_{(N,D,D)}^{B, \; \chi = -1} = &
 \frac{(q^{1+N_f r} a^{2N_f}; q)_{\infty} \prod_{\alpha=1}^{N_f} (-q^{1+(N_f - 1)r/2} a^{N_f - 1} x_{\alpha}^{-1}; q)_{\infty}}{\prod_{\alpha \le \beta}^{N_f} (q^r a^2 x_{\alpha} x_{\beta}; q)_{\infty}}
\end{align}
which is not trivially related to the $\chi = +1$ case.

For the case of $N=2$, $N_f = 1$ we have a simple derivation of this identity
\begin{align}
    \mathbb{II}_{(\mathcal{N},N,D)}^{A, \; \chi = -1} =
    \frac{(-q;q)_{\infty} (\pm q^{1+r/2} a; q)_{\infty}}{(\pm q^{r/2} a; q)_{\infty}}
    = & \frac{(-q;q)_{\infty}}{(1 - q^{r/2} a) (1 + q^{r/2} a)}
     = \frac{(-q;q)_{\infty}}{(1 - q^r a^2)}
    \nonumber \\
    = & \frac{(q^{1 + r} a^2; q)_{\infty} (-q; q)_{\infty}}{(q^r a^2; q)_{\infty}}
     = \mathbb{II}_{(N,D,D)}^{B, \; \chi = -1} \; .
\end{align}

For the case of $N=4$, $N_f = 3$ we have
\begin{align}
\label{so4nf3chi_M1_halfA}
\mathbb{II}_{(\mathcal{N},N,D)}^{A, \; \chi = -1}
 = &
\frac{(q)_{\infty} (-q; q)_{\infty}}{2} \oint \frac{ds}{2\pi i s}
(s^{\pm}; q)_{\infty} (-s^{\pm}; q)_{\infty}
 \nonumber \\
 &\times
\frac{(\pm q^{1+3r/2} a^3; q)_{\infty} (-q^{1+3r/2} s^{\pm} a^3; q)_{\infty}}{\prod_{\alpha = 1}^3 (\pm q^{r/2} a x_{\alpha}; q)_{\infty} \left( (q^{r/2} s^{\pm} a x_{\alpha}; q)_{\infty} \right)},
 \\
\label{so4nf3chi_M1_halfB}
\mathbb{II}_{(N,D,D)}^{B, \; \chi = -1} = &
 \frac{(q^{1+3r} a^6; q)_{\infty} \prod_{\alpha=1}^3 (-q^{1+r} a^2 x_{\alpha}^{-1}; q)_{\infty}}{\prod_{\alpha \le \beta} (q^r a^2 x_{\alpha} x_{\beta}; q)_{\infty}}. 
\end{align}

For example, turning off the fugacity $x_{\alpha}$ and setting $r=\frac14$, 
the half-indices (\ref{so4nf3chi_M1_halfA}) and (\ref{so4nf3chi_M1_halfB}) are expanded as 
\begin{align}
&\mathbb{II}_{(\mathcal{N},N,D)}^{A, \; \chi = -1}
=
\mathbb{II}_{(N,D,D)}^{B, \; \chi = -1}
\nonumber\\
&=1+6a^2q^{1/4}+21a^4q^{1/2}+56a^6q^{3/4}
+126a^8q+9(a^2+28a^{10})q^{5/4}
+6a^4(9+77a^8)q^{3/2}
\nonumber\\
&+4a^6(47+198a^8)q^{7/4}
+3a^8(166+429a^8)q^2
+a^2(9+1113a^8+2002a^{16})q^{9/4}+\cdots
\end{align}

\subsection{$SU(N)$ with rank-$2$ antisymmetric, $4$ fundamental and $N-1 (+ 1)$ antifundamental chirals}
\label{sec_New_SUN_integral_AS}
In this subsection we consider theory A with $SU(N)$ gauge group, 
an antisymmetric chiral $\Phi$, $N_f = 4$ fundamental chirals $Q_{I}$ and $N_a = N - 1$ antifundamental chirals $\overline{Q}_{\alpha}$ with R-charge $0$. 
These all have Neumann boundary conditions. 
There is an additional antifundamental chiral $\widetilde{Q}$ with Dirichlet boundary condition. 
We have the boundary 't Hooft anomaly 
\begin{align}
\label{bdy_SUN_AS_4_Nm1_anomA}
\Acal^A & = \underbrace{N\Tr(s^2) + \frac{N^2 - 1}{2}r^2}_{\textrm{VM}, \; \Ncal}
 - \underbrace{\left( 2\Tr(s^2) + \frac{N}{2}\Tr(x^2) + 2N(a-r)^2 \right)}_{Q_I, \; N}
  \nonumber \\
 & - \underbrace{\left( \frac{N-1}{2}\Tr(s^2) + \frac{N}{2}\Tr(\tilde{x}^2) + \frac{N(N-1)}{2}(b-r)^2 \right)}_{\overline{Q}_{\alpha}, \; N}
  \nonumber \\
 & - \underbrace{\left( \frac{N-2}{2}\Tr(s^2) + \frac{N(N-1)}{4}(A - r)^2 \right)}_{\Phi, \; \Ncal}
  \nonumber \\
 & + \underbrace{\left( \frac{1}{2}\Tr(s^2) + \frac{N}{2}\big( (2 - N)A - 4a - (N-1)b + r \big)^2 \right)}_{\widetilde{Q}, \; D}
  \nonumber \\
  &=  - \frac{N}{2} \Tr(x^2) - \frac{N}{2} \Tr(\tilde{x}^2) + 6N a^2 + 4N(N-2) Aa + N(N-2) (N-1) Ab
  \nonumber \\
  & - \frac{1}{2} (N-3) N Ar + 4N (N-1) ab + \frac{1}{4} N(N-3)(2N-3))A^2
  \nonumber \\
  & + \frac{1}{2} N (N-1)(N-2)b^2 - \frac{1}{4} (N+1)(N+2) r^2
\end{align}
for theory A.
 
The half-index is evaluated as
\begin{align}
\label{alt_SUN_even_AS_hindexA}
&
\mathbb{II}_{(\mathcal{N},N,N,N,D)}^{A}
=
\frac{(q)_{\infty}^{N-1}}{N!} \prod_{i=1}^{N-1} \oint \frac{ds_i}{2\pi i s_i}
\prod_{i \ne j}^N (s_i s_j^{-1}; q)_{\infty} 
\nonumber \\
&\times \prod_{i = 1}^N \frac{(q^{((N-2)r_A + 4r_a + (N-1)r_b)/2} A^{N-2} a^4 b^{N-1} s_i; q)_{\infty}}{\left( \prod_{\alpha = 1}^{N-1} (q^{r_b/2} b s_i^{-1} \tilde{x}_{\alpha}; q)_{\infty} \right) \left( \prod_{I = 1}^4 (q^{r_a/2} a s_i x_I; q)_{\infty} \right) \left( \prod_{i < j}^N (q^{r_A/2} A s_i {s_j}; q)_{\infty} \right)}
\end{align}
where $\prod_{i=1}^N s_i = \prod_{I = 1}^4 x_I = \prod_{\alpha = 1}^{N-1} \tilde{x}_{\alpha} = 1$.
The Neumann half-index (\ref{alt_SUN_even_AS_hindexA}) generalizes the 
Gustafson-Rakha integral (\ref{bdy_SUN_AS_3_N_hindexA}) with a different combination of fundamental and antifundamental chirals. 

For $N = 2n$ the half-index coincides with 
\begin{align}
\label{alt_SUN_even_AS_hindexB1}
&
\frac{\left( q^{((N-2)r_A + 4r_b)/2} A^{N-2} a^4; q \right)_{\infty} \left( \prod_{\alpha = 1}^{N-1} \left( q^{((N-3)r_A + 4r_a + (N-2)r_b)/2} A^{N-3} a^4 b^{N-2} \tilde{x}_{\alpha}^{-1};q \right)_{\infty} \right)}
 {\left( q^{nr_A/2} A^n; q \right)_{\infty} \left( q^{((N-2)r_A + 4r_a)/2} A^{N-2} a^4; q \right)_{\infty} \prod_{I = 1}^4 \prod_{\alpha = 1}^{N-1} (q^{(r_a + r_b)/2} ab x_I \tilde{x}_{\alpha}; q)_{\infty}}
\nonumber \\
&\times \frac{\left( \prod_{I = 1}^4 \left( q^{((N-2)r_A + 3r_a + (N-1)r_b)/2} A^{N-2} a^3 b^{N-1} x_I^{-1};q \right)_{\infty} \right)}
 {\left( \prod_{I < J}^4 \left( q^{((n-1)r_A + 2r_a)/2} A^{n-1} a^2 x_I x_J; q \right)_{\infty} \right) \prod_{\alpha < \beta}^{N-1} (q^{(r_A + 2r_b)/2} A b^2 \tilde{x}_{\alpha} \tilde{x}_{\beta}; q)_{\infty}}. 
\end{align}

For $N=2n+1$ it agrees with 
\begin{align}
\label{alt_SUN_even_AS_hindexB2}
&
\frac{\left( q^{((N-2)r_A + 4r_a)/2} A^{N-2} a^4; q \right)_{\infty} \left( \prod_{\alpha = 1}^{N-1} \left( q^{((N-3)r_A + 4r_a + (N-2)r_b)/2} A^{N-3} a^4 b^{N-2} \tilde{x}_{\alpha}^{-1};q \right)_{\infty} \right)}
 {\left( \prod_{I = 1}^4 \left( q^{(nr_A + r_a)/2} A^n a x_I; q \right)_{\infty} \right) \prod_{I = 1}^4 \prod_{\alpha = 1}^{N-1} (q^{(r_a + r_b)/2} ab x_I \tilde{x}_{\alpha}; q)_{\infty}}
 \nonumber \\
&\times \frac{\left( \prod_{I = 1}^4 \left( q^{((N-2)r_A + 3r_a + (N-1)r_b)/2} A^{N-2} a^3 b^{N-1} x_I^{-1};q \right)_{\infty} \right)}
 {\left( \prod_{I = 1}^4 \left( q^{((n-1)r_A + 3r_a)/2} A^{n-1} a^3 x_I^{-1}; q \right)_{\infty} \right) \prod_{\alpha < \beta}^{N-1} (q^{(r_A + 2r_b)/2} A b^2 \tilde{x}_{\alpha} \tilde{x}_{\beta}; q)_{\infty}}. 
\end{align}
For $SU(2)$, $SU(3)$ and $SU(4)$, the identity is equivalent to previous cases so the
first new example is $SU(5)$ with the identity
\begin{align}
&
\frac{(q)_{\infty}^{4}}{5!} \prod_{i=1}^{4} \oint \frac{ds_i}{2\pi i s_i}
\prod_{i \ne j}^5 (s_i s_j^{-1}; q)_{\infty} 
\nonumber \\
&\times \prod_{i = 1}^5 \frac{(q^{(3r_A + 4r_a + 4r_b)/2} A^3 a^4 b^4 s_i; q)_{\infty}}{\left( \prod_{\alpha = 1}^4 (q^{r_b/2} b s_i^{-1} \tilde{x}_{\alpha}; q)_{\infty} \right) \left( \prod_{I = 1}^4 (q^{r_a/2} a s_i x_I; q)_{\infty} \right) \left( \prod_{i < j}^5 (q^{r_A/2} A s_i {s_j}; q)_{\infty} \right)}
\nonumber \\
&= 
\frac{\left( q^{(3r_A + 4r_a)/2} A^3 a^4; q \right)_{\infty} \left( \prod_{\alpha = 1}^{4} \left( q^{(2r_A + 4r_a + 3r_b)/2} A^{2} a^4 b^{3} \tilde{x}_{\alpha}^{-1};q \right)_{\infty} \right)}
 {\left( \prod_{I = 1}^4 \left( q^{(nr_A + r_a)/2} A^n a x_I; q \right)_{\infty} \right) \prod_{I = 1}^4 \prod_{\alpha = 1}^{4} (q^{(r_a + r_b)/2} ab x_I \tilde{x}_{\alpha}; q)_{\infty}}
\nonumber \\
&\times \frac{\left( \prod_{I = 1}^4 \left( q^{(3r_A + 3r_a + 4r_b)/2} A^{3} a^3 b^{4} x_I^{-1};q \right)_{\infty} \right)}
 {\left( \prod_{I = 1}^4 \left( q^{((n-1)r_A + 3r_a)/2} A^{n-1} a^3 x_I^{-1}; q \right)_{\infty} \right) \prod_{\alpha < \beta}^{4} (q^{(r_A + 2r_b)/2} A b^2 \tilde{x}_{\alpha} \tilde{x}_{\beta}; q)_{\infty}}
\label{New_AS_su5}
\end{align}
For example, for $r_a=r_b=1/2$, $r_A=1/3$ we have checked that the identity (\ref{New_AS_su5}) holds as we have the following $q$-series expansion: 
\begin{align}
&1+16abq^{1/2}+4aA^2q^{7/12}+6Ab^2q^{2/3}+4a^3Aq^{11/12}
\nonumber\\
&+136a^2b^2q+64a^2A^2bq^{13/12}+
(10a^2A^4+96Ab^3)q^{7/6}+\cdots
\end{align}
Note that these half-indices should be convergent for any positive values for $r_A$, $r_a$ and $r_b$.

From the expressions (\ref{alt_SUN_even_AS_hindexB1}) and (\ref{alt_SUN_even_AS_hindexB1}) 
we obtain the confining dual descriptions as theory B. 
Theory B has no gauge group and the chirals depend on whether $N$ is even, i.e.\ $N = 2n$, or odd, i.e.\ $N = 2n + 1$. Note that in both cases there is a
global $SU(4) \times SU(N-1)$ flavor symmetry in theory A, and we can label the
chirals by their representations of this group along with several Abelian charges. 

For the case of $N = 2n$ we have
\begin{align}
\hspace*{-1cm}
\label{alt_SUN_even_AS_charges}
\begin{array}{c|c|c|c|c|c|c|c|c}
& \textrm{bc} & SU(N = 2n) & SU(N_f = 4) & SU(N_a = N - 1) & U(1)_A & U(1)_a & U(1)_b & U(1)_R \\ \hline
\textrm{VM} & \mathcal{N} & {\bf Adj} & {\bf 1} & {\bf 1} & 0 & 0 & 0 & 0 \\
\Phi & \textrm{N} & {\bf N(N-1)/2} & {\bf 1} & {\bf 1} & 1 & 0 & 0 & 0 \\
Q_I & \textrm{N} & {\bf N} & {\bf 4} & {\bf 1} & 0 & 1 & 0 & 0 \\
\overline{Q}_{\alpha} & \textrm{N} & {\bf \overline{N}} & {\bf 1} & {\bf N_a} & 0 & 0 & 1 & 0 \\
\widetilde{Q} & \textrm{D} & {\bf \overline{N}} & {\bf 1} & {\bf 1} & 2 - N & -4 & 1 - N & 2 \\
 \hline
M_{I \alpha} & \textrm{N} & {\bf 1} & {\bf 4} & {\bf N_a} & 0 & 1 & 1 & 0 \\
B & \textrm{N} & {\bf 1} & {\bf 1} & {\bf 1} & n-2 & 4 & 0 & 0 \\
\phi & \textrm{N} & {\bf 1} & {\bf 1} & {\bf 1} & n & 0 & 0 & 0 \\
\overline{M}_{\alpha \beta} & \textrm{N} & {\bf 1} & {\bf 1} & {\bf N_a(N_a-1)/2} & 1 & 0 & 2 & 0 \\
B_{IJ} & \textrm{N} & {\bf 1} & {\bf 6} & {\bf 1} & n-1 & 2 & 0 & 0 \\
\widetilde{M}_I & \textrm{D} & {\bf 1} & {\bf 4} & {\bf 1} & 2 - N & -3 & 1-N & 2 \\
\widetilde{M}_{\alpha} & \textrm{D} & {\bf 1} & {\bf 1} & {\bf N_a} & 3-N & -4 & 2-N & 2 \\
\widetilde{B} & \textrm{D} & {\bf 1} & {\bf 1} & {\bf 1} & 2-N & -4 & 0 & 2 \\
\end{array}
\end{align}

The boundary 't Hooft anomaly for theory B with $N=2n$ is given by
\begin{align}
\label{bdy_SUN_AS_4_Nm1_anomB1}
&\Acal^{B, N=2n} 
\nonumber\\
&=  - \underbrace{\left( \frac{N}{2} \Tr(x^2) + 2\Tr(\tilde{x}^2) + 2N(a+b-r)^2 \right)}_{M_{I \alpha}, \; N}
  - \underbrace{\frac{1}{2}\big( (n-2)A + 4a - r \big)^2}_{B, \; N}
  - \underbrace{\frac{1}{2}(nA-r)^2}_{\phi, \; N}
  \nonumber \\
 & - \underbrace{\left( \frac{N-3}{2} \Tr(\tilde{x}^2) + \frac{(N-1)(N-2)}{4}(A + 2b - r)^2 \right)}_{\overline{M}_{\alpha \beta}, \; N}
  \nonumber \\
 & - \underbrace{\left( \Tr(x^2) + 3\big( (n-1)A + 2a - r \big)^2 \right)}_{B_{IJ}, \; N}
   \nonumber \\
 & + \underbrace{\frac{1}{2}\big( (1-N)A - 4a + r \big)^2}_{\widetilde{B}, \; D}
 + \underbrace{\left( \frac{1}{2} \Tr(\tilde{x}^2) + \frac{N-1}{2}\big( (3-N)A - 4a +(2-N)b + r)^2 \right)}_{\widetilde{M}_{\alpha}, \; D}
  \nonumber \\
  & + \underbrace{\left( \frac{1}{2} \Tr(x^2) + 2\big( (2-N)A - 3a + (1-N)b + r \big)^2 \right)}_{\widetilde{M}_I, \; D}. 
\end{align}
This is equal to the boundary anomaly (\ref{bdy_SUN_AS_4_Nm1_anomA}) for theory A.

We have the operator mapping $M_{I \alpha} \sim Q_I \overline{Q}_{\alpha}$, $B \sim \epsilon \Phi^{n-2} Q_1 Q_2 Q_3 Q_4$, $\phi \sim \epsilon \Phi^n$, $\overline{M}_{\alpha \beta} \sim \Phi \overline{Q}_{\alpha} \overline{Q}_{\beta}$, $B_{IJ} \sim \epsilon \Phi^{n-1} Q_I Q_J$, $\widetilde{M}_I \sim Q_I \widetilde{Q}$, $\widetilde{M}_{\alpha} \sim \Phi \overline{Q}_{\alpha} \widetilde{Q}$ and $\widetilde{B} \sim \epsilon \overline{Q}_1 \cdots \overline{Q}_{N-1} \widetilde{Q}$.

In the case of $N = 2$ the antisymmetric chiral $\Phi$ in theory A is a singlet and we can remove it along with the singlet $\phi$ in theory B. This leaves the same duality as discussed in section~\ref{sec_su2nf3na3}. Also, the case of $N = 4$ is equivalent to the $N = 4$ case in section~\ref{sec_GR_integral_AS} after exchanging the gauge group representations ${\bf 4} \leftrightarrow {\bf \overline{4}}$.

For $N=2n+1$ the field content and charges are 
\begin{align}
\hspace*{-1cm}
\label{alt_SUN_odd_AS_charges}
\begin{array}{c|c|c|c|c|c|c|c|c}
& \textrm{bc} & SU(N = 2n+1) & SU(N_f = 4) & SU(N_a = N - 1) & U(1)_A & U(1)_a & U(1)_b & U(1)_R \\ \hline
\textrm{VM} & \mathcal{N} & {\bf Adj} & {\bf 1} & {\bf 1} & 0 & 0 & 0 & 0 \\
\Phi & \textrm{N} & {\bf N(N-1)/2} & {\bf 1} & {\bf 1} & 1 & 0 & 0 & 0 \\
Q_I & \textrm{N} & {\bf N} & {\bf 4} & {\bf 1} & 0 & 1 & 0 & 0 \\
\overline{Q}_{\alpha} & \textrm{N} & {\bf \overline{N}} & {\bf 1} & {\bf N_a} & 0 & 0 & 1 & 0 \\
\widetilde{Q} & \textrm{D} & {\bf \overline{N}} & {\bf 1} & {\bf 1} & 2 - N & -4 & 1-N & 2 \\
 \hline
M_{I \alpha} & \textrm{N} & {\bf 1} & {\bf 4} & {\bf N_a} & 0 & 1 & 1 & 0 \\
B_I & \textrm{N} & {\bf 1} & {\bf 4} & {\bf 1} & n & 1 & 0 & 0 \\
\overline{M}_{\alpha \beta} & \textrm{N} & {\bf 1} & {\bf 1} & {\bf N_a(N_a-1)/2} & 1 & 0 & 2 & 0 \\
B^I & \textrm{N} & {\bf 1} & {\bf \overline{4}} & {\bf 1} & n-1 & 3 & 0 & 0 \\
\widetilde{M}_I & \textrm{D} & {\bf 1} & {\bf 4} & {\bf 1} & 2-N & -3 & 1-N & 2 \\
\widetilde{M}_{\alpha} & \textrm{D} & {\bf 1} & {\bf 1} & {\bf N_a} & 3-N & -4 & 2-N & 2 \\
\widetilde{B} & \textrm{D} & {\bf 1} & {\bf 1} & {\bf 1} & 2-N & -4 & 0 & 2 \\
\end{array}
\end{align}
In the case of $N = 3$ the antisymmetric chiral $\Phi$ in theory A is a fundamental chiral. 
This leaves the same duality as discussed in section~\ref{sec_GNR_integrals} in the case of $N = 3$.

The 't Hooft anomalies match since 
the boundary anomaly for theory B with $N=2n+1$ is calculated as
\begin{align}
&
\Acal^{B, N=2n+1} 
\nonumber\\
&=  - \underbrace{\left( \frac{N-1}{2} \Tr(x^2) + 2\Tr(\tilde{x}^2) + 2(N-1)(a+b-r)^2 \right)}_{M_{I \alpha}, \; N}
  - \underbrace{\left( \frac{1}{2}\Tr(x^2) + 2\big( nA + a - r \big)^2 \right)}_{B_I, \; N}
  \nonumber \\
 & - \underbrace{\left( \frac{N-3}{2} \Tr(\tilde{x}^2) + \frac{(N-1)(N-2)}{4}(A + 2b - r)^2 \right)}_{\overline{M}_{\alpha \beta}, \; N}
  \nonumber \\
 & - \underbrace{\left( \frac{1}{2}\Tr(x^2) + 2\big( (n-1)A + 3a - r \big)^2 \right)}_{B^I, \; N}
   \nonumber \\
 & + \underbrace{\frac{1}{2}\big( (2-N)A - 4a + r \big)^2}_{\widetilde{B}, \; D}
 + \underbrace{\left( \frac{1}{2} \Tr(\tilde{x}^2) + \frac{N-1}{2}\big( (3-N)A - 4a +(2-N)b + r)^2 \right)}_{\widetilde{M}_{\alpha}, \; D}
  \nonumber \\
  & + \underbrace{\left( \frac{1}{2} \Tr(x^2) + 2\big( (2-N)A - 3a + (1-N)b + r \big)^2 \right)}_{\widetilde{M}_I, \; D}. 
  \end{align}

The operator mapping is $M_{I \alpha} \sim Q_I \overline{Q}_{\alpha}$, $B_I \sim \epsilon \Phi^n Q_I$, $\overline{M}_{\alpha \beta} \sim \Phi \overline{Q}_{\alpha} \overline{Q}_{\beta}$, $B^I \sim \epsilon \Phi^{n-1} Q_J Q_K Q_L \epsilon^{IJKL}$, $\widetilde{M}_I \sim Q_I \widetilde{Q}$, $\widetilde{M}_{\alpha} \sim \Phi \overline{Q}_{\alpha} \widetilde{Q}$ and $\widetilde{B} \sim \epsilon \overline{Q}_1 \cdots \overline{Q}_{N-1} \widetilde{Q}$.

The matching of the half-indices and the anomalies support the following boundary confining dualities: 
\begin{align}
\label{bdy_SUN_AS_4_Nm1_dual1}
&\textrm{$SU(N)$ $+$ antisym. chiral $\Phi$ $+$ $4$ fund. chirals $Q_{I}$}
\nonumber\\
&\textrm{$+$ $N-1$ antifund. chiral $\overline{Q}_{\alpha}$ $+$ $1$ antifund. chiral $\widetilde{Q}$ with b.c. $(\mathcal{N},N,N,N,D)$}
\nonumber\\
&\Leftrightarrow 
\textrm{
an $SU(4)\times SU(N-1)$ bifund. $M_{I \alpha}$ $+$ a singlet $B$ $+$ a singlet $\phi$}
\nonumber\\
&\textrm{$+$ an $SU(N-1)$ antisym. chiral $\overline{M}_{\alpha\beta}$ $+$ an $SU(4)$ antisym. chiral $B_{IJ}$ 
$+$ an $SU(N)$ fund. chiral $\widetilde{M}_{\alpha}$}
\nonumber\\
&\textrm{
$+$ a singlet $\widetilde{B}$ $+$ an $SU(4)$ fund. chiral $\widetilde{M}_I$ with b.c. $(N,N,N,N,N,D,D,D)$}. 
\end{align}
for $N=2n$ and
\begin{align}
\label{bdy_SUN_AS_4_Nm1_dual2}
&\textrm{$SU(N)$ $+$ antisym. chiral $\Phi$ $+$ $4$ fund. chirals $Q_{I}$}
\nonumber\\
&\textrm{$+$ $N-1$ antifund. chiral $\overline{Q}_{\alpha}$ $+$ $1$ antifund. chiral $\widetilde{Q}$ with b.c. $(\mathcal{N},N,N,N,D)$}
\nonumber\\
&\Leftrightarrow 
\textrm{
an $SU(4)\times SU(N-1)$ bifund. $M_{I \alpha}$ $+$ an $SU(4)$ fund. $B_I$}
\nonumber\\
&\textrm{$+$ an $SU(N-1)$ antisym. chiral $\overline{M}_{\alpha\beta}$ $+$ an $SU(4)$ antifund. $B^I$ 
$+$ an $SU(N-1)$ fund. chiral $\widetilde{M}_{\alpha}$}
\nonumber\\
&\textrm{
$+$ an $SU(4)$ fund. $\widetilde{M}_I$ $+$ a singlet $\widetilde{B}$ with b.c. $(N,N,N,N,D,D,D)$}. 
\end{align}
for $N=2n+1$.

These boundary confining dualities (\ref{bdy_SUN_AS_4_Nm1_dual1}) and (\ref{bdy_SUN_AS_4_Nm1_dual2}) are extended to the bulk with superpotentials
\begin{align}
    W^{N = 2n} = & \widetilde{M}_{\alpha} \left( B \overline{M}_{\alpha_1 \beta_1} - \epsilon^{IJKL} M_{I \alpha_1} M_{J \beta_1} B_{KL} \right) \overline{M}_{\alpha_2 \beta_2} \cdots \overline{M}_{\alpha_{n-1} \beta_{n-1}} \epsilon^{\alpha \alpha_1 \beta_1 \cdots \alpha_{n-1} \beta_{n-1}}
    \nonumber \\
    & + \widetilde{M}_I \epsilon^{IJKL} M_{J \alpha} \left( B_{KL} \overline{M}_{\alpha_1 \beta_1} - \phi M_{K \alpha_1} M_{L \beta_1} \right) \overline{M}_{\alpha_2 \beta_2} \cdots \overline{M}_{\alpha_n \beta_n} \epsilon^{\alpha \alpha_1 \beta_1 \cdots \alpha_{n-1} \beta_{n-1}}
    \nonumber \\
    & + \widetilde{B} \left( B \phi - \epsilon^{IJKL} B_{IJ} B_{KL} \right)
    \\
    W^{N = 2n+1} = & \widetilde{M}_I \left( B^I \overline{M}_{\alpha_1 \beta_1} - \epsilon^{IJKL} B_J M_{K \alpha_1} M_{L \beta_1} \right) \overline{M}_{\alpha_2 \beta_2} \cdots \overline{M}_{\alpha_n \beta_n} \epsilon^{\alpha_1 \beta_1 \cdots \alpha_n \beta_n}
    \nonumber \\
    & + \widetilde{M}_{\alpha_1} \left( B^I M_{I \beta_1} \overline{M}_{\alpha_2 \beta_2} - \epsilon^{IJKL} B_I M_{J \beta_1} M_{K \alpha_2} M_{L \beta_2} \right) \overline{M}_{\alpha_3 \beta_3} \cdots \overline{M}_{\alpha_n \beta_n} \epsilon^{\alpha_1 \beta_1 \cdots \alpha_n \beta_n}
    \nonumber \\
    & + \widetilde{B} B^I B_I
\end{align}

In fact, this boundary duality corresponds to the same bulk duality described in section~\ref{sec_GR_integral_AS} subject to a redefinition of the global charges. The labelling of the chiral multiplets looks different but only because we presented the chirals in terms of the global symmetries with the specific boundary conditions.

\subsection*{Acknowledgements}
The authors would like to thank Stefano Cremonesi for useful discussions and comments. 
The work of T.O. was supported by the Startup Funding no. 4007012317 of the Southeast University. 
The work of D.S. was supported in part by the STFC Consolidated grant ST/T000708/1.

\bibliographystyle{utphys}
\bibliography{ref}

\end{document}